%% file: main.tex
\documentclass[twocolumn]{aastex701}
 
\usepackage{rotating}
\usepackage{xcolor}
\usepackage{amsmath}
\usepackage{xspace}
\usepackage{tabulary}
\usepackage{multirow}
\usepackage{hyperref}
\usepackage{caption}
\usepackage{makecell}
\usepackage{multirow}

\usepackage[utf8]{inputenc}
\usepackage{booktabs}
\usepackage{textcomp}
\usepackage{enumitem}

\usepackage{tabularx}
\usepackage{array}
\usepackage{etoolbox}
\makeatletter
\def\countcols#1{%
  \ifcat#1c%
    \global\advance\colcount by 1\relax
  \fi
  \def\lookforD{#1}%
  \ifx\lookforD\xD
    \global\advance\howmanyDs by 1\relax
    \global\advance\colcount by 1\relax
  \fi
  \futurelet\next\lookatnext}
\patchcmd{\@array@array@new}{\def\catchpreamble{#2\relax} }{\def\catchpreamble{#2\relax}}{}{\PackageError{table-spacing}{Cannot patch catchpreamble}{}}
\patchcmd{\@array@array@new}{\gdef\tabfullpreamble{#2\relax} }{\gdef\tabfullpreamble{#2\relax}}{}{\PackageError{table-spacing}{Cannot patch tabfullpreamble}{}}
\patchcmd{\@array@array@new}{\def\two{#2} }{\def\two{#2}}{}{\PackageError{table-spacing}{Cannot patch column comparison}{}}
\patchcmd{\@array@array@new}
  {{\colcount=0 \def\two{#2}\ifx\matrixpreamble\two\else\expandafter\countcols\catchpreamble\global\totalcolumns=\colcount\fi} }
  {{\colcount=0 \def\two{#2}\ifx\matrixpreamble\two\else\expandafter\countcols\catchpreamble\global\totalcolumns=\colcount\fi}}
  {}{\PackageError{table-spacing}{Cannot patch column-count group}{}}
\makeatother
\usepackage{booktabs}
\usepackage{tikz}
\usetikzlibrary{shapes.geometric, arrows.meta, positioning, calc}

\usepackage{natbib}
\defcitealias{Lin2025_localLRD}{Paper~I}

\newcommand{\ha}{H$\alpha$}
\newcommand{\hb}{H$\beta$}

\newcommand{\oiii}{[\ion{O}{3}]$\lambda\lambda$ 4960, 5008}
\newcommand{\nii}{[\ion{N}{2}]$\lambda\lambda$6549, 6585}

\usepackage[encapsulated]{CJK}
\usepackage[section]{placeins}

\newcommand{\sedpair}[2]{%
    \noindent
    \includegraphics[width=0.49\textwidth]{#1}%
    \hfill
    \includegraphics[width=0.49\textwidth]{#2}%
    \par
    \vspace{0.5em}
}

\received{{2026 May 20}}
\revised{{2026 August 27}}
\accepted{{2026 September 22}}

\journalinfo{Accepted for publication in The Astrophysical Journal}

\shorttitle{DESI DR1 LRDs}
\shortauthors{Lin et al.}

\graphicspath{{./}{figures/}}

\begin{document}
    
\title{(LRDs)$^2$: The Low-ReDshift Little Red Dots Survey. \\ \uppercase\expandafter{\romannumeral2}.   DESI DR1 Sample}

\input{00_Authorship}

\begin{abstract}
JWST has revealed a substantial population of “Little Red Dots” (LRDs) at $z>4$, challenging conventional {active galactic nuclei (AGN)} frameworks. However, the low-redshift regime remains largely unexplored. In the second paper of the (LRDs)$^2$ series, we present a systematic selection from DESI DR1 and identify 27 LRDs at $z=0.2$–$0.9$, yielding a number density lower limit  of $7.5 \times 10^{-10}\,\mathrm{cMpc}^{-3}$. We conducted near-IR spectroscopic follow-up observations for 21 of them, revealing their full {spectral energy distribution (SED)} shapes and emission lines. These low-$z$ LRDs share the hallmark properties of their high-$z$ counterparts: compact morphology, V-shaped UV-optical continua, broad Balmer emission with extreme decrements (median ${\rm H\alpha/H\beta}\sim15$), frequent Balmer absorption (70\%), and blackbody-like optical-to-near-IR continua. All have low metallicity, occupy the same regions in the BPT diagram as high-$z$ LRDs, and have softer ionizing spectra than typical AGNs.  The consistency between low-$z$ and high-$z$ LRD properties indicates the same physical processes at work. The correlation between broad-line Balmer luminosity and $L_{\rm 5100}$ deviates from that of local type-1 AGNs, limiting the direct application of local BH mass calibrations. Ionized [\ion{O}{3}] outflows are ubiquitous (78\%). One LRD at $z=0.196$, J1717+3807, shows robust long-term variability in $i$ and {the Wide-field Infrared Survey Explorer (WISE)} bands. The {optical-to-near-IR} continua of LRDs reveal a wide range of temperatures $\sim$2000–4700 K (peak $0.6$–$1.4\,\micron$), with a subset showing cooler and larger envelopes than those at high $z$. Low-$z$ LRDs serve not only as proximate laboratories for probing the nature of LRDs, but also trace the cosmic evolution of this population from the cosmic dawn to the present day.
\end{abstract}


\input{01_Introduction}

\input{02_Selection_and_Sample}

\input{03_Measurement}

\input{04_Results}

\input{05_Summary}

\section*{Data Availability}

The DESI DR1 spectra and the measurements presented in this work are publicly available on Zenodo at
{\url{https://doi.org/10.5281/zenodo.20309302}}.

\begin{acknowledgments}

We thank Xihan Ji and Franz Bauer for constructive and friendly discussions. We thank the LCO, LBT, and Keck staff for their kind support during our observing runs.  We thank Jenny Powers for her expert assistance with LBT observations and Greg Doppmann for his expert assistance with Keck observations. 

This research used data obtained with the Dark Energy Spectroscopic Instrument (DESI). DESI construction and operations are managed by the Lawrence Berkeley National Laboratory. This material is based upon work supported by the U.S. Department of Energy, Office of Science, Office of High-Energy Physics, under {contract No.} DE–AC02–05CH11231, and by the National Energy Research Scientific Computing Center, a DOE Office of Science User Facility under the same contract. Additional support for DESI was provided by the U.S. National Science Foundation (NSF), Division of Astronomical Sciences under {contract No.} AST-0950945 to the NSF’s National Optical-Infrared Astronomy Research Laboratory; the Science and Technology Facilities Council of the United Kingdom; the Gordon and Betty Moore Foundation; the Heising-Simons Foundation; the French Alternative Energies and Atomic Energy Commission (CEA); the National Council of Humanities, Science and Technology of Mexico (CONAHCYT); the Ministry of Science and Innovation of Spain (MICINN), and by the DESI Member Institutions: www.desi.lbl.gov/collaborating-institutions. The DESI collaboration is honored to be permitted to conduct scientific research on I’oligam Du’ag (Kitt Peak), a mountain with particular significance to the Tohono O’odham Nation. Any opinions, findings, and conclusions or recommendations expressed in this material are those of the author(s) and do not necessarily reflect the views of the U.S. National Science Foundation, the U.S. Department of Energy, or any of the listed funding agencies.

The Legacy Surveys consist of three individual and complementary projects: the Dark Energy Camera Legacy Survey (DECaLS; Proposal ID \#2014B-0404; PIs: David Schlegel and Arjun Dey), the Beijing-Arizona Sky Survey (BASS; NOAO Prop. ID \#2015A-0801; PIs: Zhou Xu and Xiaohui Fan), and the Mayall \textit{z}-band Legacy Survey (MzLS; Prop. ID \#2016A-0453; PI: Arjun Dey). DECaLS, BASS, and MzLS together include data obtained, respectively, at the Blanco telescope, Cerro Tololo Inter-American Observatory, NSF’s NOIRLab; the Bok telescope, Steward Observatory, University of Arizona; and the Mayall telescope, Kitt Peak National Observatory, NOIRLab. Pipeline processing and analyses of the data were supported by NOIRLab and the Lawrence Berkeley National Laboratory (LBNL). The Legacy Surveys project is honored to be permitted to conduct astronomical research on Iolkam Du’ag (Kitt Peak), a mountain with particular significance to the Tohono O’odham Nation.

NOIRLab is operated by the Association of Universities for Research in Astronomy (AURA) under a cooperative agreement with the National Science Foundation. LBNL is managed by the Regents of the University of California under contract to the U.S. Department of Energy.

This project used data obtained with the Dark Energy Camera (DECam), which was constructed by the Dark Energy Survey (DES) collaboration. Funding for the DES Projects has been provided by the U.S. Department of Energy, the U.S. National Science Foundation, the Ministry of Science and Education of Spain, the Science and Technology Facilities Council of the United Kingdom, the Higher Education Funding Council for England, the National Center for Supercomputing Applications at the University of Illinois at Urbana-Champaign, the Kavli Institute of Cosmological Physics at the University of Chicago, the Center for Cosmology and Astro-Particle Physics at the Ohio State University, the Mitchell Institute for Fundamental Physics and Astronomy at Texas A\&M University, Financiadora de Estudos e Projetos, Fundação Carlos Chagas Filho de Amparo à Pesquisa do Estado do Rio de Janeiro, Conselho Nacional de Desenvolvimento Científico e Tecnológico, and the Ministério da Ciência, Tecnologia e Inovação, the Deutsche Forschungsgemeinschaft, and the Collaborating Institutions in the Dark Energy Survey. The Collaborating Institutions are Argonne National Laboratory, the University of California at Santa Cruz, the University of Cambridge, Centro de Investigaciones Energéticas, Medioambientales y Tecnológicas-Madrid, the University of Chicago, University College London, the DES-Brazil Consortium, the University of Edinburgh, the Eidgenössische Technische Hochschule (ETH) Zürich, Fermi National Accelerator Laboratory, the University of Illinois at Urbana-Champaign, the Institut de Ciències de l’Espai (IEEC/CSIC), the Institut de Física d’Altes Energies, Lawrence Berkeley National Laboratory, the Ludwig-Maximilians-Universität München and the associated Excellence Cluster Universe, the University of Michigan, NSF’s NOIRLab, the University of Nottingham, the Ohio State University, the University of Pennsylvania, the University of Portsmouth, SLAC National Accelerator Laboratory, Stanford University, the University of Sussex, and Texas A\&M University.

BASS is a key project of the Telescope Access Program (TAP), which has been funded by the National Astronomical Observatories of China, the Chinese Academy of Sciences (the Strategic Priority Research Program ``The Emergence of Cosmological Structures'', {grant \#}XDB09000000), and the Special Fund for Astronomy from the Ministry of Finance. BASS is also supported by the External Cooperation Program of Chinese Academy of Sciences ({grant \#}114A11KYSB20160057), and the Chinese National Natural Science Foundation ({grant \#}12120101003, \#11433005).

The Legacy Survey team makes use of data products from the Near-Earth Object Wide-field Infrared Survey Explorer (NEOWISE), which is a project of the Jet Propulsion Laboratory/California Institute of Technology. NEOWISE is funded by the National Aeronautics and Space Administration.

The Legacy Surveys imaging of the DESI footprint is supported by the Director, Office of Science, Office of High Energy Physics of the U.S. Department of Energy under {contract No.} DE-AC02-05CH1123, by the National Energy Research Scientific Computing Center, a DOE Office of Science User Facility under the same contract, and by the U.S. National Science Foundation, Division of Astronomical Sciences under {contract No.} AST-0950945 to NOAO.

This research uses services or data provided by the Astro Data Lab, which is part of the Community Science and Data Center (CSDC) Program of NSF NOIRLab. NOIRLab is operated by the Association of Universities for Research in Astronomy (AURA), Inc. under a cooperative agreement with the U.S. National Science Foundation.

This research is based on observations made with the Galaxy Evolution Explorer, obtained from the MAST data archive at the Space Telescope Science Institute, which is operated by the Association of Universities for Research in Astronomy, Inc., under NASA contract NAS 5–26555.

This work has made use of data from the European Space Agency (ESA) mission
{\it Gaia} (\url{https://www.cosmos.esa.int/gaia}), processed by the {\it Gaia}
Data Processing and Analysis Consortium (DPAC,
\url{https://www.cosmos.esa.int/web/gaia/dpac/consortium}). Funding for the DPAC
has been provided by national institutions, in particular the institutions
participating in the {\it Gaia} Multilateral Agreement.

This work is based in part on data obtained as part of the UKIRT Infrared Deep Sky Survey.

The Pan-STARRS1 Surveys (PS1) and the PS1 public science archive have been made possible through contributions by the Institute for Astronomy, the University of Hawaii, the Pan-STARRS Project Office, the Max-Planck Society and its participating institutes, the Max Planck Institute for Astronomy, Heidelberg and the Max Planck Institute for Extraterrestrial Physics, Garching, The Johns Hopkins University, Durham University, the University of Edinburgh, the Queen's University Belfast, the Harvard-Smithsonian Center for Astrophysics, the Las Cumbres Observatory Global Telescope Network Incorporated, the National Central University of Taiwan, the Space Telescope Science Institute, the National Aeronautics and Space Administration under {grant No.} NNX08AR22G issued through the Planetary Science Division of the NASA Science Mission Directorate, the National Science Foundation {grant No.} AST-1238877, the University of Maryland, Eotvos Lorand University (ELTE), the Los Alamos National Laboratory, and the Gordon and Betty Moore Foundation.

This publication makes use of data products from the Spectro-Photometer for the History of the Universe, Epoch of Reionization and Ices Explorer (SPHEREx), which is a joint project of the Jet Propulsion Laboratory and the California Institute of Technology, and is funded by the National Aeronautics and Space Administration.

This paper includes data gathered with the 6.5 meter Magellan Telescopes located at Las Campanas Observatory, Chile.

The LBT is an international collaboration among institutions in the United States, {Italy, and Germany}. LBT Corporation partners are: The University of Arizona on behalf of the Arizona university system; Istituto Nazionale di Astrofisica, Italy; LBT Beteiligungsgesellschaft, Germany, representing the Max-Planck Society, the Astrophysical Institute Potsdam, and Heidelberg University; The Ohio State University, and The Research Corporation, on behalf of The University of Notre Dame, University of Minnesota and University of Virginia.

Some of the data presented herein were obtained at Keck Observatory, which is a private 501(c)3 {nonprofit} organization operated as a scientific partnership among the California Institute of Technology, the University of California, and the National Aeronautics and Space Administration. The Observatory was made possible by the generous financial support of the W. M. Keck Foundation.

The authors wish to recognize and acknowledge the very significant cultural role and reverence that the summit of Maunakea has always had within the Native Hawaiian community. We are most fortunate to have the opportunity to conduct observations from this mountain.
\end{acknowledgments}

\facility{Mayall (DESI), Mayall (Mosaic-3), Blanco
(DECam), Bok (90Prime), GALEX, {PS1}, Gaia, WISE, NEOWISE,  Magellan:Baade (FIRE), LBT (LUCI), Keck:II (NIRES), SPHEREx, Astro Data Lab}

\input{99_Appendix}

\bibliography{main}{}
\bibliographystyle{aasjournalv7}

\end{document}

%% file: 00_Authorship.tex
\author[0000-0001-6052-4234]{Xiaojing Lin}
\affiliation{Department of Astronomy, Tsinghua University, Beijing 100084, {People\textquotesingle s Republic of China}}
\email[show]{xiaojinglin.astro@gmail.com}

\author[0000-0003-3310-0131]{Xiaohui Fan}
\affiliation{Steward Observatory, University of Arizona, 933 N. Cherry {Avenue}, Tucson, AZ 85721, USA}
\email[]{xfan@arizona.edu}

\author[0000-0001-8467-6478]{Zheng Cai}
\affiliation{Department of Astronomy, Tsinghua University, Beijing 100084, {People\textquotesingle s Republic of China}}
\email[]{zcai@tsinghua.edu.cn}

\author[0000-0003-4247-0169]{Yichen Liu}
\affiliation{Steward Observatory, University of Arizona, 933 N. Cherry {Avenue}, Tucson, AZ 85721, USA}
\email{yichenliu@arizona.edu}

\author[0000-0002-4622-6617]{Fengwu Sun}
\affiliation{Center for Astrophysics $|$ Harvard \& Smithsonian, 60 Garden {Street}, Cambridge, MA 02138, USA}
\email[]{}

\author[0000-0002-1620-0897]{Fuyan Bian}
\affiliation{European Southern Observatory, Alonso de C\'ordova 3107, Casilla 19001, Vitacura, Santiago 19, Chile}
\affiliation{Chinese Academy of Sciences South America Center for Astronomy, National Astronomical Observatories, CAS, Beijing 100101, {People\textquotesingle s Republic of China}}
\email[]{}

\author[0000-0001-6251-649X]{Mingyu Li}
\affiliation{Department of Astronomy, Tsinghua University, Beijing 100084, {People\textquotesingle s Republic of China}}
\email[]{}

\author[0000-0001-7557-9713]{Junjie Mao}
\affiliation{Department of Astronomy, Tsinghua University, Beijing 100084, {People\textquotesingle s Republic of China}}
\email[]{}

\author[0000-0002-5612-3427]{Jenny E. Greene}
\affiliation{Department of Astrophysical Sciences, Princeton University, 4 Ivy Lane, Princeton, NJ 08544, USA}
\email[]{}

\author[0000-0003-2488-4667]{Hanpu Liu}
\affiliation{Department of Astrophysical Sciences, Princeton University, 4 Ivy Lane, Princeton, NJ 08544, USA}
\email[]{}

\author[0000-0001-9592-4190]{Jiaxuan Li (\begin{CJK*}{UTF8}{gbsn}李嘉轩\ignorespacesafterend\end{CJK*})}
\affiliation{Department of Astrophysical Sciences, Princeton University, 4 Ivy Lane, Princeton, NJ 08544, USA}
\email[]{}

\author[0000-0003-3762-7344]{Weizhe Liu \begin{CJK}{UTF8}{gbsn}(刘伟哲)\end{CJK}}
\affiliation{Steward Observatory, University of Arizona, 933 N. Cherry {Avenue}, Tucson, AZ 85721, USA}
\email[]{}

\author[0000-0002-0463-9528]{Yilun Ma (\begin{CJK*}{UTF8}{gbsn}马逸伦\ignorespacesafterend\end{CJK*})}
\affiliation{Department of Astrophysical Sciences, Princeton University, 4 Ivy Lane, Princeton, NJ 08544, USA}
\email[]{}

\author[0000-0002-8246-7792]{Ze{c}hang Sun}
\affiliation{Department of Astronomy, Tsinghua University, Beijing 100084, {People\textquotesingle s Republic of China}}
\email[]{}

\author[0000-0002-2420-5022]{Zijian Zhang}
\affiliation{Kavli Institute for Astronomy and Astrophysics, Peking University, Beijing 100871, {People\textquotesingle s Republic of China}}
\affiliation{Department of Astronomy, School of Physics, Peking University, Beijing 100871, {People\textquotesingle s Republic of China}}
\email[]{}

%% file: 01_Introduction.tex
\section{Introduction}

``Little Red Dots” (LRDs), first identified at $z>4$ during early JWST operations \citep[e.g.,][]{Labbe2025,  Matthee2024, Greene2024}, have emerged as one of the most intriguing and puzzling discoveries in recent years. These sources exhibit compact morphologies and broad Balmer emission lines with FWHMs exceeding $1000\ \mathrm{km\ s^{-1}}$, which likely originate from broad-line regions (BLRs) in the vicinity of black holes (BHs) in the galactic nuclei {\citep[e.g.,][]{Matthee2024, Lin2024, Lin2025, Zhang2025, Kocevski2025, Hviding2025}}. They also show unusual V-shaped spectral energy distributions (SEDs), characterized by blue UV continua and red optical continua \citep[e.g., ][]{Greene2024, Furtak2024}, declining near-infrared continua {\citep[e.g.,][]{Perez-Gonzalez2024, Akins2024, Setton2025}}, frequent Balmer absorption superimposed on the broad emission lines \citep[e.g.,][]{Lin2024, Kocevski2024}, and, in some cases, extremely strong Balmer breaks that cannot be explained by normal stellar populations {\citep[e.g.,][]{Setton2024, Wang2024, deGraaff2025, Naidu2025}}. Together, these enigmatic properties have sparked intense debate on the physical nature of this population. LRDs constitute a significant fraction of the galaxy population at {high redshift} {\citep{Lin2024, Lin2025, Zhang2025}}. Their number densities are 10–100 times higher than the faint-end extrapolations of the high-$z$ quasar luminosity function at $z
\gtrsim5$ {\citep[e.g.,][]{Kokorev2024, Kocevski2025, Akins2024}}. If LRDs are truly powered by BHs, they must represent a common and distinct phase of BH growth in the early Universe.

The distinctive features observed in the UV–optical spectra of LRDs have motivated extensive multiwavelength follow-up observations. These studies further reveal that LRDs differ markedly from previously known {active galactic nuclei (AGNs)}. LRDs are generally faint in the mid and far infrared {\citep{Akins2024, Perez-Gonzalez2024, Williams2024}}, submillimeter {\citep{Casey2024, Casey2025, Setton2025}}, X-ray \citep{Yue2024, Ananna2024, Sacchi2025}, and radio  \citep{Mazzolari2024, Perger2025}, and most show little variability on rest-frame timescales of months {\citep[][]{Kokubo2024, ZZhang2025, ZLiu2026}}. Only a small fraction exhibits detectable variability, either on year-long or even century-long timescales {\citep{Tee25, Furtak2025, Ji2025, ZZhang2025_century}}. Significant efforts have also been devoted to constraining their total bolometric luminosities {\citep{Greene2025}}, host-galaxy properties {\citep[e.g.,][]{CH-Chen2025, Zhang2025_cweb_host, Sun2026_bhstar}}, and halo masses \citep[e.g.,][]{Arita2025, Carranza-Escudero2025, Pizzati2025, Matthee2025, Lin2026}. Current evidence suggests that these quantities are modest compared to those of quasars at similar redshifts.  
One of the most debated topics concerns BH mass measurements. It remains uncertain whether scattering in dense gas contributes to the broadening of the Balmer lines {\citep{Rusakov2025, Brazzini2025, Chang2026}}, and whether empirical single-epoch relations can reliably estimate BH masses (typically yielding $10^6-10^9~M_\odot$). All of these issues are crucial for uncovering the nature of the central engine in LRDs.

To explain the puzzling observed properties of LRDs, a variety of theoretical models have been proposed. Early interpretations invoked {ultramassive} host galaxies \citep{Labbe2024}, dusty AGNs \citep{Kocevski2024, Greene2024, Li2025}, or young nuclear stellar clusters \citep{Baggen2024, Carranza-Escudero2025}. More recent models explore accretion disks that emit blackbody-like continua {\citep{Inayoshi2025b, Zwick2025, CX-Zhang2026,YX-Chen2026}} or BHs embedded in dense gas. The latter scenario has evolved rapidly, from a simple dense shell of hydrogen gas in front of an AGN {\citep{Inayoshi2024, Ji2025}}, to a cocoon-like or atmosphere-like envelope that enshrouds BHs {\citep{Rusakov2025, Kido2025, Liu2025, Inayoshi2025, Asada2026}}, often referred to as a ``BH-star'' {\citep{Naidu2025, deGraaff2025_all}}, and further to clumpy envelope structures \citep{Tang2026_SPURS_LRDs, Ji2026_SPURS_LRDs}. Many theoretical studies interpret LRDs as undergoing super-Eddington accretion, representing a crucial phase in the growth of supermassive BHs {\citep[e.g.,][]{Pacucci2024, Liu2025, Vaida2025, Begelman2026, Madau2026}}.

 With extensive JWST observations of $z>4$ LRDs now accumulated, in-depth demographic analyses are underway to characterize their population-level properties {\citep[e.g.,][]{Barro2025, deGraaff2025_all, Perez-Gonzalez2026}}. At the same time, searches for LRDs at cosmic noon and lower redshifts have yielded significant new discoveries. Systematic searches using wide-field imaging surveys from both space and the ground have provided strong constraints on the LRD luminosity function and number density from $z=5$ down to $z \sim 1$ {\citep{Ma2025, Ma2025_cutoff, LRD_EuclidCollaboration2025}}. The discovery of local LRDs at $z\lesssim0.3$ suggests that the conditions giving rise to LRDs, while rare in the local Universe, are not exclusive to the early Universe {\citep{RLin2025, Lin2025_localLRD, Ji2025_lord}}. Thanks to their proximity and brightness, these local systems enable observations that are challenging to obtain at high redshift, including high-resolution spectroscopy spanning the full UV-to-mid-IR wavelength range, as well as resolved emission-line profiles and absorption features that directly probe the structure and physical conditions of the circum-BH environment.  Transition-phase candidates that may bridge LRDs and conventional UV-luminous AGN have been identified at $z \sim 3$, though their interpretation remains debated {\citep{Fu2025_transition, Hviding2026_xdot}}. Most recently, two cosmic noon LRDs have been reported to exhibit water absorption features, indicating cool BH envelopes ($\lesssim$3000–4000 K) and motivating the development of new theoretical models \citep{Wang2026}. 
 
 All these low-$z$ advances begin to reveal substantial diversity within the LRD population, spanning a wide range of SED shapes, BH envelope temperatures, and host-galaxy environments. These findings extend beyond what current theoretical frameworks have anticipated. They also open a unique window into the cosmic evolution of LRDs. Only by tracing their properties from $z > 6$ to $z \sim 0$ can we determine whether LRDs represent a transient yet universal phase of BH growth, how the physical conditions in their circum-BH environments evolve with cosmic time, and whether they eventually transition into conventional AGNs. However, the current low-$z$ sample remains small. Most low-$z$ LRDs in the JWST archive and those selected and confirmed from wide-field surveys lie at $z=2$–$3$ {\citep{Juodzbalis2024, Ma2025, Loiacono2025, Wang2026}}.  Only a few LRDs are reported at $z<0.5$ {\citep{RLin2025, Lin2025_localLRD, Ji2025_lord, Park2026}}. Some other low-$z$ AGNs exhibiting partially LRD-like features have been proposed as LRD analogs {\citep[e.g.,][]{Ding2026, Bao2026}},  but several of their properties still differ markedly from those of JWST LRDs (e.g., morphology and emission-line properties).  Expanding the low-$z$ sample, particularly at $z<2$ with characteristics fully consistent with those of JWST-discovered LRDs, is therefore crucial for bridging the gap between the local and high-redshift Universe.

This series of papers aims to conduct \textit{{Low-ReDshift LRD surveys (LRDs)$^2$}} with systematic follow-up observations. 
In Paper $\mathrm{I}$, \cite{Lin2025_localLRD} conducted pilot studies on a small sample of {Sloan Digital Sky Survey (SDSS)-selected} local LRDs. Built upon \citetalias{Lin2025_localLRD}, we aim to construct a well-characterized sample of low-$z$ LRDs that captures the diversity and demographics of this population, complementing the rich high-$z$ JWST LRD archive. Our goal is to characterize the cosmic evolution of LRDs, revealing their nature and the growth pathways of their BHs. As demonstrated by the SDSS selection in \citetalias{Lin2025_localLRD}, it is highly efficient to select samples from existing large spectroscopic databases with additional SED constraints from wide-field imaging surveys from near-UV to near-IR, e.g., GALEX \citep{Bianchi2017}, Legacy Survey \citep{Dey2019}, WISE \citep{Wright2010}, etc. In addition to SDSS, the Dark Energy Spectroscopic Instrument (DESI)
 provides another ideal opportunity \citep{DESICollaboration2024, DESI_DR1}. DESI is a five-year spectroscopic redshift survey conducted on the Mayall 4-meter telescope at Kitt Peak National Observatory. As the largest cosmological spectroscopic survey currently underway, DESI aims to target millions of galaxies, quasars, and stars to study dark energy and the expansion history of the Universe. DESI's extensive spectroscopic database, covering optical wavelengths from {3600–9825} \AA\ with high spectral resolution ($R=2000$–$5500$), is well suited for the selection of  LRDs at $z<1$. Although DESI covers the optical continuum and emission lines, it does not capture the full optical–near-IR continuum shape, which is crucial for diagnosing blackbody-like emission in BH envelope models. We therefore conducted systematic near-IR follow-up spectroscopy to characterize the near-IR continuum and access key emission lines not covered by DESI. This is particularly important for LRDs at $z>0.5$, where \ha\ is redshifted beyond the DESI wavelength range and falls within the near-IR spectral coverage.

In this paper, we present our selection from the DESI DR1\footnote{\url{https://data.desi.lbl.gov/doc/releases/dr1/}} spectral database and introduce the first sample of 27 LRDs at $z=0.2$–$0.9$. We conducted follow-up near-IR observations for 21 of them using LBT/LUCI, Magellan/FIRE, and Keck/NIRES, complementing their DESI optical spectra. The observing campaign is ongoing for the remaining targets. The paper is organized as follows.  In Section \ref{sec:data}, we present the datasets used in this work. In Section \ref{sec:selection}, we describe the selection criteria applied to DESI DR1, which yields 27 LRDs. In Section~\ref{sec:observation}, we describe the near-IR follow-up observations. We outline the measurement methodology for the selected LRDs in Section \ref{sec:measurement}, and in Section \ref{sec:result}, we present their properties. In Section \ref{sec:discussion}, we briefly discuss the implications of our results.  More detailed analyses and physical modeling will be presented in future work. Throughout this paper, we adopt the AB magnitude system for all photometry. All wavelengths and line identifications are given in the vacuum frame. All equivalent widths (EWs) are reported in the rest frame. $L_{\rm 5100}$ is defined as the monochromatic luminosity $\lambda L_\lambda$ evaluated at rest-frame 5100\,\AA. We adopt the cosmological parameters from \cite{PlanckCollaboration2020}.

%% file: 02_Selection_and_Sample.tex
\section{DATA}\label{sec:data}

\subsection{DESI DR1}

We refer to \cite{DESI_DR1} for details on DR1. In brief, DESI DR1 includes more than 18 million spectra of galaxies, quasars, and stars from its Main Survey observations between May 2021 and {2022 June}, along with 1.6 million from the Survey Validation program. In total, DR1 yields about 16 million extragalactic objects over 9700\,deg$^2$. DESI's spectra cover 3600–9825\,\AA\ at a resolution of $R = 2000$–5500. Notably, at wavelengths $\lambda \gtrsim 6000$\,\AA, DESI achieves $R > 4000$. The high spectral resolution enables it to clearly resolve not only the {emission-line} profiles but also the Balmer absorption features in LRDs.

In addition to the reduced spectra, DR1 also releases  \textsc{emline} catalogs as an accompanying product of the pipeline. The \textsc{emline} catalog contains Gaussian fits to the major emission lines. DR1 also includes a set of value-added catalogs. Among these, the FastSpecFit Spectral Synthesis and Emission-Line Catalog\footnote{\url{https://data.desi.lbl.gov/doc/releases/dr1/vac/fastspecfit/}} \citep[][hereafter \textsc{FastSpecFit}]{FastSpecFit} and Stellar Mass and Emission Line Catalog\footnote{\url{https://data.desi.lbl.gov/doc/releases/dr1/vac/stellar-mass-emline/}} \citep[][hereafter \textsc{stellar-mass-emline}]{Zou2024} provide modeled continua and {emission-line} measurements. In the \textsc{FastSpecFit} catalog, the fitting is performed by \textsc{FastSpecFit}, a stellar continuum and emission-line modeling code that uses stellar population synthesis and emission-line templates to jointly model DESI optical spectrophotometry and ultraviolet-to-infrared broadband photometry. In the \textsc{stellar-mass-emline} catalog,  emission lines are measured by single Gaussian fits, with absorption correction through continuum fitting performed by \textsc{starlight}\footnote{\url{http://www.starlight.ufsc.br/}}. In the following selection, we use the {emission-line} fits in the three catalogs,   \textsc{emline}, \textsc{FastSpecFit}, and  \textsc{stellar-mass-emline}, to pre-select candidates. However, we note that these codes are designed for simplicity and speed rather than for precise spectral modeling. In particular, for populations such as LRDs, whose physical nature is still poorly understood, the modeled spectra are inaccurate, and discrepancies between the modeled and observed spectra are always present. For our survey, the DESI pipeline fits are used only for the initial selection. All spectral properties are re-measured and re-fitted using more tailored approaches, as described in Section \ref{sec:measurement}.

\subsection{Photometry}\label{sec:photometry}

We compile photometric data from the following wide-field sky surveys:  {far-UV (FUV) and near-UV (NUV)} from GALEX DR6 \citep{Bianchi2017}; \textit{ugriz} {point-spread function (PSF)} photometry from SDSS DR17 \citep{Abazajian2009, Abdurrouf2022};  $grizy$ PSF photometry from Pan-STARRS DR2 \citep{Flewelling2020}; $G$, $G_{\mathrm{BP}}$, and $G_{\mathrm{RP}}$ mean photometry from GAIA DR3 \citep{Gaia2023}; $griz$ model photometry from Legacy Surveys DR10 \citep{Dey2019} where PSF models are used for the three objects;  $YJHK$ PSF photometry from UKIDSS DR11PLUS \citep{Lawrence2007}; and W1, W2, W3, W4 photometry from WISE  \citep{Wright2010}. PSF photometry is adopted when available, as all DESI DR1 LRDs are compact and unresolved in ground-based imaging. GALEX photometry is taken from the GUV catalog \citep{GUVcat_DOI}, while WISE photometry is incorporated into the Legacy Survey catalog based on unWISE images \citep{Schlafly2019}. All other data are retrieved from Astro Data Lab\footnote{\url{https://datalab.noirlab.edu/}} \citep{Fitzpatrick_SPIE, NIKUTTA2020100411}.

For the selected targets, we retrieve {multiepoch} optical $gri$ photometry from the Zwicky Transient Facility Data Release 24 \citep[ZTF DR24; ][]{Bellm2019, Masci2019}, which spans from {2018 March 21 to 2025 October 21} ($\sim$7.6 yr baseline). We also obtain {multiepoch} WISE W1, W2 photometry from the AllWISE Multiepoch Photometry Database \citep{Cutri2021} and the NEOWISE-R Single Exposure Source Table \citep{Mainzer2014}, covering {2010 January–2024 July} (with a {2011 March–2013 December} hibernation gap between the cryogenic AllWISE mission and the NEOWISE-Reactivation phase). For faint sources with low {signal-to-noise ratio (S/N)} in individual exposures, we retrieve forced-photometry WISE light curves from the Legacy Survey, which are up to year 6 of NEOWISE Reactivation.

\subsection{SPHEREx}
SPHEREx is a 20\,cm all-aluminum space telescope equipped with six linear variable filters that enable spectrophotometric imaging in 102 spectral channels spanning $0.75$--$5.0\,\micron$ across the full sky \citep{Bock2026}. SPHEREx features a wide, dichroic field of view of $3.5\degr \times 11\degr$, a spatial resolution of $6.2\arcsec$ per pixel, and spectral resolution of $R = 35$--$41$ at $0.75$--$3.82\,\micron$ and $R = 110$--$130$ at $3.82$--$5.0\,\micron$. We retrieve the SPHEREx spectrophotometry for our sample from the NASA/IPAC Infrared Science Archive (IRSA), where the spectra are extracted via PSF-weighted forced photometry. We cross-validate the SPHEREx spectra against existing near-infrared and WISE photometry. We retain only the SPHEREx spectrophotometry that is well consistent with the existing IR photometry.  For SPHEREx spectrophotometry with low S/N, we bin the spectra using a flux-conserving method to achieve $\mathrm{S/N} > 5$ per binned channel.

\section{Sample selection}\label{sec:selection}

\subsection{Selection Guidelines for Low-$z$ LRDs}\label{sec:selection_guideline}

We aim to select sources whose observed characteristics strictly match those of the majority of high-$z$ LRDs. Our selection prioritizes high completeness and efficiency for objects with observed properties fully consistent with representative JWST LRDs. It may exclude the most extreme or atypical LRDs, as well as transitional systems that share only part of the typical LRD characteristics. The resulting low-redshift sample serves as a laboratory for investigating the physical conditions that give rise to the LRD phenomenon. The empirical criteria defining LRDs are summarized below.

First, we require the selected objects to satisfy the defining characteristics of LRDs {\citep[e.g.,][]{Greene2024, Hviding2025}}:

\begin{itemize}
\item Broad Balmer {emission lines.}
\item Compact morphology in the {rest-frame optical.}
\item A {V-shaped SED,} characterized by {a blue UV-continuum slope} ($\beta_{\rm UV}<0$) and {a red optical-continuum slope} ($\beta_{\rm optical}>0$).
\end{itemize}

We further include additional observational characteristics of LRDs, empirically concluded from the majority of high-$z$ LRDs reported in the literature.

\begin{itemize}
\item The inflection point of the V-shaped continuum occurs near or redward of the Balmer break, and at any wavelength blueward of \ha.  

This is empirically motivated by most high-$z$ LRDs, as assembled in {\citet{Setton2024, deGraaff2025_all}}. A small fraction of LRDs show inflection points significantly redder than the Balmer limit, extending to $\sim$5000\,\AA\ \cite[e.g., UNCOVER-A2744-20698;][]{Wang2026}. 

\item A declining rest-frame near-infrared continuum in $f_\lambda$ space. 

This is motivated by the near-IR spectral shape of the majority of JWST LRDs, as revealed by either NIRSpec/Prism spectroscopy or NIRCam and MIRI photometry. Such a continuum shape has also motivated theoretical models invoking blackbody-like thermal emission to describe the overall SED.

\item Weak [\ion{N}{2}] $\lambda6585$ emission. 

This is motivated by the weak or undetected [\ion{N}{2}] emission in high-$z$ LRDs.  The weakness of [\ion{N}{2}] in LRDs likely results from low metallicity in the host galaxies or narrow-line regions. Motivated by the limited number of JWST LRDs with reported [\ion{N}{2}] detections \citep{Maiolino2024, Juodzbalis2024}, we require selected sources to either have undetected [\ion{N}{2}] emission or, if detected, to lie within the galaxy/composite region of the [\ion{N}{2}]-BPT diagram.

\item Negligible [\ion{Ne}{5}] $\lambda3427$ emission.  

This is motivated by the lack of strong high-ionization lines in the rest-frame optical spectra of most high-$z$ LRDs, and their ionizing spectra that are softer than those of typical type-1 AGNs and luminous quasars {\citep{Ji2025_lord, Wang2025_heii}}.
[\ion{Ne}{5}], with an ionization potential of 97.1 eV, traces a very hard radiation field, typically associated with photons from the AGN accretion disk or corona. At the time of writing, [\ion{Ne}{5}] has not been reported in LRDs. We thus exclude sources exhibiting [\ion{Ne}{5}] emission.

\end{itemize}

The observational features described above guide the selection criteria adopted in this work. We note that relaxing any of the observational requirements above may introduce candidates that resemble, but are not fully consistent with, the bulk of high-$z$ LRDs. For example, the X-ray Dot {\citep{Hviding2026_xdot}} exhibits a V-shaped continuum with an inflection point around 2000 \AA\ and has been proposed as an LRD in a transitional phase. Some V-shaped continuum sources with strong [\ion{N}{2}] emission have been reported as LRD analogs {\citep{Rinaldi2025, Ding2026}}. Such systems have been proposed as transitional objects between LRDs and typical type-1 AGNs or luminous quasars, or as a later evolutionary stage of the LRD population {\citep{Fu2025_transition}}.  These subsets are beyond the scope of this work.

Furthermore, some high-ionization UV lines have been reported in deep JWST/NIRSpec spectra of LRDs \citep{Tang2025_high_ionization, Tang2026_SPURS_LRDs, Ji2026_SPURS_LRDs}. At the time of writing, the lines detected in LRDs with the {highest-ionization potential} are \ion{N}{5} $\lambda1240$ (77.5 eV) and [\ion{Ne}{4}]$\lambda\lambda$2422, 2424 (63.5 eV), while [\ion{Ne}{5}] $\lambda3427$ remains undetected {\citep{Akins2024, Tang2025_high_ionization, Tang2026_SPURS_LRDs}}. Although models invoking specific geometries of the {circum-BH} environment in LRDs may potentially explain high-EW [\ion{Ne}{5}] emission \citep[e.g.,][]{Tang2025_high_ionization, Tang2026_SPURS_LRDs, Ji2026_SPURS_LRDs}, we do not include such sources in the final sample analyzed in this work. This subset is retained in the parent sample and excluded only at the final stage of selection. A detailed investigation of this subset is deferred to future studies.

\subsection{Selection criteria}

 \begin{figure*}[!t]
    \centering
\includegraphics[width=\textwidth]{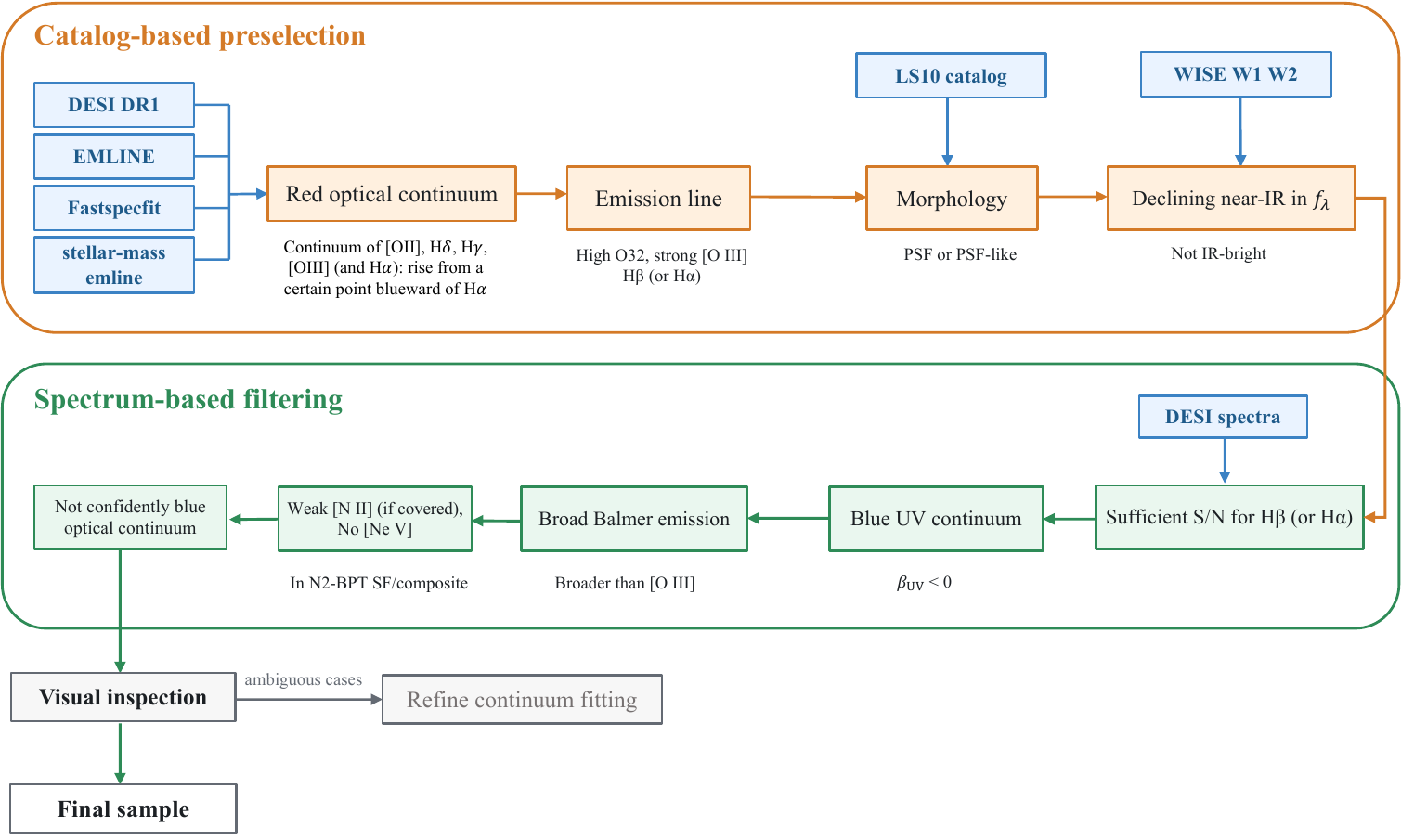}
\caption{Flowchart of the LRD selection process in DESI DR1 (Section~\ref{sec:selection}). Blue boxes denote the input datasets, orange boxes indicate catalog-based selection criteria, and green boxes indicate filtering based directly on the DESI spectra. 
}
\label{fig:selection}
\end{figure*}

Guided by the criteria defined in Section \ref{sec:selection_guideline}, we begin our selection using DESI DR1.
 We first require objects to have $z > 0.001$ to exclude stars from the DR1 catalog. We then select objects with red optical continua and subsequently apply a selection based on strong emission lines. Both steps use the \textsc{emline}, \textsc{FastSpecFit}, and \textsc{stellar-mass-emline} catalogs. Next, we require the objects to exhibit PSF-like morphology. Finally, for the selected candidates, we perform manually refined, simplified fits to their spectra, requiring them to exhibit a blue UV continuum, broad Balmer emission, and weak [\ion{N}{2}]. { The complete selection workflow is summarized in Figure~\ref{fig:selection}, while the number of sources remaining after each major selection stage is given in Table~\ref{tab:selection_accounting}.} Each step is described in detail below.

\subsubsection{Red optical continuum criteria}\label{sec:selection_red_continuum}

In this step, we select candidates exhibiting a rising optical continuum that starts at any wavelength blueward of \ha. { We use the continuum levels underlying [\ion{O}{2}]~$\lambda3727$, H$\delta$, H$\gamma$, [\ion{O}{3}]~$\lambda5008$, and, when covered, H$\alpha$. These measurements sample the continuum at progressively longer wavelengths.} 

{ The selection is designed to identify either a monotonically rising continuum or a continuum that initially declines, reaches a single minimum, and then rises monotonically toward the reddest available wavelength. We define the wavelength-ordered continuum sequence as
\begin{equation*}
\begin{array}{r@{\;}l}
(C_1,\ldots,C_5)=\bigl(
&
\mathrm{cont}_{[\mathrm{O\,II}]\,3727},
\mathrm{cont}_{\mathrm{H}\delta},
\mathrm{cont}_{\mathrm{H}\gamma},\\
&
\mathrm{cont}_{[\mathrm{O\,III}]\,5008},
\mathrm{cont}_{\mathrm{H}\alpha}
\bigr).
\end{array}
\end{equation*}
The criterion can therefore be written as
\begin{equation}
\begin{array}{ll}
C_1<C_2<\cdots<C_N, & \text{or}\\[3pt]
C_1>\cdots>C_k<C_{k+1}<\cdots<C_N,
& 2\leq k\leq N-1.
\end{array}
\label{eq:red_continuum_criterion}
\end{equation}
$N=5$ when H$\alpha$ {is covered, and $N=4$} when the continuum sequence extends only to [\ion{O}{3}]~$\lambda5008$. In the second case, $C_k$ denotes the single minimum of the sampled
continuum, which then rises monotonically toward the reddest available
wavelength.
}

The continuum measurements are taken from the \textsc{emline}, \textsc{FastSpecFit}, and \textsc{stellar-mass-emline} catalogs. {  If the full continuum sequence does not satisfy Equation~\ref{eq:red_continuum_criterion}, we apply a one-point tolerance to account for uncertainties in the DESI pipeline continuum estimates. We successively omit each intermediate continuum measurement and evaluate the corresponding reduced sequence. For each omission, the two measurements adjacent to the omitted point are compared directly. The source is retained if at least one of the resulting reduced sequences satisfies Equation~\ref{eq:red_continuum_criterion}. For example, if the H$\gamma$ estimate is omitted, the remaining sequence consists of the continuum measurements at [\ion{O}{2}]~$\lambda3727$, H$\delta$, [\ion{O}{3}]~$\lambda5008$, and H$\alpha$ when it is covered. The source is retained if this reduced sequence either rises monotonically or declines to a single minimum and then rises monotonically, e.g., $\mathrm{cont}_{[\mathrm{O~II}]~3727}>\mathrm{cont}_{\mathrm{H}\delta}<\mathrm{cont}_{[\mathrm{O~III}]~5008}<\mathrm{cont}_{\mathrm{H}\alpha}$, as defined by Equation~\ref{eq:red_continuum_criterion}. This one-point tolerance reduces the sensitivity of the selection to local variations in the automated continuum modeling while preserving the required overall continuum shape. It also permits the turnover to occur between [\ion{O}{3}]~$\lambda5008$ and H$\alpha$ when H$\alpha$ is covered.}

\subsubsection{{Emission-line} criteria}\label{sec:emline_criteria}
We further require the objects to exhibit strong Balmer and [\ion{O}{3}] emission lines. 
 Specifically, objects must satisfy the emission-line criteria listed below in either the \textsc{emline} or \textsc{FastSpecFit} fits.

\begin{itemize}
    \item \ha\ EW $> 15$ \AA\ or \hb\ EW $> 5$ \AA.
    \item \ha\ S/N $> 5$ or \hb\ S/N $> 5${.}
    \item \ha\ flux $> 10^{-16}$ erg s$^{-1}$ cm$^{-2}$ or \hb\ flux $> 5 \times 10^{-17}$ erg s$^{-1}$ cm$^{-2}$.
    \item[$\bullet$] [\ion{O}{3}] 5008 EW $> 20$  \AA{.}
    \item[$\bullet$] [\ion{O}{3}] 5008 / [\ion{O}{2}] 3727 (O32) $> 5${.}
\end{itemize}

The criteria on \ha\ or \hb\ ensure that the objects exhibit strong Balmer emission lines, thereby excluding quiescent galaxies that naturally show strong Balmer breaks or red optical continua. The criteria on [\ion{O}{3}] EW and O32 further exclude post-starburst galaxies and AGNs hosted by old stellar populations.  The threshold values adopted for the criteria above are determined through iterative optimization, aiming to be inclusive given the limited modeling power of the available catalogs.

Requiring strong [\ion{O}{3}] emission introduces an inherent selection bias. JWST-discovered LRDs with extreme Balmer breaks, exceeding those of normal stellar populations, are found to exhibit very weak [\ion{O}{3}] emission (e.g., the Cliff {\citep{deGraaff2025}}; A2744-QSO1 \citep{Furtak2024}; MoM-BH$^{*}$-1 {\citep{Naidu2025}}) and are therefore excluded by this criterion. The emission-line selection strategy is optimized for the large data volume of DESI DR1, balancing computational efficiency with the level of human inspection required. LRDs with extremely strong Balmer breaks are more efficiently identified through dedicated {Balmer-break-based} selections, which are beyond the scope of this paper and will be addressed in future work.

\subsubsection{Morphological criteria}\label{sec:morph}

For objects that satisfy the red optical continuum and {strong emission-line criteria} described above, we retrieve their morphological information from Legacy Survey DR10.

We select PSF-like sources following the quasar selection procedure in \cite{Chaussidon2023}. We first select objects classified as PSFs in the Legacy Survey catalog. To account for uncertainties in ground-based observations and morphological fitting, we also accept objects that are photometrically classified as extended but exhibit a small relative difference between the PSF and extended morphological models ($\Delta\left(\chi^2\right) / \chi^2<0.015$).

\subsubsection{Near-IR criteria}\label{sec:selection_nearIR}

For objects that satisfy the criteria above, we obtained their WISE W1 and W2 photometry from the Legacy Survey catalog, which includes unWISE measurements. We require that the W1 and W2 flux densities in $f_\lambda$ space lie below the reddest continuum level of their DESI spectra. { Although no spectral coverage is available between the {reddest optical-continuum point} and W1 during the selection, the lower W1 flux requires either a monotonic decline or a turnover followed by a decline across this interval. In either case, the infrared continuum must be declining by the W1 wavelength.
} For sources undetected in WISE, we retain them in the sample and pass them on to subsequent selection steps. %

Up to this stage, all selections are based on publicly available catalogs. { In the next stage, we use the DESI spectra for direct spectral measurements and filtering.}

\subsubsection{Criteria on UV continuum, {broad lines, and} [\ion{N}{2}] }\label{sec:manual_criteria}

For the candidates selected above, we perform a refined selection by applying simplified fits to their DESI spectra.

First, we {recompute} the S/N of \ha\ or \hb\ by directly integrating the line flux in the reduced spectra, requiring an S/N $>15$ for the integrated flux within 4000 km s$^{-1}$. This measurement, independent of the overall spectral modeling, provides an additional constraint beyond the relaxed emission-line requirements in Section~\ref{sec:emline_criteria}.

We then apply median filtering to derive the rest-frame continuum blueward of 3500\,\AA. We require objects to have {a UV-continuum slope $<0$} at rest-frame $<$3500\,\AA.

Thirdly, we compare the line profile of \hb\ or \ha\ with that of [\ion{O}{3}] $\lambda5008$. We require the Balmer lines to exhibit significant broad emission in excess of the [\ion{O}{3}] profile, with a characteristic width $>1000$ km s$^{-1}$. This criterion removes compact, V-shaped emission-line galaxies that show broad outflow components in both [\ion{O}{3}] and the Balmer lines but lack a distinct Balmer component indicative of a BLR. { Operationally, we subtract the local continuum, rebin the [\ion{O}{3}] and Balmer profiles into 100 km s$^{-1}$ intervals, and smooth them with a 150 km s$^{-1}$ Gaussian kernel. We retain sources whose Balmer emission extends beyond 1000 km s$^{-1}$ from line center, exceeds the corresponding [\ion{O}{3}] edge by $>300$ km s$^{-1}$, and has an integrated outer-wing S/N $>3$. We adopt $\sigma_{\rm [O\,III]}=200$ km s$^{-1}$ as the maximum narrow-line reference and use this value when the width cannot be reliably measured, {for example, due} to bad pixels. This provides a fast, {nonparametric} identification of broad Balmer emission without full profile fitting.}

We then apply a cut on [\ion{N}{2}] $\lambda6585$ and H$\alpha$ when H$\alpha$ is within the spectral coverage. If [\ion{N}{2}] is detected, we require that the source lie within the star-forming or composite region of the [\ion{N}{2}]–BPT diagram \citep{Baldwin1981}. We further require {nondetection} or absence of [\ion{Ne}{5}] $\lambda3427$ emission.

{ Finally, we apply an additional spectrum-based rejection of monotonically blue optical continua, independent of the value-added-catalog-based {pre-selection} (Section \ref{sec:selection_red_continuum}). This step removes sources that pass the catalog-based continuum criteria because of pipeline uncertainties but show clearly monotonically blue optical continua in the spectra. We measure the rest-frame 4200--7000\,\AA\ continuum in {400\,\AA\ wide} sliding-median windows separated by 150\,\AA\ and reject sources with continuum S/N $>3$ for which $\gtrsim70\%$ of consecutive measurements decrease toward longer wavelengths. For computational efficiency, this direct spectral check is applied only to sources surviving the preceding selection stages.}

After applying the selection criteria described above, we visually inspect the remaining objects ($N\sim5000$). { Because the selection workflow is intentionally designed to be inclusive and maximize completeness, some sources can pass the automated selection owing to measurement uncertainties or limitations in the catalog and spectral measurements. The primary contaminants at the final visual-inspection stage include noisy spectra that satisfy the selection criteria within these tolerances, as well as type-1 AGNs whose optical continua redden toward longer wavelengths likely owing to the increasing contribution from an evolved host-galaxy stellar population. The latter commonly show strong [\ion{N}{2}] emission that is blended with very broad Balmer lines, allowing them to evade the automated [\ion{N}{2}] veto.
}
 For sources whose optical continua remain ambiguous by visual inspection, we perform the refined continuum fitting described in Section~\ref{sec:continuum_param} and retain only those with robust $\beta_{\rm UV}<0$ and $\beta_{\rm optical}>0$.

{ In practice, this additional fitting was required for only two sources, J004529.15+133908.73 and J071635.74+543322.10. We obtained $\beta_{\rm optical}=-0.07^{+0.06}_{-0.03}$ and $-0.07^{+0.05}_{-0.02}$, respectively, and therefore excluded both sources. J004529.15+133908.73 is a metal-poor AGN reported by \citet{Izotov2008}, while J071635.74+543322.10 was reported as one of the eight $z<0.45$ DESI LRDs by \citet{Park2026}.}

\subsection{Final Sample}

In the end, we obtain 27 objects that satisfy all definitions and selection criteria of LRDs at high confidence levels. {Table~\ref{tab:sample_summary} summarizes basic properties of the 27 DESI DR1 LRDs.} Figure \ref{fig:example1} shows two examples from our sample. 
Figure \ref{fig:spec_zoomin} presents an example of zoomed-in spectra that reveal abundant emission lines.
The SEDs and emission lines of the full sample are presented in Appendix \ref{appendix:sample}.

Hereafter, all objects are abbreviated in the form Jhhmm+ddmm for simplicity in the subsequent analysis.

\begin{table*}[ht!]
    \centering
    \small
    \setlength{\tabcolsep}{3.0pt}
    \renewcommand{\arraystretch}{1.10}

    \noindent\makebox[\linewidth][c]{%
    \resizebox{\textwidth}{!}{\input{table/sample_summary.tex}\unskip}%
    }\par

    \caption{{
    DESI DR1 LRD sample summary. {R.A. and decl.} are the DESI targeting
    coordinates in decimal degrees, and $z$ is the systemic redshift from
    the joint narrow-line fit. The follow-up columns list the near-IR
    instrument(s), total exposure time, and observing date for each source.
    The machine-readable version is available on Zenodo as
    \texttt{master\_table.ecsv}.}}
    \label{tab:sample_summary}
\end{table*}

\subsubsection{DESI target bits}\label{sec:desi_target_bits}

For the 27 LRDs selected above, 16 sources are assigned to the primary DESI targeting programs (with 11 also included in secondary programs), and 11 are exclusively selected by secondary targeting programs.  Their DESI target bits, which encode selections from different targeting criteria, are diverse, including primary program selections such as \verb|QSO|, \verb|LRG|, and \verb|ELG|, as well as secondary program selections including \verb|WISE_VAR_QSO| and \verb|PSF_OUT_DARK| \citep{Myers2023}. For the primary program target bits, {13 of the targets} carry the \verb|QSO| bit, having passed one of the multiple quasar selection criteria \citep{Chaussidon2023}. Five objects carry the \verb|LRG| bit, consistent with the luminous red galaxy selection  \citep{Zhou2023}, and two carry the \verb|ELG| bit following the emission-line galaxy selection  \citep{Raichoor2023}.

For the secondary program target bits, 21 objects carry the \verb|WISE_VAR_QSO| bit, which flags quasar candidates selected from variability in their {10\,yr} W1 and W2 light curves. However, this selection adopts a loose random forest probability cut \citep[$p > 0.1$,][]{Chaussidon2023}. This selection was tested during DESI Survey Validation as a complementary method to see if it can recover quasars missed by the primary optical and near-infrared color-based selection, rather than as a high-purity quasar sample.  We present a more detailed variability analysis in Section~\ref{sec:variability}.  Eight objects carry the \verb|PSF_OUT_BRIGHT| or \verb|PSF_OUT_DARK| bits, which are assigned by a program targeting outlier point sources with unusual colors to recover missed quasars and identify rare or peculiar objects \citep{Myers2023}.

\section{Spectroscopic follow-up observations}\label{sec:observation}

\begin{figure*}[ht!]
    \centering
    \includegraphics[width=0.95\linewidth]{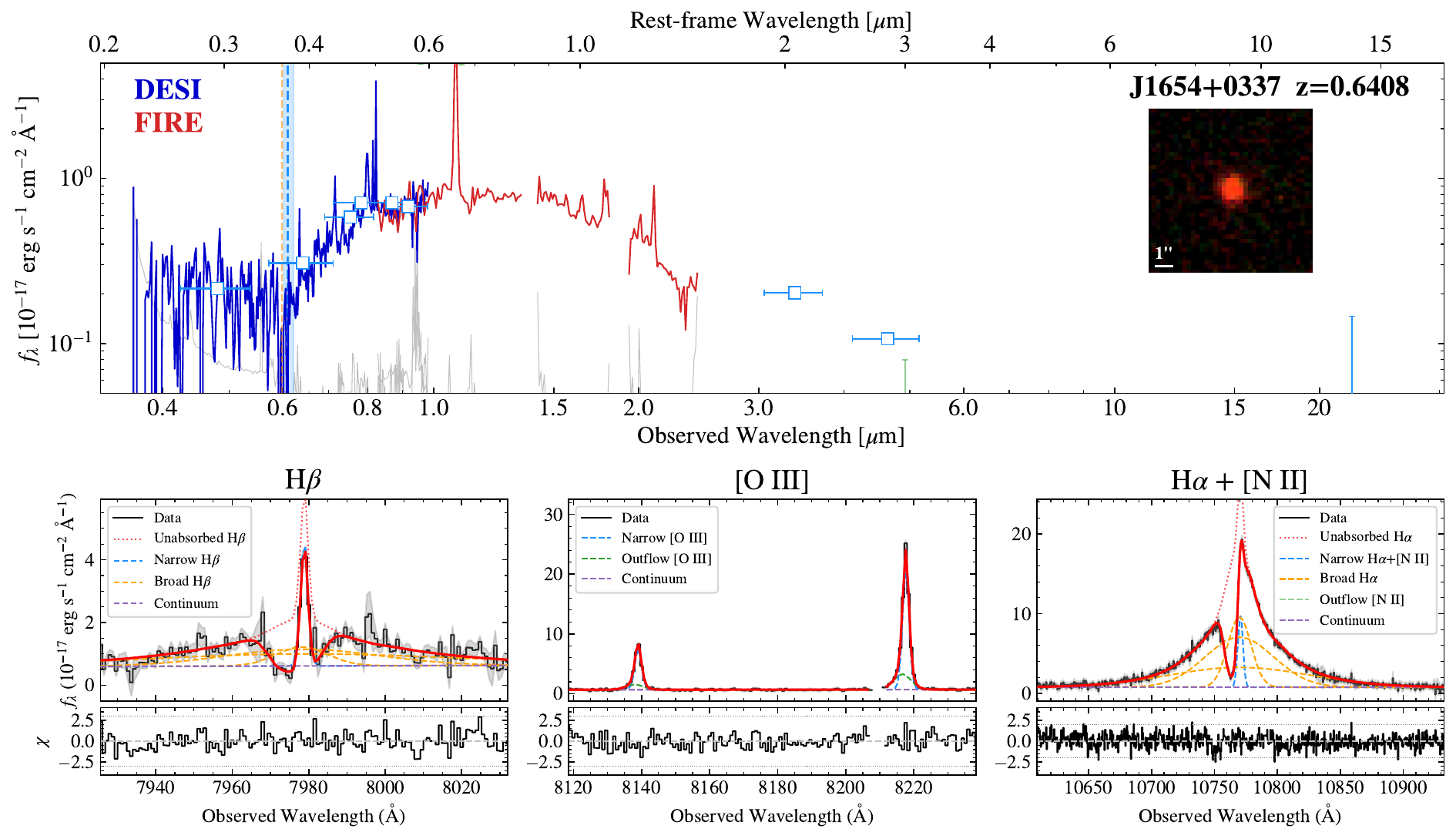}
    \bigskip
    \includegraphics[width=0.95\linewidth]{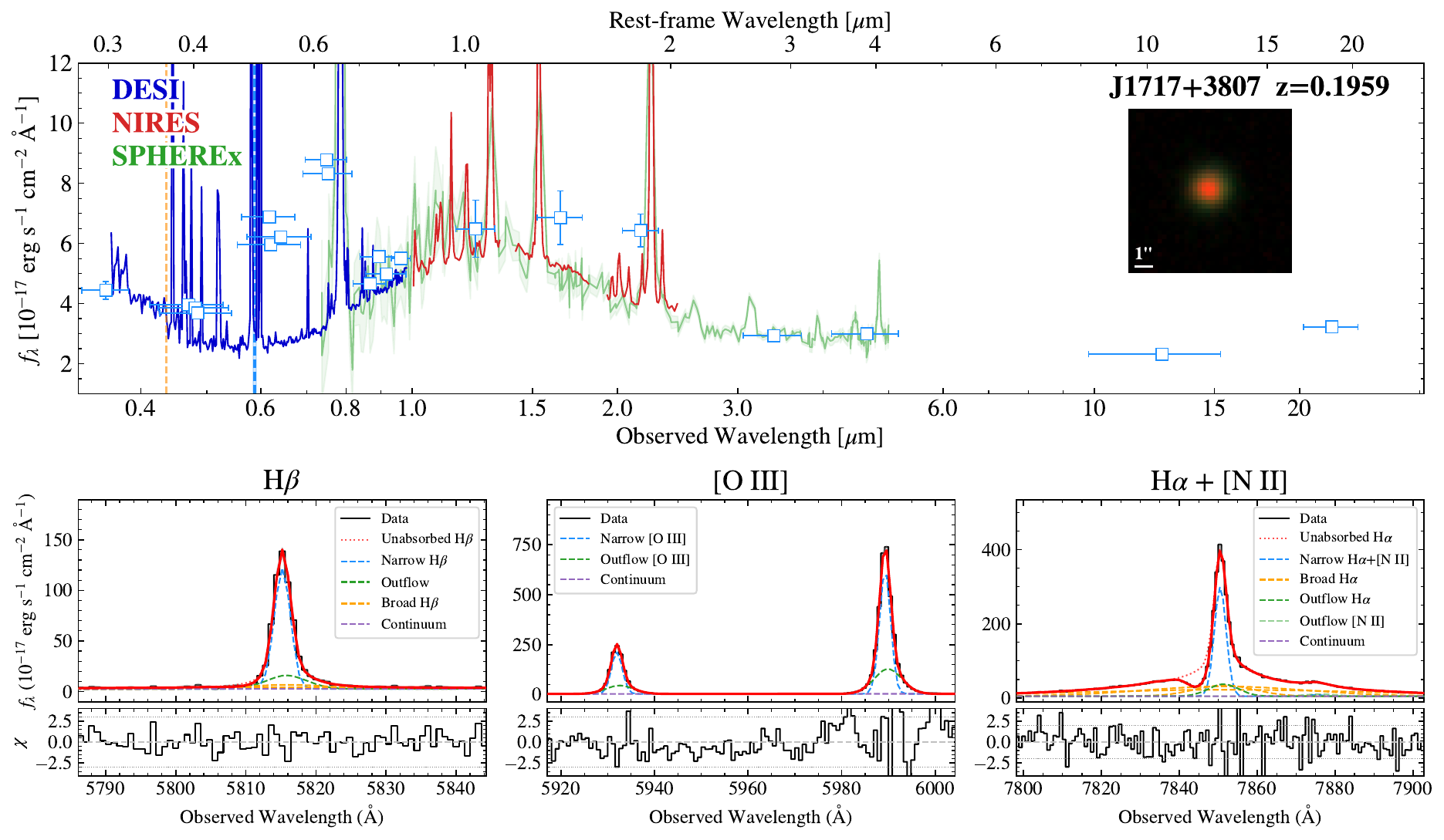}
    \caption{Example DESI DR1 LRDs presented in this work. The top panel shows the overall SEDs, with the smoothed DESI spectra plotted as blue lines and the photometry as blue squares. The gray lines indicate the spectral uncertainties. The orange dashed lines mark the Balmer-limit wavelengths, and the blue dashed lines indicate the inflection-point wavelengths. The bottom panel shows the \hb\, \oiii\ and \ha\ line profiles, with the red curves denoting the best-fit total models and the dashed lines representing the best-fit individual components.}
    \label{fig:example1}
\end{figure*}

\begin{figure*}
    \centering
    \includegraphics[width=\linewidth]{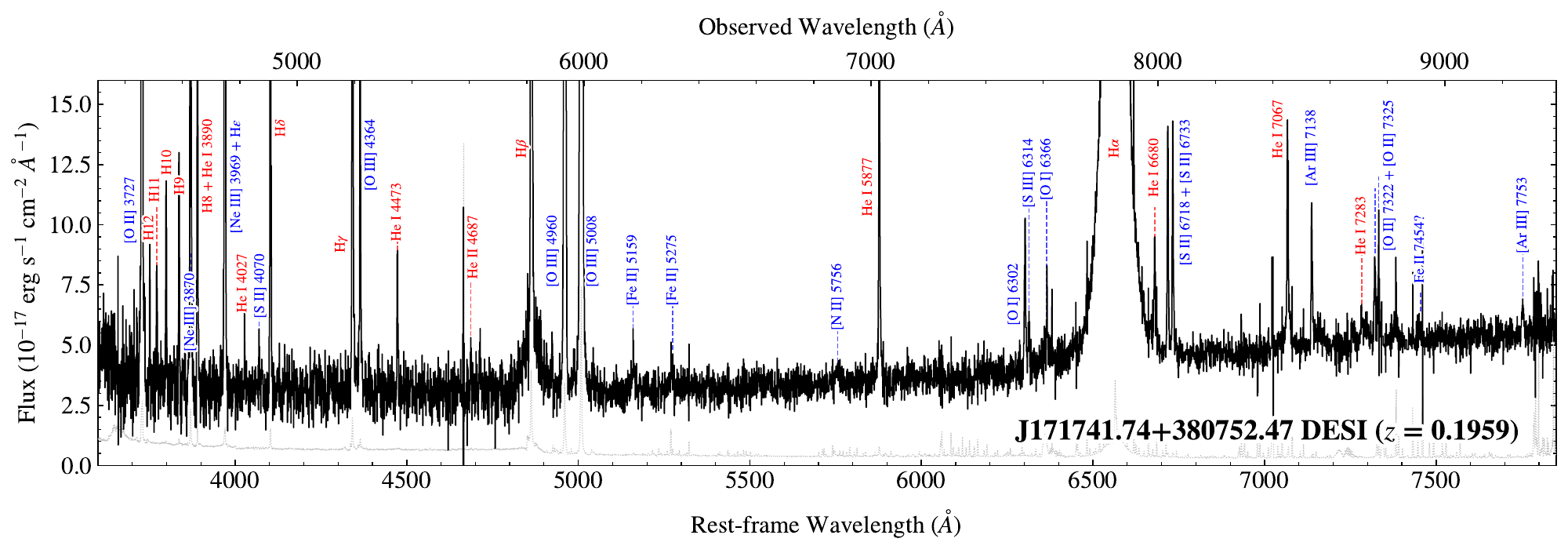}
    \includegraphics[width=\linewidth]{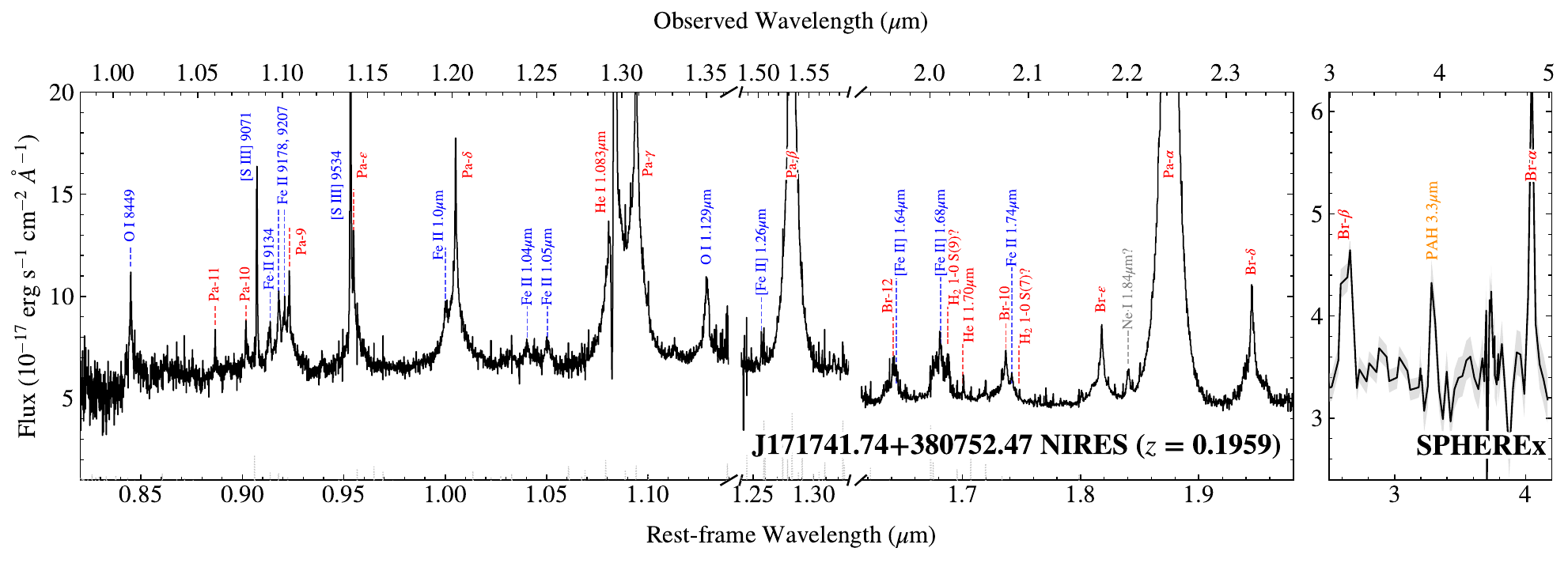}
    \caption{Zoomed-in view of the DESI, Keck/NIRES, and SPHEREx spectra of J171741.74+380752.47. 
    The spectra are shown as black lines, and the uncertainties are indicated by gray dotted lines or shaded regions. Emission lines are labeled.}
    \label{fig:spec_zoomin}
\end{figure*}

We obtained near-IR follow-up spectra for {21} of the 27 DESI DR1 LRDs using three spectrographs: Magellan/FIRE, LBT/LUCI, and Keck/NIRES. {Fourteen targets were observed with FIRE, six with LUCI, and two with NIRES; J1611+0917 was observed with both FIRE and LUCI. The observations span 2025 October 3 through 2026 June 7.}

Magellan/FIRE \citep{Simcoe2013} is a near-IR echelle spectrograph on the 6.5-m Magellan Baade Telescope covering 0.8--2.5\,\micron. {The 14 FIRE targets were observed on 2025 October 3 and 17, 2025 November 22, 2026 March 15, 2026 April 3--4, and 2026 May 28 (UTC). Each target was observed for 0.50--2.51 hr.} Slit widths of 0\farcs75 or 1\arcsec\ were used. The seeing was 0\farcs6--0\farcs9 for the runs through March, 0\farcs9--2\arcsec on April 3--4, and {0\farcs5--1\farcs0 under cloudy sky on May 28.} The FIRE spectra reach $R\sim3000$--6000.

LBT/LUCI \citep{Seifert2003} comprises two independent {multimode} near-IR instruments mounted on the 2$\times$8.4-m Large Binocular Telescope. {Six targets were observed on 2025 December 23, 2026 February 16, March 27, May 15 and 28, and June 4 and 7 (UTC).} We used the G200 grating with zJspec on LUCI1 (0.9--1.2\,\micron) and HKspec on LUCI2 (1.50--2.40\,\micron), generally simultaneously, with a 1\arcsec\ slit and $R\sim500$--1200. {The total integration per arm and visit ranged from {0.33–2.67 hr}. The seeing was approximately 0\farcs8 on {2025 December 23}, and {1\arcsec--2\arcsec} on other nights.}

Keck/NIRES \citep{Wilson2004} is a near-infrared echellette spectrograph mounted on the 10-m Keck\,II telescope. Two targets were observed on {2026 April 27 (UTC)} with seeing of 0\farcs6$\pm$0\farcs2. With its fixed 0\farcs55 slit, NIRES produces simultaneous {$J$-, $H$-, and $K$-band} spectra in five orders from {0.94–2.45}\,\micron\ at a characteristic resolution of $R\sim2700$.

The Magellan/FIRE, LBT/LUCI, and Keck/NIRES spectra were reduced by \textsc{pypeit} \citep{Prochaska2020}. We performed flat-fielding, wavelength calibration, sky subtraction, spectral extraction, flux calibration, and telluric correction. We performed absolute flux calibration by scaling each spectrum with a constant factor or a linear polynomial to match the observed {$y$-, $J$-, $H$-, and $K$-band} flux.

{Table~\ref{tab:sample_summary} summarizes spectroscopic follow-up coverage of the 27 DESI DR1 LRDs.} The follow-up spectra reveal not only the SED shapes of the LRDs (Figure \ref{fig:example1}), but also key emission lines in the rest-frame optical and near-IR, particularly \ha\ for sources at $z\gtrsim0.5$ {(Figures \ref{fig:example1} and \ref{fig:spec_zoomin})}.

%% file: table/sample_summary.tex
\begin{tabular}{
@{}
c @{\hspace{6pt}}
c @{\hspace{6pt}}
c @{\hspace{5pt}}
c @{\hspace{5pt}}
c @{\hspace{8pt}}
c @{\hspace{5pt}}
c @{\hspace{6pt}}
c
@{}
}
\hline
\hline
Name &
DESI TARGETID &
R.A. &
Decl. &
$z$ &
Follow-up instrument(s) &
$t_{\rm exp}$ &
Observing date \\
&
&
(deg) &
(deg) &
&
&
(s) &
(UTC) \\
\hline

\multicolumn{8}{c}{\textbf{Gold sample}} \\
\hline

J012930.87+062843.32 &
39627941918802007 &
22.3786291 &
6.4787023 &
0.2467 &
Magellan/FIRE &
7186 &
2025-11-22 \\

J082606.37--010001.31 &
39627762691998320 &
126.5265493 &
-1.0003626 &
0.6273 &
Magellan/FIRE &
3593 &
2025-11-22 \\

J082921.37+131237.44 &
39628104418725302 &
127.3390417 &
13.2104000 &
0.3986 &
Magellan/FIRE &
2695 &
2025-11-22 \\

J094411.31--024908.65 &
39627720740570077 &
146.0471123 &
-2.8190691 &
0.6623 &
Magellan/FIRE &
2695 &
2025-11-22 \\

J101742.79+311459.07 &
39628506396624141 &
154.4282917 &
31.2497417 &
0.6706 &
-- &
-- &
-- \\

J102553.75+502843.24 &
39633255565102753 &
156.4739579 &
50.4786779 &
0.8824 &
LBT/LUCI &
9600 &
2026-02-16 \\

J104242.43+372147.63 &
39633024593167509 &
160.6767992 &
37.3632305 &
0.6080 &
LBT/LUCI &
3600 &
2025-12-23 \\

J132137.00--021417.04 &
39627733730332064 &
200.4041667 &
-2.2380666 &
0.2244 &
Magellan/FIRE &
2726 &
2026-03-15 \\

J142337.59+520216.05 &
39633278952540746 &
215.9066241 &
52.0377920 &
0.6236 &
Keck/NIRES &
3600 &
2026-04-27 \\

J150252.34+025027.97 &
39627854949912902 &
225.7180833 &
2.8411028 &
0.2906 &
Magellan/FIRE &
1818 &
2026-03-15 \\[3pt]

\multirow{2}{*}{J161102.44+091728.60} &
\multirow{2}{*}{39628011456174085} &
\multirow{2}{*}{242.7601667} &
\multirow{2}{*}{9.2912778} &
\multirow{2}{*}{0.6952} &
LBT/LUCI &
4800 &
2026-05-28, 2026-06-07 \\

& & & & & 
Magellan/FIRE &
5453 &
2026-04-03, 2026-05-28 \\[3pt]

J162032.32+314817.02 &
39628518019040271 &
245.1346667 &
31.8047278 &
0.4099 &
-- &
-- &
-- \\

J164102.65+070806.47 &
39627963724989207 &
250.2610417 &
7.1351306 &
0.5351 &
Magellan/FIRE &
6362 &
2026-04-03, 2026-05-28 \\

J164226.97+042632.79 &
39627897593401979 &
250.6123871 &
4.4424412 &
0.7291 &
Magellan/FIRE &
5453 &
2026-05-28 \\

J164637.91+142648.62 &
39628135796315287 &
251.6579757 &
14.4468384 &
0.7071 &
Magellan/FIRE &
9025 &
2026-04-04, 2026-05-28 \\

J165450.36+033741.74 &
39627879557893559 &
253.7098296 &
3.6282615 &
0.6408 &
Magellan/FIRE &
3635 &
2026-03-15 \\

J171741.74+380752.47 &
39633045061370969 &
259.4239167 &
38.1312417 &
0.1959 &
Keck/NIRES &
1800 &
2026-04-27 \\

J212725.88--044808.92 &
39627675429507611 &
321.8578333 &
-4.8024777 &
0.5842 &
Magellan/FIRE &
5453 &
2025-10-17 \\

J225535.58+154216.29 &
39628166473454559 &
343.8982384 &
15.7045243 &
0.4273 &
Magellan/FIRE &
3635 &
2025-10-03 \\

\hline
\multicolumn{8}{c}{\textbf{Silver sample}} \\
\hline

J000927.22+081109.96 &
39627983568242116 &
2.3634167 &
8.1861000 &
0.3720 &
-- &
-- &
-- \\

J024337.99+034915.97 &
39627875992735184 &
40.9082917 &
3.8211028 &
0.4584 &
-- &
-- &
-- \\

J105620.11+275415.87 &
39628438159494354 &
164.0837917 &
27.9044083 &
0.4617 &
-- &
-- &
-- \\

J105900.29+314951.74 &
39628516865610711 &
164.7512083 &
31.8310389 &
0.4990 &
-- &
-- &
-- \\

J111943.20+021911.32 &
39627841934986929 &
169.9300000 &
2.3198111 &
0.4682 &
Magellan/FIRE &
3635 &
2026-03-15 \\

J113734.35+552028.16 &
39633325291209396 &
174.3931250 &
55.3411556 &
0.4358 &
LBT/LUCI &
3600 &
2026-03-27 \\

J134317.81+393418.07 &
39633067916136674 &
205.8242083 &
39.5716861 &
0.2933 &
LBT/LUCI &
4500 &
2026-05-15 \\

J190954.15+583112.37 &
39633369390125151 &
287.4756250 &
58.5201028 &
0.4273 &
LBT/LUCI &
3600 &
2026-06-04 \\

\hline
\hline
\end{tabular}

%% file: 03_Measurement.tex
\section{Measurements}\label{sec:measurement}

In this section, we describe our methodology and measurements to characterize the continuum shapes and emission-line profiles of the LRDs identified in DESI DR1.

\begin{table*}
\centering
\noindent\makebox[\linewidth][c]{\input{table/continuum_properties.tex}\unskip}\par
\caption{{UV–Optical Continuum Properties} of the DESI DR1 LRD sample presented in this work. The Absorber column indicates the detected absorption superimposed on the broad emission lines: H$\alpha$, H$\beta$, or H$\alpha$+H$\beta$ (when absorption is detected in both). A dash (`--') indicates no absorption detected or no spectral coverage. {The machine-readable versions are available on Zenodo as \texttt{continuum\_table.ecsv} and \texttt{main\_lines\_table.ecsv}.} \label{tab:continuum_property}}
\end{table*}

\subsection{{Continuum parameterization}}
\label{sec:continuum_param}

We measure the UV absolute magnitude at 1500\,\AA\ ($M_{\rm UV}$) by jointly modeling the DESI spectra and broadband photometry at rest-frame wavelengths $<3500$\,\AA. The \ion{Mg}{2} lines are masked, and the continuum is fitted with a power-law model normalized at rest-frame 1500\,\AA. We measure the Balmer break strength following {\cite{deGraaff2025_all}}, defined as the ratio of the continuum flux in the rest-frame 3620–3720\,\AA\ and 4000–4100\,\AA\ bands.

\begingroup
\setlength{\abovedisplayskip}{6pt}
\setlength{\abovedisplayshortskip}{6pt}
\setlength{\belowdisplayskip}{6pt}
\setlength{\belowdisplayshortskip}{6pt}
To depict the V-shaped continua of the selected LRDs, we parameterize their {UV–optical continua} using a simple broken power-law model, following \cite{Setton2024}:
\begin{equation}\label{eq:broken_powerlaw}
f_\lambda =
\begin{cases}
f_{\lambda_{\rm v}} (\lambda/\lambda_{\rm v})^{k_{\rm blue}}, & \lambda < \lambda_{\rm v} \\
f_{\lambda_{\rm v}} (\lambda/\lambda_{\rm v})^{k_{\rm red}},  & \lambda > \lambda_{\rm v}
\end{cases}
\end{equation}
where $f_{\lambda_{\rm v}}$ is the flux at the inflection point $\lambda_{\rm v}$, and $k_{\rm blue}$ and $k_{\rm red}$ are the power-law slopes of the continuum blueward and redward of $\lambda_{\rm v}$, respectively. We note that $k_{\rm blue}$ is not equivalent to {the UV-continuum slope} ($\beta_{\rm UV}$) used in the $M_{\rm UV}$ measurement. $k_{\rm blue}$ is a locally fitted parameter describing the continuum in the vicinity (1000–1500\,\AA) of $\lambda_{\rm v}$, whereas $\beta_{\rm UV}$ characterizes the overall shape of the rest-frame UV continuum.

\par\endgroup

The continuum fitting is performed iteratively in three steps.  We first mask strong emission lines in the DESI spectrum, including \ion{Mg}{2}, [\ion{O}{2}], [\ion{Ne}{3}], [\ion{O}{3}], as well as all hydrogen and helium lines. We then bin the DESI spectrum into 300 bins and use \textsc{emcee} to fit the model to the binned continuum. In the first iteration, we apply a median filter to the binned continuum and identify the minimum flux point. We then fit the broken power-law model (Equation~\ref{eq:broken_powerlaw}) within $\pm2000$\,\AA\ of this minimum to obtain an initial estimate of $\lambda_{\rm v}$. In the second iteration, we refit the broken power-law model within $\pm2000$\,\AA\ centered on this initial $\lambda_{\rm v}$. In the final iteration, we restrict the fitting window to $\pm1000$–$1500$\,\AA\ centered on the second estimate of $\lambda_{\rm v}$ to obtain locally optimized values of $\lambda_{\rm v}$, $k_{\rm blue}$, and $k_{\rm red}$. For comparison, we also apply the same parameterization to the three local LRDs reported by \citetalias{Lin2025_localLRD}.

\subsection{{Emission Line Measurements}}\label{sec:emission_line_measurement}

To model the emission-line profiles, we perform a joint fit to \hb, \oiii, \ha, and \nii.  We assume that the narrow components of {all of these lines} share the same FWHMs. { The broad components of \hb\ and \ha\ are initially modeled using three components each, with a common redshift. The FWHMs of corresponding broad components in \hb\ and \ha\ are tied together.}  The \oiii\ and \nii\ doublet flux ratios are fixed at 0.335 and 0.327, respectively. The continuum underlying \hb\ and \oiii\ is modeled as a linear function normalized at rest-frame 5100 \AA, and the \ha\ continuum is assumed to be constant.  

Starting from the initial models, the {emission-line} profiles are adjusted iteratively during the fit to account for their diversity. The adjustments are summarized below.

\begin{itemize}
\item If \oiii\ cannot be well modeled by a single Gaussian profile, an additional Gaussian component is added to both \oiii\ and \nii, with shared FWHMs and velocity offsets. This component accounts for outflows in the ionized gas. The doublet flux ratios are fixed.

\item If a broad Gaussian component in \ha\ and/or \hb\ exhibits an FWHM close to that of the outflowing component in [\ion{O}{3}], we tie them to share the same FWHM and velocity offset. The luminosities of the outflow components in [\ion{O}{3}], \hb, and \ha\ are treated as independent.

\item If one broad Gaussian component has a luminosity consistent with zero (i.e., the posterior probability of its intensity peaks at zero), we reduce the number of broad components accordingly.

\end{itemize}

In the fitting described above, the Bayesian {information criterion} (BIC) is used to compare different models and to determine whether an additional component should be included or removed in the [\ion{O}{3}] or Balmer lines. If the BIC improves { by 10} \citep{Liddle2007}, we adopt the more complex model; otherwise, we retain the simpler model.

{ We note that the multiple-Gaussian model described above is used only as a phenomenological representation of the observed broad-line profiles. The interpretation of Balmer-line profiles in LRDs remains debated, ranging from exponential profiles attributed to electron-scattering broadening {\citep[e.g.,][]{Rusakov2025}} to multi-Gaussian models associated with stratified BLRs \citep[e.g.,][]{Scholtz2026}. Here, the multiple Gaussians serve only to reproduce the observed broad component, and we do not assign physical significance to the individual components.}

For \hb\ or \ha\ lines showing absorption, we add an absorption component. The total Balmer line profile, including absorption, is parameterized as
\begin{align}\label{eq:abs}
F_\lambda &= \left( 1 - C_f + C_f e^{-\tau(\Delta v_{\rm abs}, \,\log N,\,b)} \right) \cdot (c_{\rm cont} + \phi_{\rm broad}) \nonumber \\
&\quad + \phi_{\rm narrow} + \phi_{\rm outflow}
\end{align}
where $C_f$ is the covering factor, $c_{\rm cont}$ is the continuum, $\phi_{\rm broad}$ is the {broad emission-line profile}, $\phi_{\rm narrow}$ is the {narrow emission-line profile}, and $\phi_{\rm outflow}$ represents the outflow component, if present. $\tau$ is the optical depth of the absorber, modeled as a Voigt profile as a function of the column density $\log N$ and Doppler parameter $b$, and determined by the transition and oscillator strength of \hb\ or \ha. $\Delta v_{\rm abs}$ represents the velocity shift of the absorber relative to the {emission-line center}, with negative values indicating a blueshift. In Equation~\ref{eq:abs},  the absorber absorbs both the optical continuum $c_{\rm cont}$ and the broad emission lines. This is motivated by the fact that the absorption trough can extend below the continuum level, as seen in J165450.36+033741.74 (Figure \ref{fig:example1}). We treat  \hb\ and \ha\ absorbers, when both are present, as independent, i.e., with independent $\log N$, $b$, $C_f$, and $\Delta v_{\rm abs}$.  This assumption is motivated by the fact that they can exhibit opposite velocity shifts, as observed in The Egg (J1025+1402{;} \citetalias{Lin2025_localLRD},  {\citealt{Ji2025_lord}}), {A2744-QSO1} {\citep{Ji2025, D'Eugenio2025_z7p04}}, and Irony \citep{DEugenio2025_irony}. { When the posterior probability of $C_f$ accumulates at unity, we fix $C_f=1$ and rerun the fit. Given the physical bound $0 \leq C_f \leq 1$, such boundary accumulation indicates that the data favor complete covering while providing no two-sided constraint on $C_f$. Adopting $C_f=1$ therefore corresponds to the physically preferred limiting solution.}

The intrinsic profiles described above are then convolved to the spectral resolution. 
 For DESI spectra, we convolve the intrinsic profile with the DESI {line-spread} function (LSF) using its resolution matrices \citep{Bolton2010, Guy2023}. For near-IR spectra, we convolve the profile with a Gaussian kernel corresponding to the spectral resolution $R$, as derived from {skyline} FWHMs. During the modeling, $R$ is treated as a free parameter but is constrained within the range of skyline FWHMs across the slit/order.  It ensures that the fitted results incorporate the uncertainty in $R$.  The fitting is performed using \textsc{emcee}.

For other lines, including \ion{Mg}{2}$\lambda\lambda$2796,2803, [\ion{O}{2}]$\lambda\lambda3727,3730$, [\ion{Ne}{3}]$\lambda 3870$, {and} [\ion{O}{3}]$\lambda 4364$, we model them with simple Gaussian profiles convolved with LSFs using \textsc{lmfit} \citep{lmfit}. A linear function is adopted for the underlying continuum. Doublets or lines with closely spaced wavelengths are fitted simultaneously. For [\ion{O}{3}]$\lambda4364$, the [\ion{Fe}{2}] $\lambda4360$ line is fitted simultaneously to remove its contamination. In cases of {nondetection}, we derive 3$\sigma$ upper limits by {integrating the rms} over a spectral window set by the FWHM of [\ion{Ne}{3}], typically around 100 km\,s$^{-1}$.

\subsection{Gold and Silver Samples}\label{sec:gold_silver}

Based on the continuum parameterization described in Section~\ref{sec:continuum_param}, we classify our sample into two tiers: \textsc{Gold} and \textsc{Silver}.
\begin{itemize}
    \item \textsc{Gold}: sources with $k_{\rm red} > 0$ at $\geq 3\sigma$ significance. In the DESI DR1 LRD sample in this work, 19 out of 27 sources are classified as \textsc{Gold}.
    \item \textsc{Silver}: sources with $k_{\rm red} > 0$ at $< 3\sigma$ significance.  {Eight} out of 27 sources are classified as \textsc{Silver}.
\end{itemize}

The \textsc{Gold}/\textsc{Silver} classification reflects both the intrinsically {bluer optical-continuum slopes} (while still $k_{\rm red} > 0$) and the S/N of the measurements. All \textsc{Silver} objects meet our LRD selection criteria. Notably, four of the eight \textsc{Silver} sources exhibit clear Balmer absorption (e.g., J1909+5831), a feature commonly observed among established LRDs. This classification is primarily used to prioritize follow-up observations. {Higher-S/N} data will place tighter constraints on their optical slopes.

We summarize the measured continuum properties and Balmer absorption identification of our sample in Table~\ref{tab:continuum_property}. The full fitting results, including best-fit parameters and associated uncertainties, are provided in Appendix~\ref{appendix:fitting_results}. {The emission-line properties are listed in Table~\ref{tab:emission_line_properties} and the ionized gas outflow properties in Table~\ref{tab:outflow}.} We note that the two tiers of samples occupy the same distribution in parameter space. In the following analysis, we treat them as a whole.

%% file: table/continuum_properties.tex
\begin{tabular}{lccccccc}
\hline
Name & $M_{\rm UV}$ & $\lambda L_{5100}$ & $\lambda_{\rm v}$ & $k_{\rm blue}$ & $k_{\rm red}$ & Balmer break & Absorber \\
 & (mag) & ($10^{43}\,\mathrm{erg\,s^{-1}}$) & (\AA) &  &  & strength &  \\
\hline
\multicolumn{8}{c}{\textbf{Gold sample}} \\
\hline
J0129+0628 & -16.4$\pm$0.4 & {0.43$_{-0.04}^{+0.04}$} & 6196$_{-124}^{+142}$ & {-0.8$_{-0.4}^{+0.4}$} & 1.7$_{-0.5}^{+0.4}$ & 1.0$\pm$0.1 & H$\alpha$ \\
J0826--0100 & -18.9$\pm$0.2 & 5.3$_{-0.3}^{+0.3}$ & 3749$_{-144}^{+117}$ & {-1.3$_{-0.4}^{+0.4}$} & 1.6$_{-0.3}^{+0.3}$ & 1.5$\pm$0.3 & H$\alpha$+H$\beta$ \\
J0829+1312 & -18.6$\pm$0.2 & 1.8$_{-0.1}^{+0.1}$ & 4795$_{-153}^{+105}$ & {-1.7$_{-0.2}^{+0.1}$} & 0.9$_{-0.1}^{+0.1}$ & 0.7$\pm$0.1 & H$\alpha$ \\
J0944--0249 & -20.1$\pm$0.1 & 12.4$_{-0.3}^{+0.3}$ & 3653$_{-6}^{+17}$ & {-2.3$_{-0.2}^{+0.2}$} & 2.1$_{-0.2}^{+0.2}$ & 1.7$\pm$0.2 & H$\alpha$+H$\beta$ \\
J1017+3114 & -17.9$\pm$0.2 & 3.2$_{-0.1}^{+0.1}$ & 3958$_{-80}^{+83}$ & {-1.5$_{-0.2}^{+0.2}$} & 1.8$_{-0.2}^{+0.2}$ & 1.3$\pm$0.2 & -- \\
J1025+5028 & -20.3$\pm$0.1 & 15.5$_{-0.9}^{+0.9}$ & 3660$_{-11}^{+28}$ & {-2.5$_{-0.2}^{+0.2}$} & 3.0$_{-0.3}^{+0.3}$ & 1.5$\pm$0.2 & H$\alpha$+H$\beta$ \\
J1042+3721 & -19.1$\pm$0.2 & 6.0$_{-0.3}^{+0.3}$ & 3532$_{-134}^{+119}$ & {-2.2$_{-0.4}^{+0.4}$} & 1.9$_{-0.3}^{+0.3}$ & 1.3$\pm$0.3 & -- \\
J1321--0214 & -16.8$\pm$0.6 & {0.21$_{-0.04}^{+0.04}$} & 4376$_{-172}^{+138}$ & {-2.8$_{-0.6}^{+0.5}$} & 2.2$_{-0.3}^{+0.4}$ & 0.6$\pm$0.1 & H$\alpha$ \\
J1423+5202 & -16.2$\pm$0.5 & 4.8$_{-0.2}^{+0.2}$ & 3714$_{-46}^{+62}$ & {-0.5$_{-0.6}^{+0.4}$} & 2.4$_{-0.1}^{+0.1}$ & 3.2$\pm$1.7 & H$\alpha$+H$\beta$ \\
J1502+0250 & -18.1$\pm$0.4 & 0.8$_{-0.1}^{+0.1}$ & 3759$_{-132}^{+162}$ & {-2.6$_{-0.6}^{+0.6}$} & 0.9$_{-0.1}^{+0.1}$ & 0.9$\pm$0.1 & -- \\
J1611+0917 & -20.0$\pm$0.1 & 6.5$_{-0.5}^{+0.5}$ & 3695$_{-31}^{+48}$ & {-2.7$_{-0.2}^{+0.2}$} & 2.8$_{-0.4}^{+0.4}$ & 1.4$\pm$0.2 & H$\alpha$+H$\beta$ \\
J1620+3148 & -14.9$\pm$1.0 & 0.5$_{-0.1}^{+0.1}$ & 4089$_{-286}^{+296}$ & {-2.4$_{-1.5}^{+1.4}$} & 1.7$_{-0.5}^{+0.5}$ & 1.2$\pm$1.1 & -- \\
J1641+0708 & -17.7$\pm$0.5 & -- & 3961$_{-120}^{+148}$ & {-1.6$_{-0.6}^{+0.7}$} & 3.2$_{-0.2}^{+0.2}$ & 2.0$\pm$1.3 & H$\alpha$ \\
J1642+0426 & -20.5$\pm$0.1 & 11.9$_{-0.4}^{+0.4}$ & 3663$_{-13}^{+16}$ & {-2.2$_{-0.2}^{+0.2}$} & 1.7$_{-0.2}^{+0.2}$ & 1.3$\pm$0.1 & H$\alpha$ \\
J1646+1426 & -18.6$\pm$0.2 & 11.0$_{-0.4}^{+0.4}$ & 3728$_{-51}^{+63}$ & {-0.9$_{-0.3}^{+0.3}$} & 2.3$_{-0.3}^{+0.4}$ & 1.4$\pm$0.2 & H$\alpha$+H$\beta$ \\
J1654+0337 & -18.3$\pm$0.2 & 10.2$_{-0.3}^{+0.3}$ & 3722$_{-51}^{+69}$ & {-1.4$_{-0.4}^{+0.4}$} & 3.8$_{-0.4}^{+0.4}$ & 2.6$\pm$0.8 & H$\alpha$+H$\beta$ \\
J1717+3807 & -19.1$\pm$0.1 & {1.82$_{-0.03}^{+0.03}$} & 4913$_{-23}^{+22}$ & {-1.48$_{-0.05}^{+0.05}$} & {1.56$_{-0.03}^{+0.03}$} & {0.84$\pm$0.04} & H$\alpha$+H$\beta$ \\
J2127--0448 & -17.5$\pm$0.5 & 4.0$_{-0.2}^{+0.2}$ & 3451$_{-174}^{+195}$ & {-2.1$_{-1.0}^{+0.8}$} & 2.6$_{-0.8}^{+0.8}$ & 1.8$\pm$0.8 & H$\alpha$+H$\beta$ \\
J2255+1542 & -17.4$\pm$0.7 & 0.8$_{-0.1}^{+0.1}$ & 3955$_{-216}^{+249}$ & {-2.7$_{-1.0}^{+1.0}$} & 2.4$_{-0.4}^{+0.4}$ & 1.2$\pm$0.9 & H$\alpha$ \\
\hline
\multicolumn{8}{c}{\textbf{Silver sample}} \\
\hline
J0009+0811 & -17.5$\pm$0.3 & 1.1$_{-0.1}^{+0.1}$ & 3739$_{-100}^{+140}$ & {-2.5$_{-0.4}^{+0.5}$} & 0.2$_{-0.2}^{+0.2}$ & 1.1$\pm$0.2 & -- \\
J0243+0349 & -18.2$\pm$0.4 & 0.8$_{-0.1}^{+0.1}$ & 3945$_{-225}^{+209}$ & {-2.9$_{-0.7}^{+0.7}$} & 0.5$_{-0.3}^{+0.4}$ & 1.0$\pm$0.4 & -- \\
J1056+2754 & -18.6$\pm$0.3 & 1.2$_{-0.1}^{+0.1}$ & 4235$_{-217}^{+343}$ & {-2.3$_{-0.3}^{+0.4}$} & 0.0$_{-0.1}^{+0.1}$ & 1.2$\pm$0.3 & -- \\
J1059+3149 & -18.4$\pm$0.4 & 4.0$_{-0.5}^{+0.5}$ & 3829$_{-114}^{+174}$ & {-1.6$_{-0.7}^{+0.6}$} & 0.6$_{-0.5}^{+0.7}$ & 1.5$\pm$0.6 & -- \\
J1119+0219 & -18.5$\pm$0.2 & 0.7$_{-0.1}^{+0.1}$ & 5397$_{-141}^{+78}$ & {-0.9$_{-0.4}^{+0.4}$} & 0.8$_{-0.4}^{+0.4}$ & 0.9$\pm$0.2 & H$\alpha$ \\
J1137+5520 & -15.3$\pm$0.7 & 0.9$_{-0.1}^{+0.1}$ & 4667$_{-376}^{+336}$ & {-1.1$_{-0.5}^{+0.5}$} & 0.5$_{-0.4}^{+0.5}$ & 0.8$\pm$0.4 & H$\alpha$ \\
J1343+3934 & -17.5$\pm$0.4 & 0.5$_{-0.1}^{+0.1}$ & 5459$_{-95}^{+32}$ & {-1.0$_{-0.4}^{+0.4}$} & 0.5$_{-0.4}^{+0.4}$ & 0.8$\pm$0.2 & H$\alpha$ \\
J1909+5831 & -18.8$\pm$0.5 & 0.7$_{-0.1}^{+0.1}$ & 3933$_{-242}^{+220}$ & {-2.9$_{-0.5}^{+0.5}$} & 0.3$_{-0.2}^{+0.2}$ & 1.3$\pm$0.7 & H$\alpha$ \\
\hline
\end{tabular}

%% file: 04_Results.tex
\section{Results}\label{sec:result}

In this section, we quantify the properties of the DESI LRDs. Both the \textsc{Gold} and \textsc{Silver} samples are included in the analysis. The two tiers occupy the same regions of parameter space across all diagnostic diagrams.

\subsection{Evidences of LRD Nature}

We begin by establishing that the DESI sources are consistent with high-redshift LRDs. The principal observational signatures commonly associated with the majority of JWST LRDs are present in our sample, spanning morphology, continuum shape,  Balmer-line and narrow-line properties. The consistency is summarized below and examined in detail in the subsections indicated.

\begin{itemize}
    \item[(1)] {Compact morphologies.}

    \item[(2)] Continuum properties:
    \begin{itemize}[leftmargin=1.5em, itemsep=1pt, topsep=1pt]
        \item V-shaped rest-frame {UV–optical continua} ($\beta_{\rm UV} < 0$, $\beta_{\rm optical} > 0$) (Section~\ref{sec:continuum_luminosity}){.}
        
        \item Inflection points near or redward of the Balmer limit (Section~\ref{sec:continuum_luminosity}){.}
        \item {Declining rest-frame near-IR continua} (see Section~\ref{sec:hr_diagram}){.}
    \end{itemize}
    
    \item[(3)]  Balmer-line properties:
    \begin{itemize}[leftmargin=1.5em, itemsep=1pt, topsep=1pt]
        \item Broad Balmer emission (Section \ref{sec:balmer_emission}){.}
        \item Extremely high broad-line Balmer decrements (Section~\ref{sec:balmer_decrement}){.}
        \item Frequent Balmer absorption (Section~\ref{sec:balmer_absorption}){.}
    \end{itemize}
    
    \item[(4)]  Narrow emission-line properties:
    \begin{itemize}[leftmargin=1.5em, itemsep=1pt, topsep=1pt]
        \item Low metallicity, similar to that of high-redshift galaxies (Section~\ref{sec:metallicity_BPT}){.}
        \item Occupying the same region as high-$z$ LRDs in the BPT diagram (Section~\ref{sec:metallicity_BPT}){.}
        \item Softer ionizing spectra than typical quasars and type~1 AGNs (Section \ref{sec:He2}){.}
        \end{itemize}
\item[(5)] Variability:
\begin{itemize}[leftmargin=1.5em, itemsep=1pt, topsep=1pt]
\item The majority show no significant optical variability on rest-frame month timescales.
 (Section \ref{sec:optical_variability}).

\end{itemize}  

\end{itemize}

All of these properties are consistent with those of high-$z$ LRDs and occupy the same distributions (e.g., the continuum and Balmer line luminosities). 
Note that although some of the commonalities between low- and high-$z$ populations (e.g., continuum shape, morphology, broad emission lines) resulted from our adopted low$-z$ LRD selection criteria (\ref{sec:selection_guideline}), others (e.g., Balmer absorption, Balmer decrements, low metallicity, variability) are not part of the LRD selection. 
This overall consistency from continuum to line properties indicates that the features observed in both high- and low-$z$ LRDs are governed by similar underlying physical processes. Therefore, these DESI LRDs are representative of the LRD population, but just reside at $z<1$.

\subsection{LRD number density at $z<1$}\label{sec:number_density}

We estimate the number density of LRDs from our DESI DR1 sample. The survey covers {9739} deg$^{2}$, and our selection is designed to identify LRDs with \hb\ and/or \ha\ lines observable by DESI, i.e., at $z<1$. This corresponds to a total survey volume of $3.6\times10^{10}$ cMpc$^3$. Based on the full sample of 27 DESI DR1 LRDs, we derive a number density of $7.5\times10^{-10}$ cMpc$^{-3}$.  
The estimate is in good agreement with the $5\times10^{-10}$ cMpc$^{-3}$ estimated from SDSS-selected LRDs at $z=0$–0.5 \citepalias{Lin2025_localLRD}. According to {\cite{Ma2025}}, the number density at $z<1$ is about 4.6 orders of magnitude lower than at $z>4$, 3.8 orders of magnitude lower than at $z \approx 2.7$–$3.7$, and 3.4 orders of magnitude lower than at $1.7 < z < 2.7$.

However, the number density derived above should be considered a conservative lower limit. As noted by \citetalias{Lin2025_localLRD}, LRDs identified in wide-field spectroscopic surveys such as SDSS and DESI are highly incomplete. In this work, the selection is limited by both the photometric pre-selection of DESI targets and the survey’s cosmology-driven targeting strategy (see Section \ref{sec:desi_target_bits}). Future surveys beyond the limitation of SDSS/DESI pre-selection are needed to establish a complete sample of low-$z$ LRDs.

\subsection{Continuum shape and luminosity }\label{sec:continuum_luminosity}

\begin{figure*}
    \centering
\includegraphics[width=\linewidth]{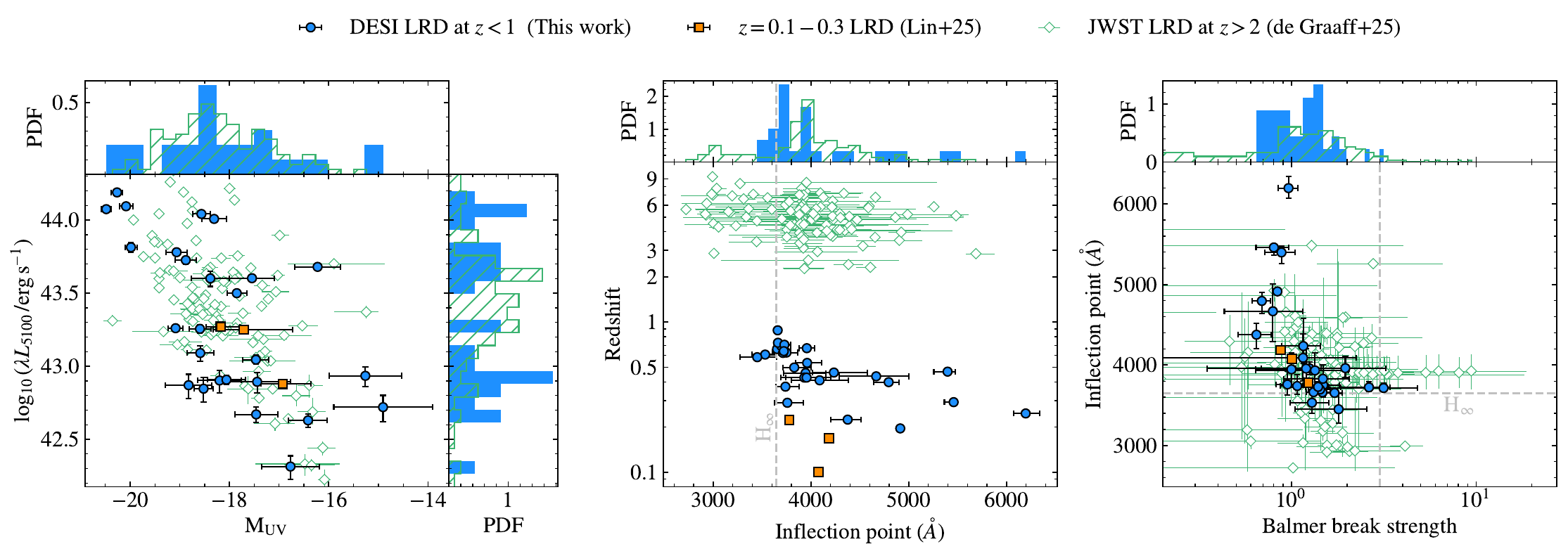}
    \caption{\textit{Left}: {the} distribution of $L_{\rm 5100}$ and $M_{\rm UV}$ of LRDs across cosmic time.  \textit{Middle}: {the} distribution of V-shaped SED inflection points in LRDs. \textit{Right}: {the} distribution of inflection points and Balmer break strengths. In each panel, LRDs at $z<1$ from DESI DR1  are shown as blue circles. Local LRDs at $z=0.1$–$0.3$ in \citetalias{Lin2025_localLRD} are orange squares, and JWST-discovered LRDs at $z>2$ {\citep{deGraaff2025_all}} are green. In the histogram, blue bars show the DESI DR1 LRD distribution, while green bars indicate JWST LRDs at $z>2$. In the \textit{middle} and \textit{right} panels, the wavelengths of the Balmer limit are marked with gray dashed lines. In the \textit{right} panel, we also include a gray dashed vertical line indicating a Balmer break strength of 3, the maximum expected for a dust-free stellar population. }
    \label{fig:continnum_distribution}
\end{figure*}

We demonstrate the distribution of the continuum luminosity of DESI DR1 LRDs in the left panel of Figure \ref{fig:continnum_distribution}. Their $L_{\rm 5100}$ and $M_{\rm UV}$ values span the same range as JWST-discovered LRDs at $z>2$, with $L_{\rm 5100}$ between $2\times10^{42}$ and $2\times10^{44}$ erg s$^{-1}$, and $M_{\rm UV}$ from –20.5 to approximately –15 mag.

The middle panel of Figure~\ref{fig:continnum_distribution} shows the distribution of V-shaped SED inflection points in LRDs. While many inflection points lie near the Balmer limit, a notable fraction occur at longer wavelengths. Six objects in our sample show inflection points $> 4500$\,\AA. Such behavior is also seen in certain $z>2$ LRDs {\citep{Setton2025, deGraaff2025_all, Barro2025}}. The subset of long-wavelength inflection points benefits from our more {relaxed optical-continuum selection} criteria in Section~\ref{sec:selection_red_continuum}. A clear example is J1717+3807, which was targeted by SDSS eBOSS but missed in the selection of  \citetalias{Lin2025_localLRD}. This is because \citetalias{Lin2025_localLRD} required the continuum under [\ion{O}{2}] to be lower than that under [\ion{O}{3}] 5008 for the inflection point to lie near the Balmer limit, but such a condition is not met by this object. In J1717+3807, the inflection point occurs near [\ion{O}{3}] 5008, and our more relaxed {red optical-continuum criteria} allow it to be recovered in the sample.

For most DESI DR1 LRDs with inflection points near the Balmer limit, the Balmer break strength exceeds unity, as shown in the right panel of Figure~\ref{fig:continnum_distribution}. The maximum Balmer break strength in our sample reaches $\sim$3, observed in J1423+5202 and J1654+0337. While we do not identify sources with break strengths as extreme as the most prominent high-$z$ LRDs (e.g., {The Cliff; }{\citealt{deGraaff2025}}, MoM-BH$^*$, {\citealt{Naidu2025}}), we caution that this is likely a selection effect arising from our requirement of strong [\ion{O}{3}] emission (The Cliff with [\ion{O}{3}] EW 7.3 \AA, MoM-BH$^*$ with [\ion{O}{3}] EW 3 \AA). Additionally, sources with inflection points at longer wavelengths naturally exhibit weaker Balmer break strengths. In this case, the Balmer break alone loses its diagnostic power.

\subsection{Balmer lines}\label{sec:balmer_lines}

\subsubsection{Balmer emission}\label{sec:balmer_emission}

\begin{figure*}
    \centering
\includegraphics[width=\linewidth]{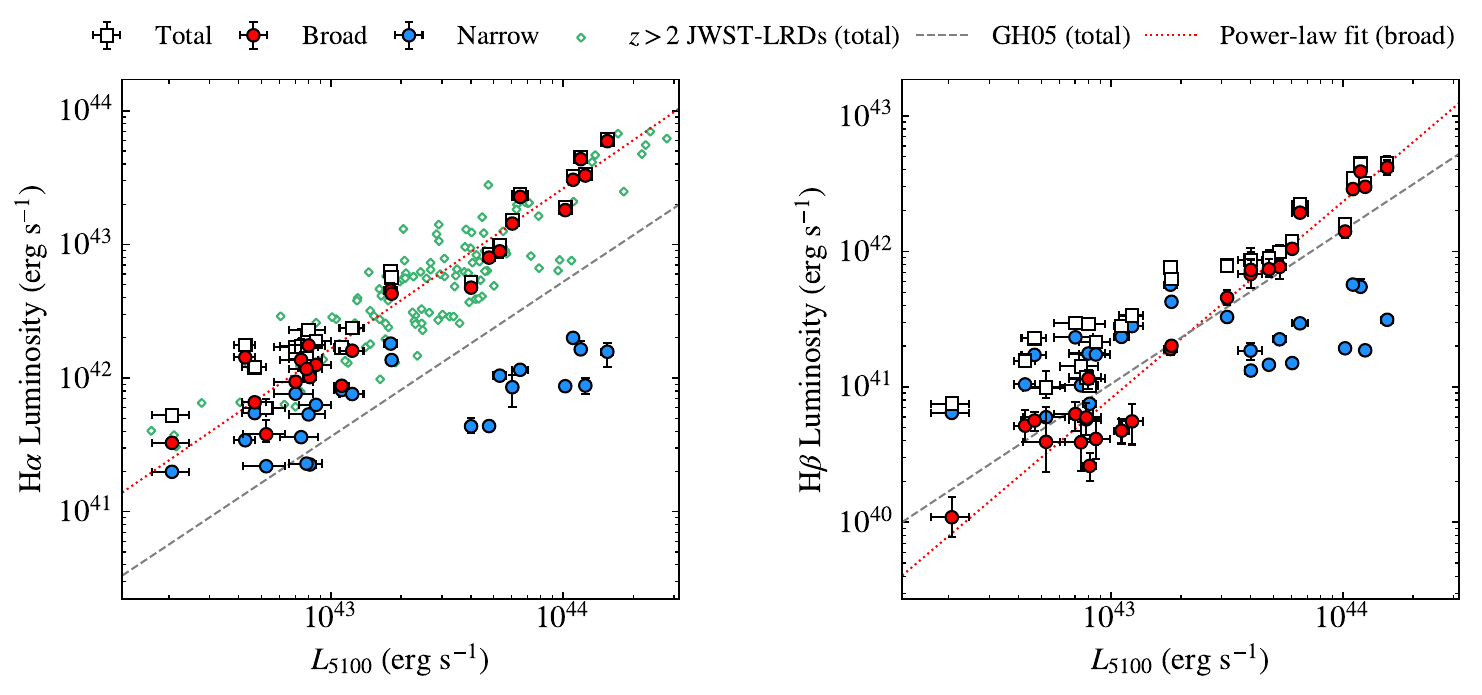}
    \caption{\textit{Left}: H$\alpha$ luminosity of DESI DR1 LRDs {vs.} $L_{\rm 5100}$. Total (broad+narrow), broad, and narrow H$\alpha$ luminosities are shown as white squares, red circles, and blue circles, respectively. The total \ha\ luminosity of $z>2$ LRDs from {\cite{deGraaff2025_all}} is shown as green diamonds. The correlation between total H$\alpha$ luminosity and $L_{\rm 5100}$ for type-1 AGNs from \cite{Greene2005} is shown as the gray dashed line. The correlation between broad H$\alpha$ luminosity and $L_{\rm 5100}$ is shown as the red dashed line.
\textit{Right}:  Same as the \textit{Left} panel, but for \hb. 
}
\label{fig:balmer_luminosity}
\end{figure*}

The left panel of Figure~\ref{fig:balmer_luminosity} shows the relation between H$\alpha$ luminosity and $L_{\rm 5100}$. The broad component ($L_{\rm H\alpha, broad}$) tightly correlates with $L_{\rm 5100}$. The correlation between total H$\alpha$ luminosity ($L_{\rm H\alpha, total}$) and $L_{\rm 5100}$ is similar to that observed in high-$z$ LRDs, but is primarily dominated by the broad-line emission. In contrast, the narrow H$\alpha$ luminosity does not exhibit a strong increase with $L_{\rm 5100}$ at $L_{\rm 5100} > 3 \times 10^{42}$~erg~s$^{-1}$. We fit the correlation between $L_{\rm H\alpha, broad}$ and $L_{\rm 5100}$, yielding

\begin{align}\label{eq:Lha_L5100}
\log \left( \frac{L_{\rm H\alpha,broad}}{10^{42}~{\rm erg~s^{-1}}} \right) 
&= (1.42 \pm 0.03) \nonumber\\
&\quad + (1.20 \pm 0.06) \cdot 
\log \left( \frac{L_{\rm 5100}}{10^{44}~{\rm erg~s^{-1}}} \right)
\end{align}

For H$\beta$ (right panel of Figure~\ref{fig:balmer_luminosity}), although the total H$\beta$ luminosity ($L_{\rm H\beta, total}$) correlates with $L_{\rm 5100}$, the correlation is dominated by the narrow component at $\log(L_{\rm 5100}/{\rm erg~s^{-1}})<43.2$, but by the broad component at higher $L_{\rm 5100}$. While the narrow H$\beta$ increases moderately with $L_{\rm 5100}$, the broad H$\beta$ luminosity ($L_{\rm H\beta, broad}$) exhibits a stronger correlation and can be parameterized as

\begin{align}\label{eq:Lhb_L5100}
\log \left( \frac{L_{\rm H\beta,broad}}{10^{42}~{\rm erg~s^{-1}}} \right) 
&= (0.37 \pm 0.02) \nonumber\\
&\quad + (1.45 \pm 0.06) \cdot 
\log \left( \frac{L_{\rm 5100}}{10^{44}~{\rm erg~s^{-1}}} \right)
\end{align}
The $L_{\rm H\alpha, broad}$–$L_{\rm 5100}$ and $L_{\rm H\beta, broad}$–$L_{\rm 5100}$ relations for LRDs (Equations~\ref{eq:Lha_L5100} and \ref{eq:Lhb_L5100}) do not follow the trends observed for type-1 AGNs in \cite{Greene2005} as shown in %
Figure~\ref{fig:balmer_luminosity}. For the $L_{\rm H\alpha, broad}$–$L_{\rm 5100}$ relation, the slope is close to that reported by \cite{Greene2005} ($1.157\pm0.005$), but the normalization is {5} times higher. In contrast, the $L_{\rm H\beta, broad}$–$L_{\rm 5100}$ relation exhibits a steeper slope than H$\alpha$ (right panel of Figure~\ref{fig:balmer_luminosity}).

First, the two relations deviate from those of local type-1 AGNs in \cite{Greene2005}, indicating that single-epoch BH mass estimators \citep{Greene2005, Reines2015} are not directly applicable to LRDs. These estimators rely on two key assumptions: (1) that $L_{\rm 5100}$ traces {the BLR size} as in local type-1 AGNs, and (2) that the broad-line luminosity scales with $L_{\rm 5100}$ as in local type-1 AGNs. Neither assumption could hold in LRDs, particularly in the context of the {BH–envelope} framework discussed in recent studies.

Furthermore, the tight correlations between $L_{\rm H\alpha, broad}$–$L_{\rm 5100}$ and $L_{\rm H\beta, broad}$–$L_{\rm 5100}$ in LRDs suggest that the broad emission lines and optical continuum are closely coupled in their physical origin. The relation points to a potential new calibration for BH mass measurements. Why the $L_{\rm H\alpha, broad}$–$L_{\rm 5100}$ relation shares a similar slope as that in \cite{Greene2005} remains an open question. Another key question is how these correlations fit within the physical picture of LRDs. In particular, within the leading BH–envelope framework, $L_{\rm 5100}$ is often located blueward of the peak of a blackbody-like continuum, whose temperature varies across sources. This question remains unclear and warrants further investigation.

\subsubsection{Balmer decrement}\label{sec:balmer_decrement}

\begin{figure*}
\includegraphics[width=\textwidth]{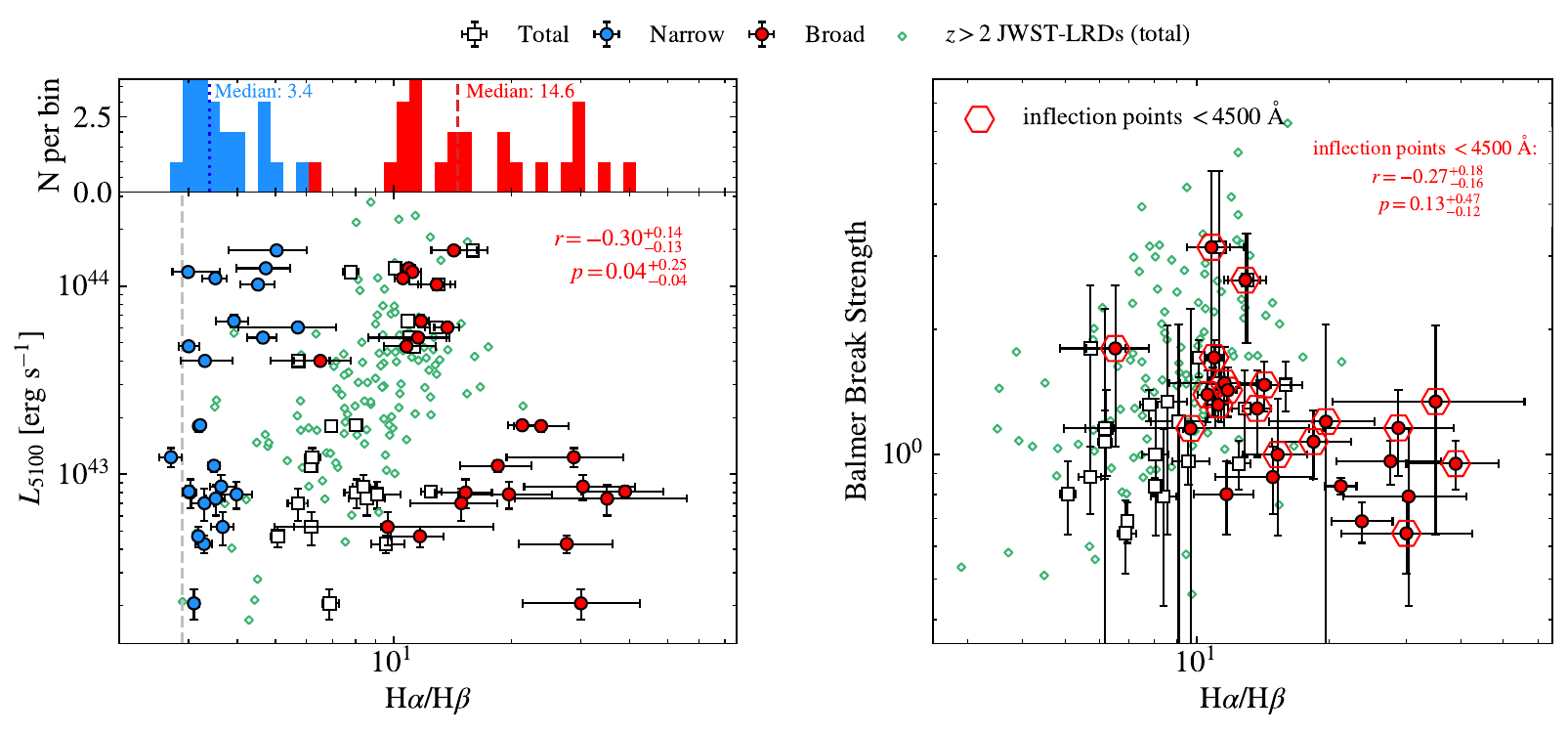}
\caption{\textit{Left}: Balmer decrement of the broad and narrow components {vs.} $L_{\rm 5100}$. Total (broad+narrow), broad, and narrow \ha/\hb\ are shown as white squares, red circles, and blue circles, respectively. The total \ha/\hb\ of $z>2$ LRDs from {\cite{deGraaff2025_all}} {is shown with green diamonds.  The top panel} shows the narrow-line H$\alpha$/H$\beta$ distribution in blue and the broad-line distribution in red, with median values labeled.  \textit{Right}: Balmer decrement {vs.} Balmer break strength. A subset of the DESI DR1 LRDs with inflection points blueward of 4500\,\AA\ {is marked with the red hexagons}.}
\label{fig:balmer_decrement}
\end{figure*}

We further examine the Balmer decrements in LRDs as shown in the left panel of 
Figure~\ref{fig:balmer_decrement}. 
The narrow lines exhibit relatively small H$\alpha$/H$\beta$ ratios, with a median value of 3.4, compared to the Case B value of 2.89. In contrast, the broad lines display much larger Balmer decrements, with a median value of 14.6. As already discussed in numerous studies of LRDs {\citep[e.g.,][]{Li2025,Setton2025,Casey2025, Chen2025}}, such extreme values are difficult to explain solely by dust attenuation. For an observed \ha/\hb\ ratio of 14.6, the implied $A_V$ under an SMC extinction law is 4.7, assuming an intrinsic ratio of 2.89. This high $A_V$ would produce substantial dust-reprocessed emission in the IR, which is in conflict with the declining IR continuum that we observe. Instead, these extreme Balmer decrements may arise from radiative transfer effects in sufficiently high-density gas ($n_{\rm H}\gtrsim10^{8}$–$10^{10} \mathrm{cm^{-3}}$), without necessarily invoking dust attenuation {\citep{Yan2025}}.

The total H$\alpha$/H$\beta$ ratios span a range comparable to that observed in high-$z$ LRDs (left panel of Figure \ref{fig:balmer_decrement}). Owing to the limited number of sources, we do not recover the statistically significant correlation between the total H$\alpha$/H$\beta$ ratio and $L_{\rm 5100}$ reported in high-$z$ LRDs. Nevertheless, Figure~\ref{fig:balmer_decrement} suggests that the apparent correlation in high-$z$ LRDs likely reflects the varying contributions of {narrow and broad-line components}. At lower $L_{\rm 5100}$, $L_{\rm H\beta, total}$ is dominated by narrow-line emission, biasing the total Balmer decrement toward smaller values. In contrast, at higher $L_{\rm 5100}$, the broad component dominates, resulting in a larger total decrement.

Sources with the highest $L_{\rm 5100}$ exhibit relatively smaller broad-line H$\alpha$/H$\beta$ ratios (but still $\sim$10), whereas the ratios tend to increase toward lower $L_{\rm 5100}$. However, due to the large uncertainties in broad-line decomposition (mainly from H$\beta$ when the lines are faint), Kendall's $\tau$\footnote{The Kendall $\tau$ correlation analysis is performed using \texttt{pymccorrelation}, considering uncertainties in both {the $x$- and $y$-values}.} test {yields $p$-values consistent with $p>0.05$ within $1\sigma$, indicating no statistically significant correlation}. Likewise, we find no significant correlation between the Balmer break strength and the broad H$\alpha$/H$\beta$ ratio (Figure~\ref{fig:balmer_decrement}, right panel). Even when restricting the analysis to objects with inflection points blueward of 4500\,\AA, where the Balmer break remains a reliable diagnostic, no clear correlation emerges.  A detailed analysis of the Balmer decrement and its connection to the SED shape, including the Balmer break, will require further modeling of the dense gas, incorporating the gas density ($n_{\rm H}$), column density ($N_{\rm H}$), and other physical properties.

\subsubsection{Balmer absorption}\label{sec:balmer_absorption}

\begin{table*}[ht]
\centering
\setlength{\tabcolsep}{4pt}
\noindent\makebox[\linewidth][c]{\input{table/balmer_absorption_properties.tex}\unskip}\par
\caption{Absorption properties of the sources with detected Balmer absorption. {The machine-readable version is available on Zenodo as \texttt{main\_lines\_table.ecsv}.} \label{tab:absorption} }
\end{table*}

In the full sample, 19 out of 27 (70\%) of the DESI LRDs exhibit absorption in either \ha\ or \hb.  Among the sources with H$\beta$ absorption, nine also have spectral coverage of H$\alpha$, and all nine display absorption in H$\alpha$. 

\begin{figure}
    \centering
    \includegraphics[width=\linewidth]{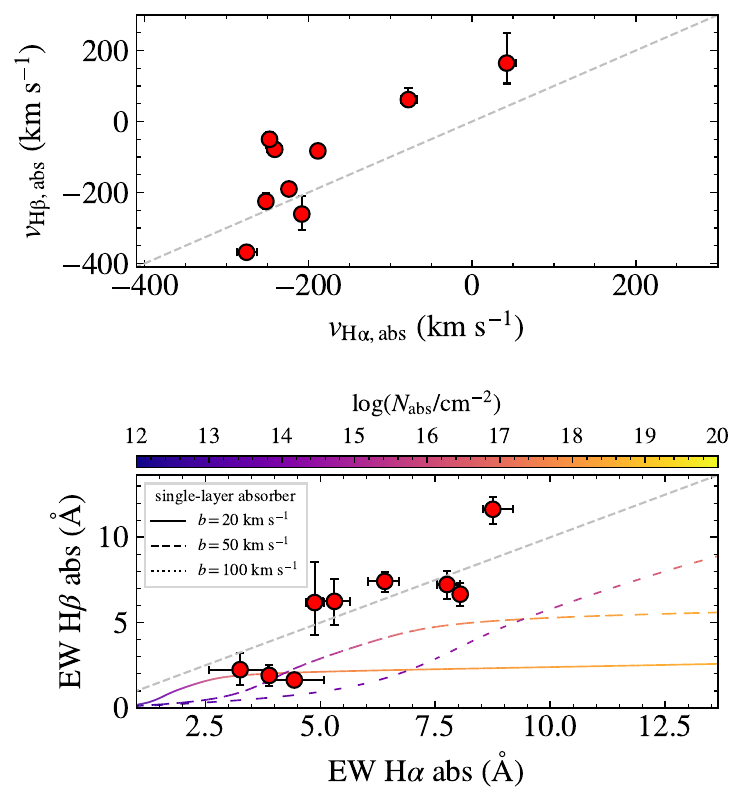}
    \caption{Velocity shifts and EWs of H$\alpha$ and H$\beta$ absorption for sources in which both absorbers are detected. The gray dashed line indicates the one-to-one relation. In the \textit{bottom} panel, the color-coded curves show the expected EWs of H$\alpha$ and H$\beta$ as a function of $\log N$ and $b$, assuming one gas cloud produces both transitions.}
    \label{fig:Abs_Ha_Hb}
\end{figure}

We examine the observed velocity shifts and EWs of the H$\alpha$ and H$\beta$ absorbers from the same sources in Figure~\ref{fig:Abs_Ha_Hb}. Although H$\alpha$ and H$\beta$ are treated as independent in the fitting procedure, their velocity shifts ($v_{\rm H\beta, abs}$ and $v_{\rm H\alpha, abs}$) and EWs are positively correlated.
This indicates that the gas producing both transitions is physically linked and kinematically coupled.   In the bottom panel of Figure~\ref{fig:Abs_Ha_Hb}, the color-coded curves show the expected relations if the H$\alpha$ and H$\beta$ absorption arises from the same gas clouds under a single-layer assumption, i.e., Voigt profiles for H$\alpha$ and H$\beta$ with shared $\log N$ and $b$, and $C_f = 1$. The H$\beta$ EWs are systematically larger than expected given the $\log N$ and $b$ inferred from H$\alpha$. 
This suggests that the full series of Balmer absorption in LRDs originates from multiple gas clouds rather than a single gas cloud.

As described in Section~\ref{sec:emission_line_measurement}, we model the optical depth of the Balmer absorbers using Voigt profiles, assuming a single-layer absorbing cloud. Here we highlight J1654+0337 and J1611+0917, shown in Figure~\ref{fig:example1} and Figure ~\ref{fig:sample_gold}. In both objects, the absorption troughs extend marginally below the continuum level. This indicates that the Balmer absorbers attenuate both the optical continuum and the broad emission lines, and that the covering fraction over the continuum source should be high. Such deep \hb\ absorption, reaching below the continuum level, is also observed in LRD ``Irony'' at $z=6.68$ \citep{DEugenio2025_irony}.  This evidence motivates the assumption adopted in Equation~\ref{eq:abs}, in which both the continuum and broad lines serve as the background source against the absorption.

Table~\ref{tab:absorption} presents the physical parameters of the Balmer absorbers. The velocity shifts of the absorbers, $v_{\rm abs}$, range from $-368$ to $+165$ km s$^{-1}$, with most systems being blueshifted and a small fraction redshifted ({two} \ha\ absorbers and {two} \hb\ absorbers). Most Balmer absorbers have covering fractions $C_f \approx 1$, while only a small subset exhibits lower values ranging from {0.4–0.9}. The column densities span $\log(N_{\rm H\beta}/\rm cm^{-2})=13.9-15.5$ and $\log(N_{\rm H\alpha}/\rm cm^{-2})=12.9-18.5$. The Doppler parameters of the Balmer absorbers $b$ range from {$\sim$21–500} km s$^{-1}$, though the smallest values (approaching the thermal limit) are driven by degenerate fits in a few high-$\log N$ \ha\ absorbers and should be treated with caution. $b$ represents the quadrature sum of thermal and {nonthermal} motions, such as macroturbulence, {microturbulence, or} bulk gas motion. For hydrogen at $T \sim 10^4$ K, the thermal contribution is only $\sim$13 km s$^{-1}$. {Using $b_{\rm nonthermal}=(b^2-b_{\rm thermal}^2)^{1/2}$, the central measurements imply $b_{\rm nonthermal}\simeq16$--500 km s$^{-1}$.} 

However, we emphasize that the conclusions above are based on Equation~\ref{eq:abs} and are subject to the caveats inherent in this formulation. Most importantly, it adopts a simplified single-layer approximation that cannot capture the complex physical gradients likely present within these absorbers. The Balmer absorbers, commonly interpreted as dense gas outflows (or inflows in redshifted systems), may have substructure along the line of sight. Analogous to stellar atmosphere modeling, the line cores and wings may originate from regions with different opacities, temperatures, and densities. A more physically realistic description would therefore require radiative transfer through multiple layers, particularly if the absorbers are associated with a dense envelope analogous to a stellar atmosphere surrounding the BH. Furthermore, Equation~\ref{eq:abs} assumes an identical covering fraction for both the continuum source and the BLR. This assumption may not hold, as the covering factor depends sensitively on the relative geometry of the emitting and absorbing regions, which remains poorly constrained. In the BH–envelope scenario, the spatial relationship among the central BH, the BLR, and the outflowing dense H I gas is unclear \citep[e.g. see the models summarized in][]{Asada2026}. Allowing different $C_f$  for the continuum and BLR in our simplified fitting framework results in strong degeneracy, yielding no meaningful constraints.

\begin{figure*}
    \includegraphics[width=\linewidth]{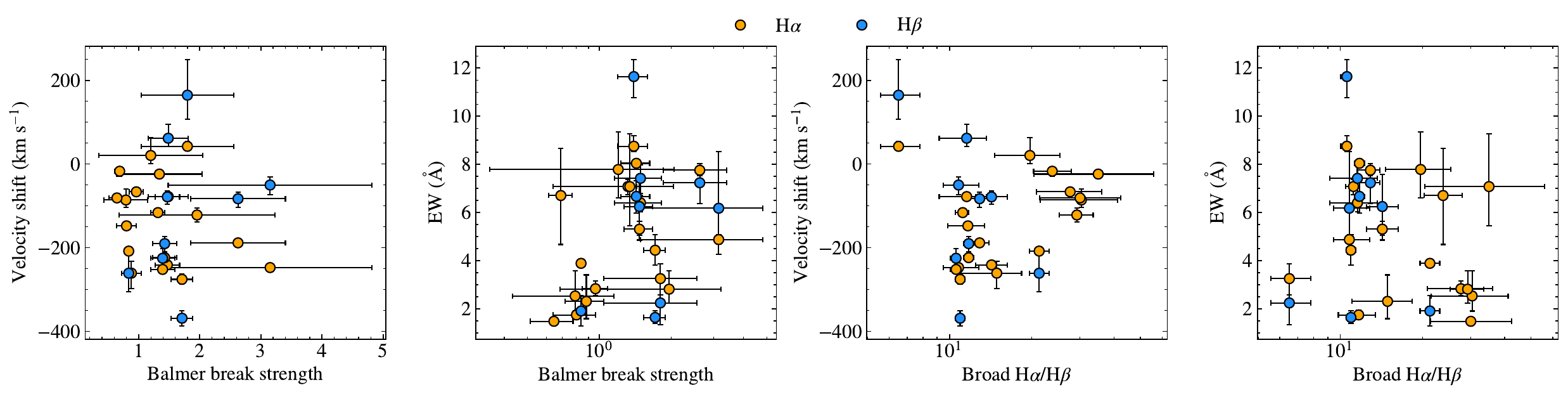}
    \caption{The velocities and EWs of Balmer absorption {vs.} their Balmer break strength and decrement. \label{fig:Balmer_absorption_correlation}}
\end{figure*}

In Figure~\ref{fig:Balmer_absorption_correlation}, we examine the relationship between the Balmer absorption features and the Balmer break strength and decrement. We characterize the absorbers using purely observational quantities: velocity offsets and EWs. We find no clear correlation between the velocity offsets or EWs of the Balmer absorbers and the Balmer break strength or decrement.   However, we also caution that this conclusion is drawn from a small sample and is subject to considerable uncertainties, primarily driven by fitting degeneracies. A larger sample will be required to robustly constrain the connection between Balmer absorption properties and the continuum shape.

\subsection{{Emission-line diagnostics}}\label{sec:emission_line_diagnostics}

\subsubsection{Metallicity and BPT diagram}\label{sec:metallicity_BPT}

\begin{figure*}
    \centering
\includegraphics[width=\linewidth]{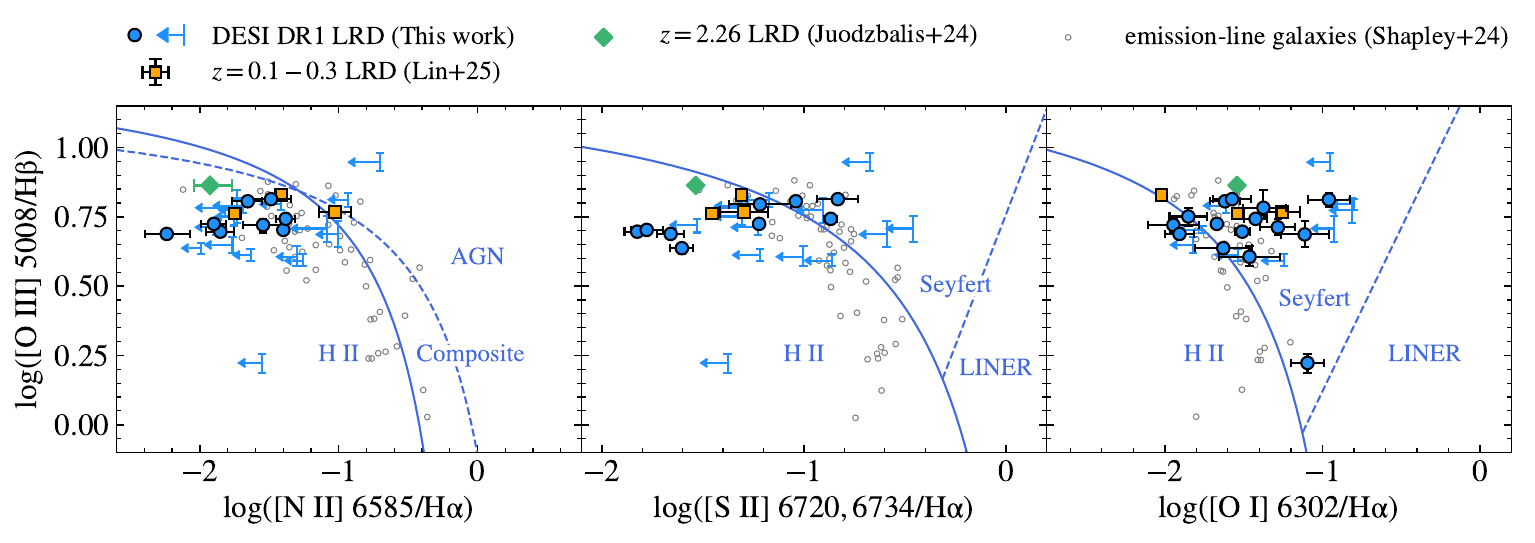}
    \caption{BPT diagram for the narrow emission lines in LRDs. DESI DR1 LRDs are shown in blue. The $z<0.3$ LRDs from \citetalias{Lin2025_localLRD} are shown as orange squares, and the $z=2.26$ LRD from \cite{Juodzbalis2024} is shown as a green diamond. The emission-line galaxies at $z = 1.4$–$7.5$ from \citet{Shapley2025} are shown as gray circles.  The classification boundaries for different ionization mechanisms \citep{Kauffmann2003, Kewley2006} are shown as blue solid and dashed lines. The corresponding regions are labeled.\label{fig:BPT}  }
\end{figure*}

We measure gas-phase metallicities using the direct $T_e$ method. [\ion{O}{3}]$\lambda4364$ line is detected at $\mathrm{S/N} > 5$ in 22 out of the 27 LRDs. We first determine the electron density ($n_e$) in the $\mathrm{O}^{+}$ zones using the [\ion{O}{2}] $\lambda\lambda3727,3730$ doublet. In our sample, 18 sources have [\ion{O}{2}] doublet ratios measured at $\mathrm{S/N} > 3$, of which 16 have line ratios that fall within the density-sensitive regime. The [\ion{O}{2}]-derived $n_e$ ranges from $32^{+508}_{-31}$ to $2907^{+4747}_{-2315}$\,cm$^{-3}$, with a median of 287\,cm$^{-3}$. Four sources in our sample have [\ion{S}{2}] $\lambda\lambda6718,6733$ doublet ratios with $\mathrm{S/N} > 3$ in the density-sensitive regime, yielding [\ion{S}{2}]-derived densities from $142^{+579}_{-141}$ to $867^{+12442}_{-866}$\,cm$^{-3}$. 

We then estimate the electron temperature ($T_e$) by adopting $n_e$ derived from  [\ion{O}{2}]. However, the measured [\ion{O}{3}] $\lambda5008/\lambda4364$ ratios yield exceptionally high $T_e$ values, ranging from approximately {15{,}000–80{,}000}\,K among the solutions within the density-sensitive regime. One possibility is that the [\ion{O}{3}] $\lambda5008,\lambda4364$ lines are from a region with much higher density, where the $n_e$ derived from [\ion{O}{2}] is not representative of the conditions. 
The critical density of [\ion{O}{3}] $\lambda5008$ is an order of magnitude higher than that of [\ion{O}{2}] $\lambda3727$, and the critical density of [\ion{O}{3}] $\lambda4364$ {is 3 orders} of magnitude higher. The $n_e$ derived from [\ion{O}{2}] traces the low-density zones, whereas the [\ion{O}{3}] lines may originate from higher-density regions. If the true $n_e$ in the $\mathrm{O}^{++}$ regions significantly exceeds that inferred from [\ion{O}{2}], the resulting $T_e$ would be systematically overestimated. 

Although the direct $T_e$ method may not be fully applicable, the weak or negligible [\ion{N}{2}] emission observed in the BPT diagram (discussed below) independently indicates that the selected LRDs are metal-poor. 
Adopting fiducial values of $T_e = 15{,}000$\,K and the median $N_e = 287$\,cm$^{-3}$ provides an order-of-magnitude metallicity estimate of $12 + \log(\mathrm{O/H}) = 7.7$--$8.1$, with a median of 7.8, corresponding to approximately $0.13\,Z_{\odot}$.

In the BPT diagram (Figure~\ref{fig:BPT}, \citealt{Baldwin1981}), DESI LRDs occupy a region characterized by their low metallicity and high ionization parameters ($\log U$), a regime that can be shared by both AGNs and galaxies with similar properties \citep{Sanders2023, Shapley2025}. In the [\ion{N}{2}] and [\ion{S}{2}] diagrams (left and middle panels of Figure \ref{fig:BPT}), all LRDs lie within the star-forming region or near the boundary between the Seyfert and \ion{H}{2} loci. In contrast, in the [\ion{O}{1}] diagram, most LRDs fall in the Seyfert region or near the Seyfert–\ion{H}{2} boundary. However, this does not necessarily imply AGN-dominated narrow-line emission. Indeed, we find that emission-line galaxies at $z>1.4$ \citep{Shapley2025} can occupy the same parameter space as the LRDs.

\subsubsection{\ion{He}{2} 4687}\label{sec:He2}

\begin{figure}
    \centering
\includegraphics[width=\linewidth]{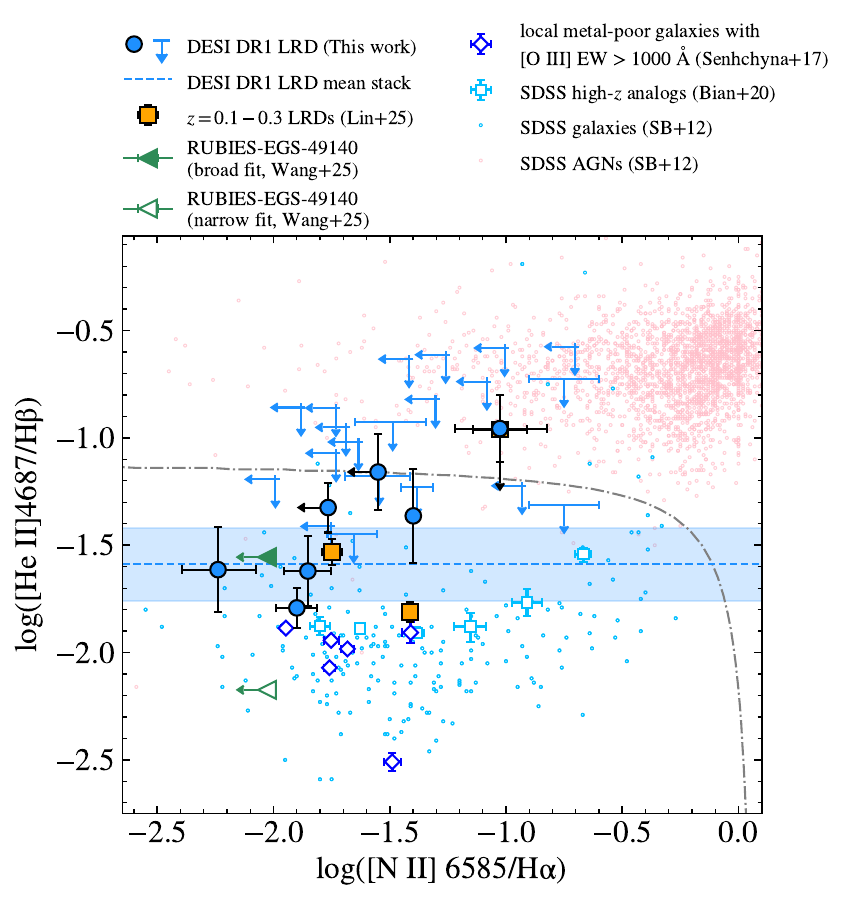}
    \caption{\ion{He}{2} 4687/\hb\ {vs.} [\ion{N}{2}] 6585/\ha\ for LRDs over a wide range of redshifts. The DESI sample in this work is shown as blue circles and triangles, where the latter denote sources for which only upper limits are available for both line ratios.  The mean stacked \ion{He}{2}/\hb\ ratio of the DESI LRDs is shown as a blue dashed horizontal line, with the shaded region as the associated uncertainty. Local LRDs at $z=0.1$–$0.3$ from \citetalias{Lin2025_localLRD} are shown as orange squares. \ion{He}{2} emission in JWST-discovered LRD RUBIES-EGS-49140 {\citep{Wang2025_heii}} is shown as a filled green triangle when measured with a narrow+broad profile, and as an unfilled green triangle when measured with only a narrow Gaussian. We also include SDSS-selected galaxies ({light-blue circles}) and AGNs (pink circles) compiled by \cite{Shirazi2012}, as well as high-$z$ galaxy analogs ({light-blue squares}) from \cite{Bian2020} and extremely metal-poor galaxies with [\ion{O}{3}] $\lambda5008$ EW $> 1000$\,\AA\ (blue diamonds) from \cite{Senchyna2017}. The {dashed–dotted} line marks the theoretical maximum starburst line from \cite{Shirazi2012}. \label{fig:heii_diagram}
    }
\end{figure}

The ionization potential of He$^{++}$ is high (54.4 eV), making \ion{He}{2} 4687 a powerful diagnostic of the ionizing spectrum.   
Figure \ref{fig:heii_diagram} shows the \ion{He}{2} 4687/\hb\ versus [\ion{N}{2}] 6585/\ha\ diagram. Low-$z$ LRDs, including the DESI LRDs presented here and those reported by \citetalias{Lin2025_localLRD}, together with the high-$z$ LRDs from {\cite{Wang2025_heii}}, consistently show \ion{He}{2} 4687/\hb\ ratios lower than those of typical type-1 AGNs. The \ion{He}{2} $\lambda4687$/\hb\ ratios span the range between star-forming galaxies and AGNs \citep{Shirazi2012, Bian2020}.

\begin{figure}
    \centering
    \includegraphics[width=\linewidth]{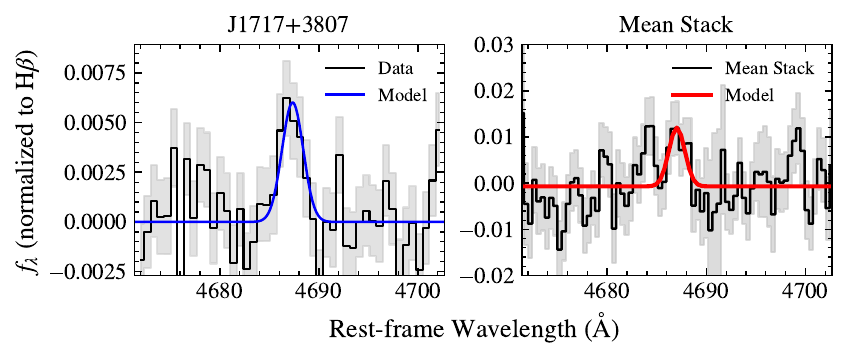}
    \caption{ \textit{Left}: {example} of an individual \ion{He}{2} 4687 detection in one of the DESI LRDs. \textit{Right}: {mean} stacked \ion{He}{2} 4687 spectrum for all the DESI LRDs. Both spectra are normalized by the narrow \hb\ flux.}
    \label{fig:HeII_individual_stack}
\end{figure}

In the left panel of Figure \ref{fig:HeII_individual_stack}, we show an example of an individual \ion{He}{2} detection from our sample.
To estimate the average \ion{He}{2} strength, we stack the \ion{He}{2} 4687 lines of all DESI LRDs after subtracting their local continua and normalizing by the narrow \hb\ flux. The right panel of Figure \ref{fig:HeII_individual_stack} shows the resulting mean stacked \ion{He}{2} line, yielding a stacked value of $\log$(\ion{He}{2}/\hb) = $-1.59 \pm 0.17$. As shown in Figure~\ref{fig:heii_diagram}, the stacked value exceeds that of most star-forming systems, including high-$z$ analogs and extremely metal-poor galaxies, but remains below that of typical AGNs. This suggests that, on average, DESI LRDs have an ionizing spectrum harder than that of most star-forming galaxies, yet softer than that of typical AGNs, as discussed in {\cite{Wang2025_heii}}.

\subsubsection{\ion{Mg}{2} $\lambda\lambda$2796, 2803}

\begin{figure}
    \centering
\includegraphics[width=\linewidth]{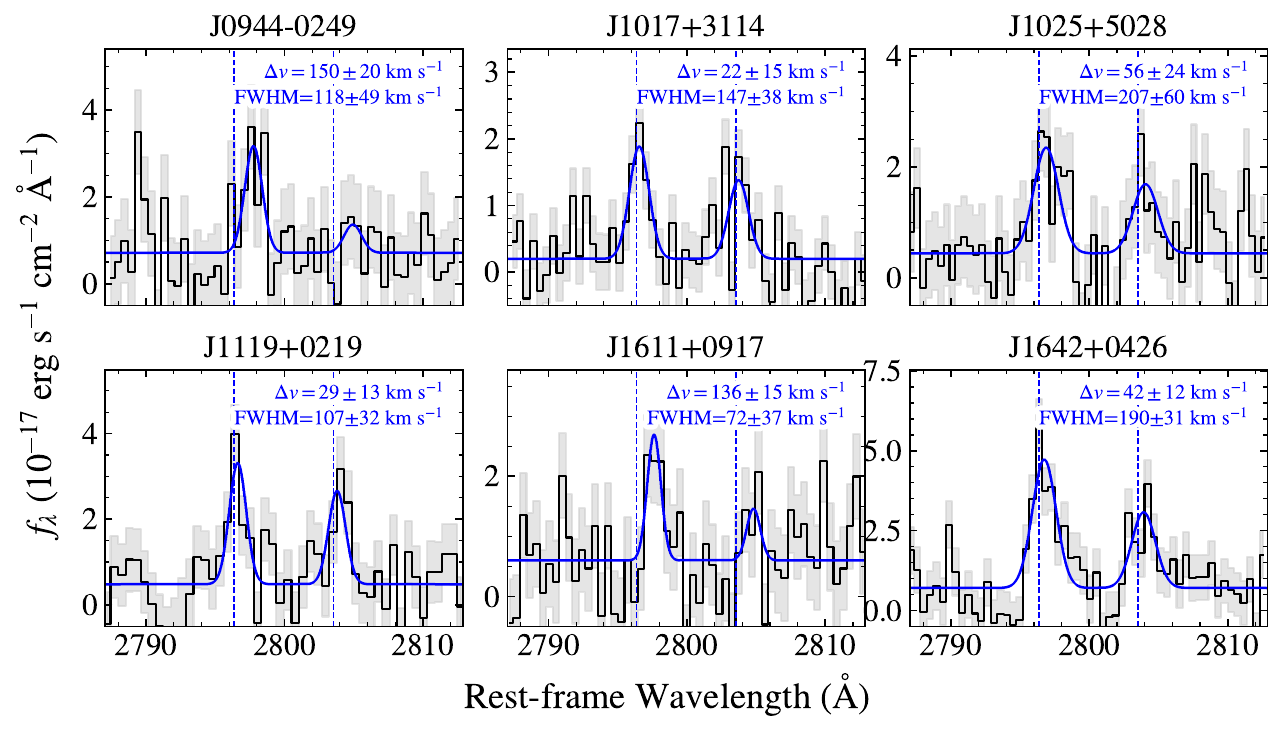}
    \caption{Six DESI DR1 LRDs are detected with \ion{Mg}{2}$\lambda\lambda$2796, 2803. The DESI spectra are shown in black, with {gray} shaded regions indicating the uncertainties. The best-fit \ion{Mg}{2} doublets are shown in blue. The rest-frame wavelengths of \ion{Mg}{2} at the redshift determined by   [\ion{O}{3}] 5008   are indicated by blue dashed vertical lines.}
    \label{fig:mgii_line}
\end{figure}

\begin{figure}
    \centering
\includegraphics[width=\linewidth]{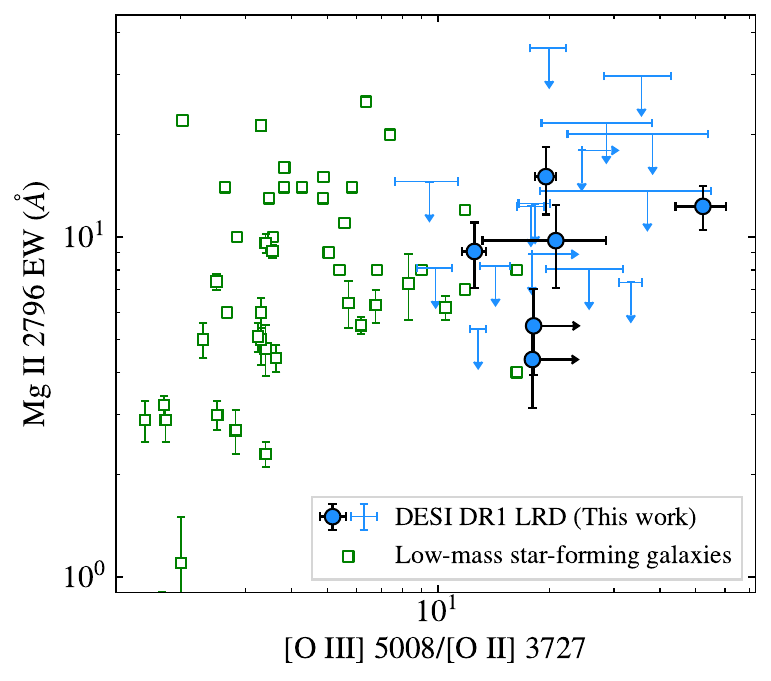}
    \caption{[\ion{O}{3}] 5008/[\ion{O}{2}] 3727 {vs.} \ion{Mg}{2} 2796 EW for DESI DR1 LRDs and low-mass star-forming galaxies at $z<1$. The blue circles represent DESI DR1 LRDs, while the blue upper limits indicate upper limits on the \ion{Mg}{2} 2796 EW. The green squares denote low-mass star-forming galaxies at $z < 1$, whose \ion{Mg}{2} 2796 EWs are drawn from Lyman continuum leakers compiled in \cite{Izotov2016, Izotov2018, Izotov2022, Izotov2025, Henry2018, Xu2023}.}
    \label{fig:o32_mgii}
\end{figure}

We detect clear \ion{Mg}{2} $\lambda\lambda$2796, 2803 emission lines in six DESI DR1 LRDs (Figure \ref{fig:mgii_line}). The \ion{Mg}{2} doublet exhibits marginally resolved narrow components with FWHMs of 73–208 km s$^{-1}$, comparable to narrow components of [\ion{O}{3}] and the Balmer lines. We do not detect broad components in the DESI spectra. The \ion{Mg}{2} emission is redshifted by 22–151 km s$^{-1}$ relative to the systemic redshift defined by [\ion{O}{3}] 5008, consistent with its resonant nature.   The EWs of \ion{Mg}{2} 2796 lines range from 4 to 15\,\AA, consistent with those of low-mass star-forming galaxies exhibiting high [\ion{O}{3}] 5008/[\ion{O}{2}] 3727 ratios (Figure \ref{fig:o32_mgii}).

\bigskip
In summary, based on the BPT diagram and the \ion{Mg}{2} $\lambda\lambda2796,2803$ doublet, we conclude that the narrow emission lines in low-$z$ LRDs, from the UV to the optical, are consistent with low metallicity and high $\log U$, which can be produced by either AGNs or galaxies. On the other hand, the \ion{He}{2} diagnostic diagram reveals an ionizing spectrum that is harder than that of most star-forming galaxies, yet still softer than that of typical AGNs.

\subsection{[\ion{O}{3}] ionized outflow}\label{sec:ionized_outflow}

In our sample, we find a remarkably high incidence rate of [\ion{O}{3}] outflows: 78\% (21 out of 27) of the LRDs exhibit significant broad [\ion{O}{3}] components, with velocity offsets to the narrow lines ranging from {$-85$–37} km s$^{-1}$ and FWHMs from {173–658} km s$^{-1}$. Similar ionized outflows have also been reported in high-redshift LRDs discovered with JWST/NIRSpec grating observations, but statistics are lacking \cite[e.g.,][]{Juodzbalis2024, DEugenio2025_irony}. The incidence rate of outflows in DESI LRDs is significantly higher than in typical high-$z$ galaxies. For galaxies at $z=3$–9, the [\ion{O}{3}] outflow incidence is reported to be roughly 30\% by \cite{Xu2025} and {25\%–40\%} by \cite{Carniani2024}. The outflow incidence rate in DESI LRDs is comparable only to that observed in the most extreme starburst galaxies. For instance, in low-mass ($M_* = 10^4$–$10^7\,M_\odot$) galaxies with high specific star-formation rates (${\rm sSFR} \approx 100$–$1000~{\rm Gyr}^{-1}$) in the local Universe, the incidence rate of ionized outflows is approximately 67\% (\citealt{Xu2022}). 

The outflows in DESI LRDs also exhibit elevated [\ion{O}{3}]$\lambda$5008/H$\beta$ ratios, which, together with their high incidence rate, point to a connection with AGN activity. Stellar-driven outflows typically exhibit relatively strong Balmer-line components, placing the outflowing gas in the non-AGN region of the BPT diagram. In contrast, in our joint fits of [\ion{O}{3}] and Balmer lines, the Balmer-line outflow is negligible in 22 objects and detected in only five. This implies that the outflowing gas has high ionization parameters, although fitting degeneracies may be present. For sources with negligible H$\beta$ outflows, the outflowing gas has very high $\log$([\ion{O}{3}]$\lambda$5008/H$\beta$) ratios, placing it firmly in the AGN region of the BPT diagram. Among the five objects with measurable Balmer outflows, $\log$([\ion{O}{3}]$\lambda$5008/H$\beta$) of the outflowing component ranges from {0.80–1.21}, again keeping them within the AGN region or close to the AGN–star-forming boundary.  This suggests that the outflowing gas is photoionized by AGN radiation.

\begin{figure}
    \centering
    \includegraphics[width=\linewidth]{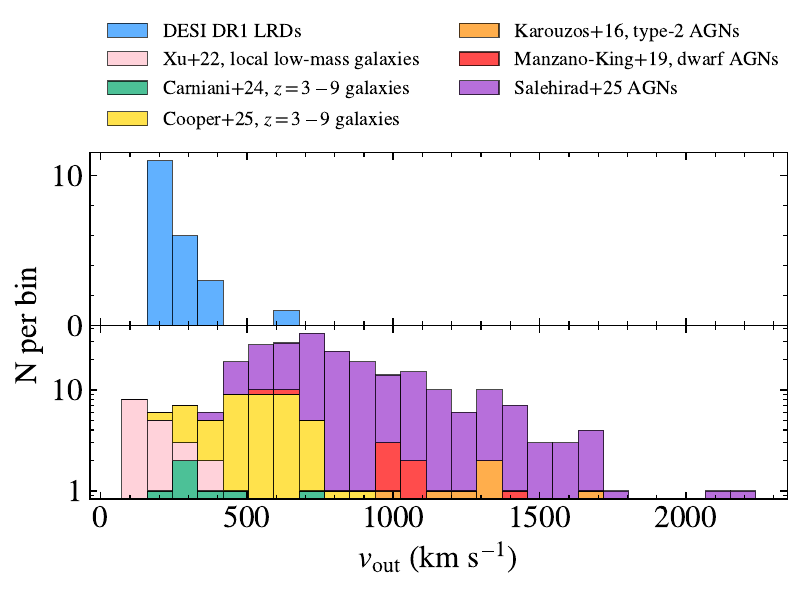}
    \caption{The distribution of outflow velocity $v_{\rm out}$ in DESI DR1 LRDs in this work, local low-mass galaxies with high sSFR \citep{Xu2022}, $z=3-9$ galaxies \citep{Carniani2024, Cooper2025}, and AGNs \citep{Karouzos2016, Manzano-King2019, Salehirad2025}. {All of the literature} $v_{\rm out}$ values are calculated with the definition in Equation~\ref{eq:vout}, using the reported velocity offsets and FWHMs. The $v_{\rm out}$ values for AGNs in \cite{Salehirad2025} correspond to sources classified as AGNs in all three BPT diagrams ([\ion{N}{2}], [\ion{S}{2}], and [\ion{O}{1}]).}
    \label{fig:vout_hist}
\end{figure}

However, the outflow velocities are relatively modest. The mechanisms responsible for their kinematics (e.g., winds, slow shocks) are still uncertain.  We compute the outflow velocity following \cite{Rupke2005}, defined as
\begin{equation} \label{eq:vout}
v_{\rm out} = |\Delta v_{\rm [O~III], out}| + 2\sigma_{\rm [O~III], out},
\end{equation}
where $\Delta v_{\rm [O~III], out}$ is the velocity offset of the outflow component relative to the narrow [\ion{O}{3}] lines, and $\sigma_{\rm [O~III], out}$ is the velocity dispersion of the outflow component. The derived $v_{\rm out}$ values for the DESI LRDs span 163–644 km s$^{-1}$, with a median of 234 km s$^{-1}$. As shown in Figure~\ref{fig:vout_hist}, these velocities are comparable to those measured in low-mass galaxies with high specific star-formation rates \citep{Xu2022}, but are lower than those of galactic outflows at $z=3$–9 \citep{Carniani2024, Cooper2025}. We caution, however, that observational limitations may bias the high-$z$ outflow statistics, as weak outflows could be missed in faint galaxies observed with JWST/NIRSpec at $R \lesssim 2700$. The $v_{\rm out}$ values in the DESI LRDs are also significantly lower than those of the AGN-driven outflows \citep{Karouzos2016, Manzano-King2019, WLiu2020, Salehirad2025, WLiu2024}, which can reach $v_{\rm out} > 1000$ km s$^{-1}$.

\subsection{Variability}\label{sec:variability}
  
\subsubsection{Optical Variability}\label{sec:optical_variability}

We assess the significance of optical variability in the DESI LRDs using ZTF $gri$ light curves. First, we rescale the per-epoch uncertainties by $f_{\rm err}=\sqrt{\langle\chi^{2}/\nu\rangle_{\rm star}}$,  the median over {nonvariable} PSF reference stars ($\chi^{2}/\nu<1.5$ against a constant-flux model) within $\pm 0.5$ mag of the target on the same ZTF readout channel ($\sim$0.7 deg$^2$). For each light curve, we perform a $\chi^2$ test against a constant-flux model.
 We then compute the fractional variability amplitude \citep{Vaughan2003},
$$
F_{\rm var}=\frac{1}{\langle f \rangle}\sqrt{S^{2}-\overline{\sigma^{2}_{\rm err}}}
$$
where $S^{2}$ is the {sample flux variance, and} $\overline{\sigma^{2}_{\rm err}}$ the mean squared photometric error. We estimate $\sigma_{F_{\rm var}}$ from 1000 bootstrap resamples. We additionally require the measurement to be leave-one-out (LOO) robust: $F_{\rm var}$ recomputed with any single epoch removed must remain at $\geq 50\%$ of the all-epoch value; light curves failing this test are flagged as outlier-driven. As a robustness check, we repeat the analysis after dropping the LOO-identified outlier from the target and from every reference simultaneously, so the comparison remains fair.  Finally, we compare each target's $F_{\rm var}$ against the empirical $F_{\rm var}$ distribution of reference stars matched to the target in both brightness and number of epochs ({within a factor of 2}).  In addition to $F_{\rm var}$, we apply a linear-trend $F$-test \citep{Bevington2003} that compares a linear fit to a constant-flux model and returns a $p$-value for the trend, following the approach used to identify long-term mid-IR fading in changing-look AGN \citep{Stern2018}. The test reveals smooth, monotonic variability that $F_{\rm var}$ alone cannot distinguish from noise.  A source is flagged as a variability candidate if it meets all of the following criteria: (a) $p$-value of the $\chi^2$ test $<0.01$; (b) S/N of $F_{\rm var}$ is $\geq3$ and LOO-robust; (c) either $F_{\rm var}$ lies above the 95th percentile of the reference-star $F_{\rm var}$ distribution, or a weighted linear-trend $F$-test yields $p_{\rm trend}<0.01$ and $p_{\rm trend}$ is smaller than the 5th percentile of the trend $p$-value distribution of the reference stars.  Together, these criteria ensure that the light curve both deviates from a constant-flux model and differs significantly from those of {nonvariable} reference sources. 

Of the 27 DESI LRDs in this work, 13 have light curves suitable for variability analysis, spanning {4.1–7.5 yr} in the observed frame (median {6.9 yr}), corresponding to {3.1–6.3 yr} in the rest frame (median {4.1 yr}). In the remaining 14, 10 lack ZTF DR24 detections due to their faintness, and {four have fewer than} 10 usable epochs in any filter.  Among the 13 sources, only the $i$-band light curve of J1717$+$3807 robustly {satisfies all of the criteria}. J1717$+$3807's $i$-band light curve (512 exposures) faded between 2018 and 2022 at a rate of $(7.5 \pm 1.0)\times 10^{-3}$ mag\,yr$^{-1}$. In contrast, its $g$ and $r$ fluxes (rest-frame $\sim$3970 and 5375 \AA) are flat over the same window. J1646$+$1426's $i$-band light curve marginally passes the criteria, but it is based on only 23 exposures. It has $F_{\rm var} = 0.33$ at $3.8\sigma$ and only marginally exceeds the 95th percentile of its reference-star distribution (95.6th percentile). We thus classify it as tentative and note that its statistical significance requires further confirmation. Its light curve is shown in Appendix \ref{appendix:lc}. The remaining 11 show no significant evidence of variability. 

\subsubsection{Infrared Variability}

 We apply the same assessment to the WISE W1 and W2 light curves. 

Among the 27 DESI LRDs, three are flagged as WISE-variable candidates (in W1 and/or W2) by the automated statistical criteria above. J1717+3807 is the most significant variable source in both bands, with $F_{\rm var} = 0.033 \pm 0.005$ in W1 ($7.2\sigma$) and $F_{\rm var} = 0.045 \pm 0.006$ in W2 ($7.9\sigma$). Both light curves exhibit coherent monotonic fading over the $\sim${14 yr} baseline, with rates of $(8.8 \pm 0.5)\times10^{-3}$ mag\,yr$^{-1}$ in W1 and $(11.3 \pm 0.5)\times10^{-3}$ mag\,yr$^{-1}$ in W2, as shown in Figure  \ref{fig:J1717_lc_sf}. J1646+1426 is classified as marginally variable only in W2, with $F_{\rm var} = 0.16 \pm 0.05$ ($3.0\sigma$), and shows stochastic variability. J1909+5831 is flagged as a variability candidate in W1, with $F_{\rm var} = 0.51 \pm 0.10$ ($5.2\sigma$). It exhibits a brightening but with {low-S/N photometry}. Its light curve is extracted from the Legacy Survey DR9 per-visit forced photometry on the unWISE coadded images, as it is too faint to be reliably detected in individual WISE single-exposure frames. The variability should be re-evaluated with more robust photometric methods to ensure a reliable measurement. After photometric-quality assessment, only J1717+3807 is considered robustly variable, while J1646+1426 and J1909+5831 remain tentative. The light curves of J1646+1426 and J1909+5831 are shown in Appendix \ref{appendix:lc}. 
 The remaining 24 sources are not flagged as variable by the automated criteria.

\subsubsection{The variability of J1717+3807}

\begin{figure*}[htbp]
    \centering
    \includegraphics[width=\linewidth]{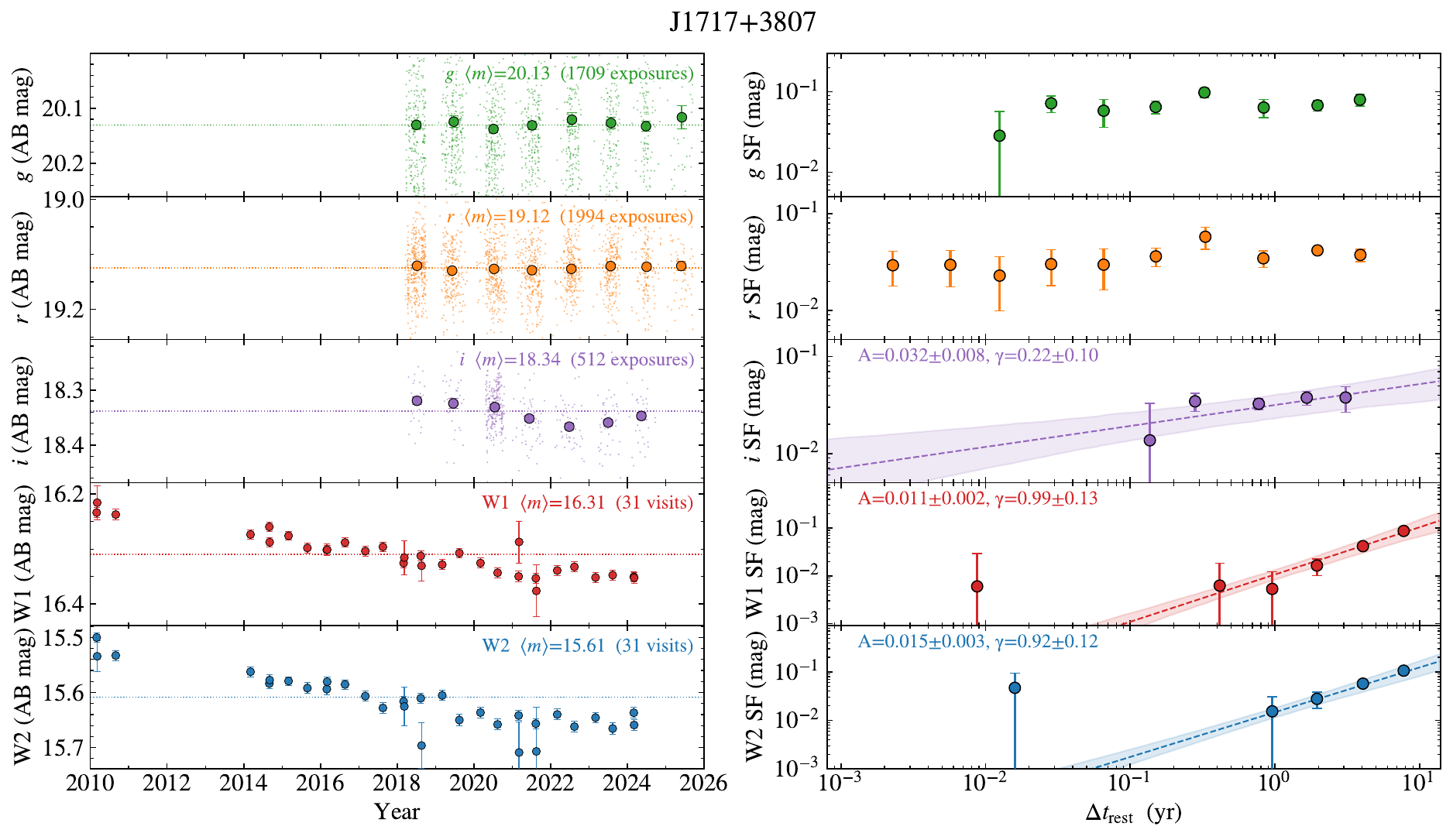}
    \caption{\textit{Left}: {the} ZTF $gri$ and WISE W1 W2 light curves of J1717+3807 over {$\gtrsim$10 yr}. \textit{Right}: {the} structure function (SF) of the light curves (light dots), the binned SF (filled dots), along with the best-fit power-law (dashed lines).}
    \label{fig:J1717_lc_sf}
\end{figure*}

From {2018–2022} (rest-frame {3.3 yr}), J1717+3807's $i$-band magnitude faded by $0.05 \pm 0.01$ mag, then rose by $0.02 \pm 0.01$ mag by 2024 (rest-frame {1.6 yr}). From {2010–2024} (rest-frame {11.7 yr}), the 31 epochs of WISE W1 show a smooth fading of $0.123\pm0.006$ mag, and W2 fades by $0.159\pm0.006$ mag. In contrast, its ZTF $g$ and $r$ light curves over the overlapping 2018–2025 window are consistent with constant flux.

We construct the structure function \citep[SF;][]{Hughes1992, Collier2001, Bauer2009, Kozlowski2010} and fit it with a power law \citep{Palanque-Delabrouille2011}:

\begin{equation}
\mathrm{SF}\!\bigl(\Delta t_{\rm rest}\bigr)
= A \left(\frac{\Delta t_{\rm rest}}{1\,\mathrm{yr}}\right)^{\gamma}.
\end{equation} 
As shown in the right panel of Figure \ref{fig:J1717_lc_sf},  J1717+3807's ZTF $i$ band shows a weakly significant rise. Both W1 and W2 exhibit an SF slope index of $\gamma \simeq 1$, significantly steeper than the expectation from a damped random walk model \citep[$\gamma \sim 0.5$ on timescales shorter than the characteristic timescales; ][]{MacLeod2010}. The SF of J1717+3807 indicates that its variability is dominated by a long-term trend with a characteristic timescale longer than the observing baseline {($\sim  10$ yr)}, while showing little variability on days-to-months timescales.

J1717+3807’s $g$ and $r$ bands are largely dominated by the blue UV component, which may be attributed to the host galaxy and explains the lack of significant variability. Its $i$ band contains its \ha\ emission and the rising red optical continuum (Figure \ref{fig:example1}). The $i$-band variability may be driven by variations in either the \ha\ emission line, the continuum emission, or a combination of both. Dedicated spectroscopic monitoring is required to distinguish between the scenarios. Furthermore, the long-term variability is reminiscent of the gravitationally lensed, multiply imaged LRD RXCJ2211-RX1 \citep{ZZhang2025_century} which shows significant color and brightness change over a timescale of $\sim20$\,yr in the rest frame. Within the {BH–envelope} framework, the $i$-band continuum of J1717+3807 is interpreted as the ascending portion of the thermalized blackbody emission from the envelope.  Monitoring over another decade-long baseline could reveal whether the $i$-band brightness continues to increase or instead exhibits periodic variability potentially associated with envelope pulsations, as discussed in \cite{ZZhang2025_century}.

J1717+3807's WISE W1 and W2 bands, corresponding to rest-frame wavelengths of $\sim$2.8 and 3.9\,\micron, probe an excess above the tail of the same blackbody component (Figure \ref{fig:example1}). At these wavelengths, emission from hot dust at $\sim$1000\,K may be significant or even dominant (see the SED fits in \citealt{Lin2025_localLRD}). The AGN-heated hot dust component could drive the observed WISE variability.

J1717+3807's significant variability contrasts with the behavior of the other three $z<0.3$ LRDs in \citetalias{Lin2025_localLRD} (see also {\citealt{Burke2025}}). We compare the spectral properties of J1717+3807 with those of the other three $z<0.3$ LRDs in Section \ref{sec:highlight_J1717}.  We defer a comprehensive analysis of the variability behavior of J1717+3807 to future work {(Z. Zhang 2026, in preparation)}.

\section{Discussion}\label{sec:discussion}

In this section, we provide a brief, qualitative discussion of the current sample and its implications. We defer more in-depth analysis and detailed modeling of LRDs to future works, pending more complete spectral data from our ongoing follow-up observations.

\subsection{{Hertzsprung–Russell} diagram of LRDs}\label{sec:hr_diagram}

\begin{figure*}[t]
\centering
    \includegraphics[width=0.65\linewidth]{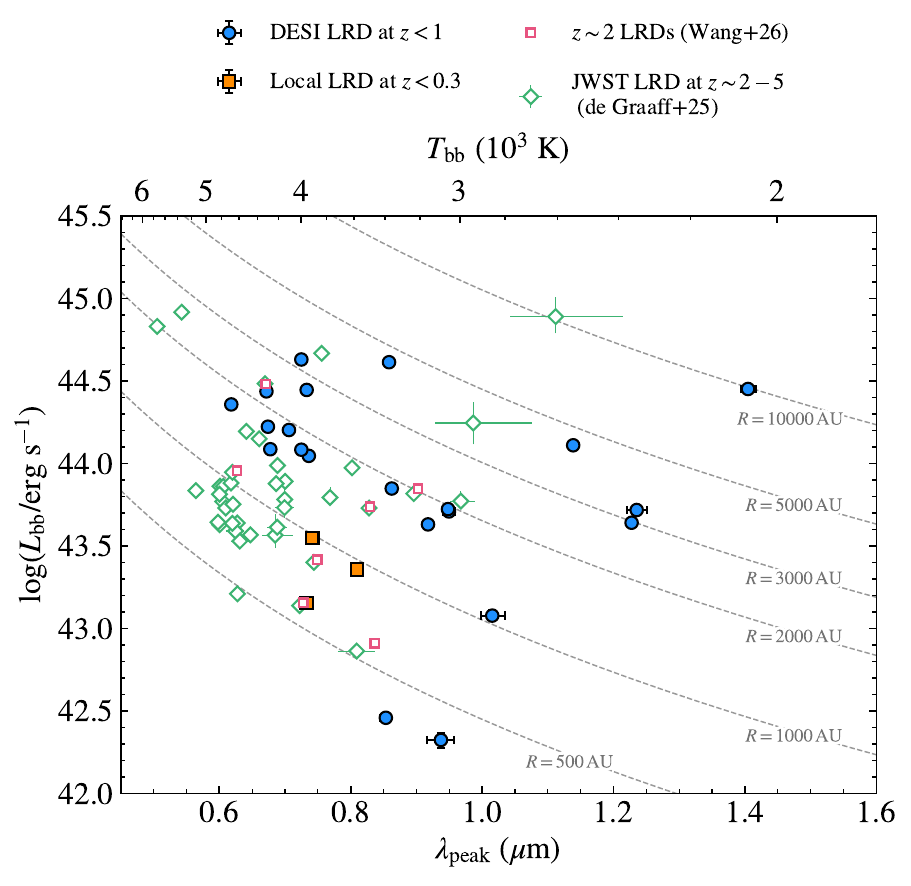}
    \caption{{Hertzsprung–Russell (H-R)} diagram of LRDs: the peak wavelength ($\lambda_{\rm peak}$) of the optical-to-near-infrared, blackbody-like continuum as a function of its luminosity ($L_{\rm bb}$).  The top axis indicates the effective temperature of a blackbody peaking at $\lambda_{\rm peak}$, based on Wien’s displacement law. Dashed gray lines mark loci of constant blackbody radius (in {astronomical units}), derived by combining $L_{\rm bb} = 4\pi R^{2}\sigma T^{4}$ with {the Stefan–Boltzmann} law. The JWST LRD sample is compiled from {\cite{deGraaff2025_all}} and \cite{Wang2026}. \label{fig:HR}}
\end{figure*}

As widely discussed in recent literature, the optical-to-near-infrared spectra of LRDs exhibit blackbody-like emission. This has motivated certain models in which BHs are enshrouded by atmosphere-like envelopes that emit thermal radiation {\citep[e.g.,][]{Inayoshi2025, Ji2025, Kido2025, Liu2025}}.

In our sample, 21 sources have near-IR spectra with sufficient S/N to cover the optical-to-IR continuum and locate the blackbody peak. We first fit the continuum blueward of the inflection points using a power-law model and subtract this component from the spectrum. The resulting residual spectrum is then fitted with a modified blackbody model.  The results are shown in Figure \ref{fig:bb_fits}, and summarized in Table \ref{tab:bb_fit}. We emphasize that this modified blackbody is used only to estimate the peak wavelength and to calculate the total luminosity of the blackbody component. We do not ascribe any physical interpretation to its modification factor.

Figure \ref{fig:HR} shows the Hertzsprung–Russell ({H-R}) diagram of LRDs: the distribution of the peak wavelength ($\lambda_{\rm peak}$) and the total luminosity of the blackbody component ($L_{\rm bb}$), with the former indicative of the temperature of the blackbody.  The value of $\lambda_{\rm peak}$ spans $0.6$–$1.4\,\micron$. Assuming a single-temperature blackbody, Wien’s displacement law yields temperatures ($T_{\rm peak}$) of $2063$–$4687\,\mathrm{K}$. The {H-R} diagram does not show a sequence or correlation between $\lambda_{\rm peak}$ ($T_{\rm peak}$) and $L_{\rm bb}$.

Five objects show $\lambda_{\rm peak}>1$\,\micron\ corresponding to temperatures below $3000$ K, which is 24\% of the 21 DESI LRDs with measured peak wavelengths. Such low temperatures fall below the range predicted by most current BH envelope models {\citep[e.g.,][]{Kido2025, Inayoshi2025,Begelman2026,Santarelli2026,Umeda2026}}. The {nonnegligible} fraction of these low-temperature envelopes poses a challenge to existing theoretical models. In contrast, at high-$z$, only one LRD at $z\approx4.4$ is reported with $\lambda_{\rm peak}>1$\,\micron\ in {\cite{deGraaff2025_all}}. Two additional sources at $z=3.5$ and $z=4.1$ are reported to be consistent with $\lambda_{\rm peak} \approx 1\,\micron$ within $1\sigma$. The three high-$\lambda_{\rm peak}$ LRDs only represent 8\% of the 40 LRDs at $z=2.3$--$4.5$ in {\cite{deGraaff2025_all}}.   While the simple statistics appear to suggest that cool envelopes are more prevalent in the low-$z$ Universe, both the high-$z$ and low-$z$ samples are subject to selection effects that have not been quantified. 

Figure \ref{fig:HR} also illustrates the loci of a constant blackbody radius as a function of $L_{\rm bb}$ and $\lambda_{\rm peak}$, assuming spherical blackbody emission. For cool LRDs with $\lambda_{\rm peak} \gtrsim 1\,\micron$ at both high and low $z$, the resulting radii are {mostly} $R \gtrsim 2000$ {au}. In this low-temperature regime, the cooler envelopes imply larger envelope radii.  If cool envelopes are indeed more common at low $z$ in the population, this would suggest that low-$z$ LRDs tend to host larger envelopes. However, before drawing any firm conclusions, selection effects should be carefully quantified.

In reality, the optical-to-near-IR continua of LRDs are unlikely to originate from a single blackbody. Many LRDs show significant deviations from a single-temperature blackbody, or can be more properly described by {multitemperature} components {\citep{deGraaff2025_all, Wang2026}}. Variations of temperature and opacity are unsurprising or even expected in the dense gas envelope, which can distort the spectral shape \citep[e.g.,][]{Liu2026}.  To measure the effective temperature, more realistic models are required to account for gas layers with different temperatures and radiative transfer effects within a genuine atmospheric structure.

\subsection{Long-wavelength inflection points}\label{sec:long_wavelength_inflection}

As noted in Section~\ref{sec:continuum_luminosity}, the inflection points do not always occur at the Balmer limit. In our sample, six of the 27 objects (22\%) exhibit inflection points at $\lambda > 4500$~\AA\ (see Table \ref{tab:continuum_property}), a higher fraction than that observed among the $z>2$ LRDs identified by {\cite{deGraaff2025_all}} (13/116 objects, $\approx$11\%). Given the currently limited sample size and incomplete selections at both high and low redshift, it is unclear whether this trend reflects selection effects or potential cosmic evolution. Nevertheless, this suggests that selecting LRDs from spectroscopic libraries such as DESI, SDSS, or JWST/NIRSpec based solely on inflection points near the Balmer limit is insufficient.  

Inflection points at wavelengths near H$\beta$+[\ion{O}{3}] are also seen in three of the four LRDs at $z\sim2$ in \cite{Wang2026}, which are thought to host cooler BH atmospheres at around 3000–4000~K.  Four of the six objects in our sample with inflection points $> 4500$~\AA\  have $\lambda_{\rm peak} > 1\,\mu$m. If converting their $\lambda_{\rm peak}$ using a single blackbody, the resulting temperature would be $T < 3000$ K, placing them as the coolest LRDs known so far. 

\begin{figure}
    \centering
    \includegraphics[width=\linewidth]{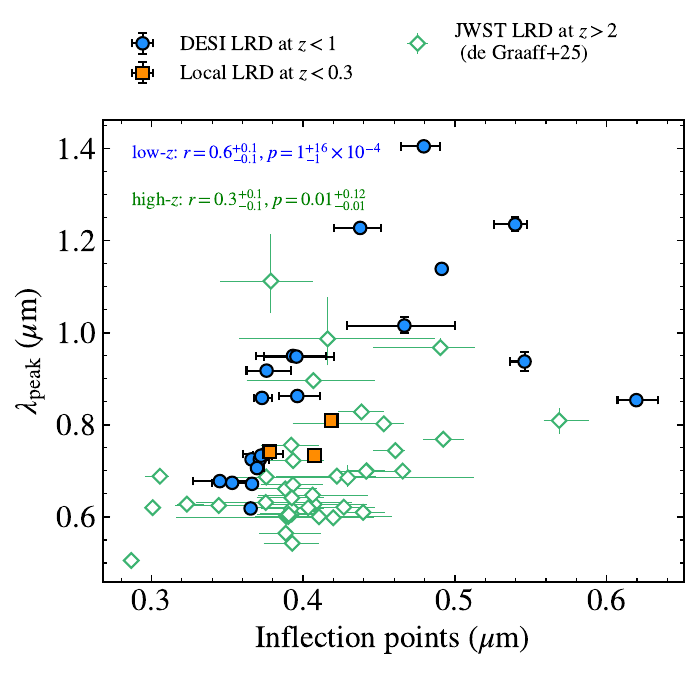}
    \caption{The correlation between the inflection points and blackbody peak wavelengths ($\lambda_{\rm peak}$) of low-$z$ and high-$z$ LRDs. The results of the Kendall $\tau$ correlation analysis for both the high-$z$ and low-$z$ samples are shown.}
    \label{fig:turnover_lambda_peak}
\end{figure}

Figure \ref{fig:turnover_lambda_peak} illustrates the correlation between the inflection points and blackbody peak wavelengths for both the low-$z$ and high-$z$ LRD samples. In the low-$z$ sample, the Kendall $\tau$ analysis reveals a clear trend in which longer inflection points correspond to redder blackbody peaks, except for the outlier J0129+0628. The longer inflection points, therefore, indicate cooler envelopes. The correlation in the high-$z$ sample is less robust.  Although the $p$-value is $0.01$, it is associated with large uncertainties. The clearer correlation seen in the low-$z$ LRDs is primarily driven by the larger fraction of cool LRDs, as discussed in Section \ref{sec:hr_diagram}.
The {observed inflection-point wavelength} is determined by a combination of the temperature of the BH envelope and the relative contribution from the host galaxy. If the envelope temperature is too low, the gas cannot produce an effective Balmer break. Variations in the relative strengths of the galaxy and envelope emission will also shift the location of the inflection points.

\subsection{Constraints on the ionizing source of narrow lines}\label{sec:ionizing_source}

In Section \ref{sec:emission_line_diagnostics}, we present diagnostics of narrow emission lines in DESI LRDs. The ionizing spectra of DESI LRDs are harder than in most star-forming galaxies, but softer than in typical type-1 AGNs. If the narrow lines originate from \ion{H}{2} regions in the host galaxies, the galaxies should be among the most extreme systems dominated by young massive stars.

However, a scenario powered solely by stellar populations poses challenges in explaining the \ion{He}{2} strength (Figure \ref{fig:heii_diagram}).   The \ion{He}{2} emission in DESI LRDs is stronger than that observed in high-$z$ analog galaxies and in metal-poor starburst galaxies.  For reference, in \textsc{BPASS v2.2} models \citep{Stanway2018}, at $Z = 0.1\,Z_\odot$ and $\log U = -2$, the maximum $\log$(\ion{He}{2}/\hb) is $-2.1$. In the most extreme case, $\log U = -1$, a metallicity of $5 \times 10^{-4}\,Z_\odot$, and an age of 25 Myr, the maximum $\log$(\ion{He}{2}/\hb) in the \textsc{BPASS} models reaches $-1.5$.  All detected \ion{He}{2} in low-$z$ LRDs have $\log$(\ion{He}{2}/\hb) $\gtrsim -1.8$, with a stacked value $\approx -1.59$ and four exceeding $-1.5$. This suggests that the observed \ion{He}{2} emission exceeds what can be produced even by the most extreme young stellar populations. A stellar origin for \ion{He}{2} would therefore require the host galaxies of DESI LRDs to have even more extreme ionization parameters. To further boost the \ion{He}{2}-ionizing photon budget, additional {nonstellar} sources, such as X-ray binaries \citep{Schaerer2019}, would be required at fractions much higher than in typical star-forming galaxies.     Shock excitation could also enhance \ion{He}{2} emission. However, the [\ion{S}{2}] BPT diagram argues against this interpretation. The [\ion{S}{2}] lines are highly sensitive to shocks: once shocks contribute more than $\sim$20\%, sources are expected to shift toward the LINER region due to enhanced [\ion{S}{2}] emission produced during the cooling of shocked gas \citep{Ho2014}. The location of LRDs in the [\ion{S}{2}] BPT diagram instead indicates that shocks do not play a significant role.

A more natural interpretation is that the narrow lines are powered by a combination of AGN and host galaxy. The presence of AGN photons could also explain the high incidence and elevated ionization parameters of [\ion{O}{3}] outflows. In the BH envelope scenario, one possibility is that photons from the central accretion disk are reprocessed by the envelope, producing a softer ionizing spectrum than that emitted directly from the disk.  Another possibility is that AGN photons escape through low-opacity channels, or that the envelope is clumpy, leading to a mixture of AGN emission, stellar light, and thermally reprocessed envelope emission. This geometry is also suggested by recent deep UV spectroscopic observations of {A2744-QSO1} \citep{Ji2026_SPURS_LRDs, Tang2026_SPURS_LRDs}.

For line diagnostics in Section \ref{sec:emission_line_diagnostics}, statistical high-$z$ LRD measurements obtained in the same way are still lacking, preventing a robust comparison between high-$z$ and low-$z$ samples. In the \ion{He}{2} diagram, the detected \ion{He}{2}/H$\beta$ ratios in the DESI LRDs are higher than the narrow-line fit of RUBIES-EGS-49140 from {\cite{Wang2025_heii}}. In the stacked NIRSpec/PRISM spectra of high-$z$ LRDs from {\cite{Wang2025_heii}}, \ion{He}{2} 4687 is not detected. In the stacked PRISM spectra of \cite{Perez-Gonzalez2026}, a broad \ion{He}{2} 4687 component (FWHM $>2000~\mathrm{km\,s^{-1}}$) is detected and attributed to either Wolf–Rayet stars or AGN activity, although the PRISM resolution is low. In contrast, we do not find significant broad \ion{He}{2} 4687 emission in our DESI LRD sample. In the absence of a statistical sample of individual high-$z$ \ion{He}{2} 4687 measurements with sufficient spectral resolution, it remains unclear whether low-$z$ LRDs exhibit systematically different \ion{He}{2} 4687 emission strengths and profiles.
Similarly, it has been reported that JWST AGNs lack prominent fast outflows \citep{Maiolino2025}, although [\ion{O}{3}] outflows have been detected in several high-$z$ LRDs \citep{Juodzbalis2024, DEugenio2025_irony,Cooper2025}. A larger sample of NIRSpec grating observations of their [\ion{O}{3}] lines is required to assess the incidence rate.

\subsection{Highlight: J1717+3807}\label{sec:highlight_J1717}

Among all the DESI DR1 LRDs presented in this work, we highlight J1717+3807 as a particularly promising target for further follow-up. Indeed, it will be observed by JWST GO 12316 (PI: Franz Bauer).

J1717+3807 is the brightest and lowest-redshift LRD in our DESI sample ($m_r \approx 19$ mag and $m_i \approx 18$ mag), comparable in brightness to the three $z<0.3$ LRDs reported in \citetalias{Lin2025_localLRD} (J1025+1402, \textit{The Egg}; J1047+0739; J1022+0841), while the remaining DESI LRDs are typically 2–3 mag fainter.

J1717+3807 shows different properties from the three bright $z<0.3$ LRDs. Its V-shaped inflection point occurs at $\sim$4910\,\AA, near \hb + [\ion{O}{3}], whereas the other three systems inflect near the Balmer break (3704–4168\,\AA).  Its optical–near-IR continuum peaks at $\sim$1.14\,\micron, while the other three objects peak at 7330–8094\,\AA. Both the inflection point and peak wavelength are significantly redder (longer) in J1717+3807. The envelope temperature $T_{\rm peak}$ is cooler ($\sim$2540\,K converted from Wien’s displacement law) than those of the other three ($\sim$4000\,K from Wien’s displacement law and $\sim$4500\,K when fitted with theoretical atmospheric models as shown in \citealt{Liu2026}). In Figure \ref{fig:HR}, J1717+3807 lies near the $R = 5000$ {au} curve, while the other three local LRDs cluster around $R \sim 1000$ {au}.

As discussed in Section \ref{sec:variability}, its $i$-band and WISE flux fade on time-scales of years. It is in contrast with the lack of significant variability seen in the other three $z<0.3$ LRDs {\citep{Burke2025}} but is reminiscent of the $\sim 20$\,yr variability observed in RXCJ2211-RX1 at $z\sim4$ \citep{ZZhang2025_century}. 
It is worth exploring whether the presence of variability is linked to the envelope temperature and therefore to its instability \citep{Cantiello2026}. A large sample of bright LRDs spanning a wide range of temperatures is needed to investigate the connection.

%% file: table/balmer_absorption_properties.tex
\begin{tabular}{lcccccccccc}
\hline
Name & $v_{\rm H\beta, abs}$ & $\log N_{\rm H\beta}$ & $b_{\rm H\beta}$ & $C_f$ & $\rm EW_{\rm H\beta}$ & $v_{\rm H\alpha, abs}$ & $\log N_{\rm H\alpha}$ & $b_{\rm H\alpha}$ & $C_f$ & $\rm EW_{\rm H\alpha}$ \\
 & (km s$^{-1}$) & ($\log$ cm$^{-2}$) & (km s$^{-1}$) &  & (\AA) & (km s$^{-1}$) & ($\log$ cm$^{-2}$) & (km s$^{-1}$) &  & (\AA) \\
\hline
\multicolumn{11}{c}{\textbf{Gold sample}} \\
\hline
J0129+0628 & -- & -- & -- & -- & -- & -66.4$_{-5.8}^{+5.2}$ & 13.3$_{-0.1}^{+0.1}$ & 72$_{-5}^{+5}$ & 1 & 2.8$_{-0.3}^{+0.3}$ \\
J0826--0100 & 62$_{-19}^{+33}$ & 15.5$_{-0.4}^{+0.7}$ & 120$_{-27}^{+34}$ & 1 & 7.4$_{-0.6}^{+0.6}$ & -77.8$_{-8.7}^{+10.2}$ & 14.5$_{-0.5}^{+0.8}$ & 84$_{-17}^{+22}$ & 0.91$_{-0.04}^{+0.04}$ & 6.4$_{-0.4}^{+0.3}$ \\
J0829+1312 & -- & -- & -- & -- & -- & -17.1$_{-12.1}^{+10.7}$ & 13.6$_{-0.2}^{+0.1}$ & 270$_{-28}^{+26}$ & 1 & 6.7$_{-2.0}^{+2.0}$ \\
J0944--0249 & -368$_{-18}^{+18}$ & 13.9$_{-0.1}^{+0.1}$ & 169$_{-22}^{+28}$ & 1 & 1.6$_{-0.2}^{+0.3}$ & -275.5$_{-11.1}^{+12.3}$ & 18.5$_{-0.1}^{+0.1}$ & 21$_{-8}^{+9}$ & 0.39$_{-0.03}^{+0.03}$ & 4.4$_{-0.6}^{+0.6}$ \\
J1025+5028 & -78$_{-18}^{+15}$ & 14.6$_{-0.1}^{+0.1}$ & 246$_{-26}^{+26}$ & 1 & 6.2$_{-1.4}^{+1.3}$ & -241.0$_{-9.6}^{+9.1}$ & 15.7$_{-1.1}^{+1.2}$ & 57$_{-10}^{+14}$ & 0.81$_{-0.03}^{+0.06}$ & 5.3$_{-0.3}^{+0.3}$ \\
J1321--0214 & -- & -- & -- & -- & -- & -81.0$_{-4.4}^{+2.6}$ & {12.93$_{-0.05}^{+0.05}$} & 49$_{-4}^{+6}$ & 1 & 1.5$_{-0.1}^{+0.1}$ \\
J1423+5202 & -50$_{-24}^{+19}$ & 14.8$_{-0.3}^{+0.3}$ & 301$_{-69}^{+65}$ & 0.70$_{-0.20}^{+0.20}$ & 6.2$_{-1.9}^{+2.4}$ & -247.4$_{-4.9}^{+5.2}$ & 13.6$_{-0.1}^{+0.1}$ & 134$_{-12}^{+9}$ & 0.84$_{-0.09}^{+0.10}$ & 4.9$_{-0.2}^{+0.2}$ \\
J1611+0917 & -190$_{-19}^{+16}$ & 14.7$_{-0.1}^{+0.1}$ & 210$_{-19}^{+19}$ & 1 & 6.7$_{-0.7}^{+0.6}$ & -223.7$_{-3.4}^{+3.4}$ & {13.80$_{-0.01}^{+0.01}$} & 176$_{-4}^{+3}$ & 1 & 8.0$_{-0.1}^{+0.1}$ \\
J1641+0708 & -- & -- & -- & -- & -- & -121.8$_{-17.2}^{+17.3}$ & 13.4$_{-0.2}^{+0.2}$ & 135$_{-28}^{+31}$ & 0.74$_{-0.20}^{+0.18}$ & 2.8$_{-0.6}^{+0.8}$ \\
J1642+0426 & -- & -- & -- & -- & -- & -116.0$_{-7.3}^{+5.1}$ & 16.2$_{-0.5}^{+0.6}$ & 85$_{-7}^{+9}$ & 0.69$_{-0.04}^{+0.03}$ & 7.1$_{-0.3}^{+0.3}$ \\
J1646+1426 & -225$_{-22}^{+23}$ & {14.83$_{-0.05}^{+0.04}$} & 500$_{-27}^{+22}$ & 1 & 11.6$_{-0.9}^{+0.7}$ & -251.8$_{-8.4}^{+5.3}$ & {13.75$_{-0.01}^{+0.02}$} & 242$_{-6}^{+14}$ & 1 & 8.8$_{-0.2}^{+0.4}$ \\
J1654+0337 & -83$_{-21}^{+16}$ & 15.0$_{-0.2}^{+0.2}$ & 212$_{-38}^{+28}$ & 0.82$_{-0.07}^{+0.08}$ & 7.2$_{-0.9}^{+0.8}$ & -188.3$_{-5.9}^{+6.2}$ & {13.80$_{-0.02}^{+0.04}$} & 167$_{-5}^{+7}$ & 0.98$_{-0.03}^{+0.01}$ & 7.8$_{-0.2}^{+0.3}$ \\
J1717+3807 & -261$_{-45}^{+50}$ & 14.0$_{-0.2}^{+0.1}$ & 182$_{-58}^{+45}$ & 1 & 1.9$_{-0.6}^{+0.6}$ & -208.0$_{-3.0}^{+3.3}$ & {13.36$_{-0.01}^{+0.01}$} & 128$_{-3}^{+4}$ & 1 & 3.9$_{-0.1}^{+0.1}$ \\
J2127--0448 & 165$_{-57}^{+85}$ & 14.0$_{-0.2}^{+0.2}$ & 184$_{-66}^{+79}$ & 1 & 2.2$_{-0.9}^{+1.0}$ & 42.3$_{-4.9}^{+11.5}$ & 14.9$_{-1.0}^{+2.0}$ & 60$_{-19}^{+21}$ & 0.56$_{-0.08}^{+0.11}$ & 3.3$_{-0.7}^{+0.6}$ \\
J2255+1542 & -- & -- & -- & -- & -- & 20.5$_{-20.2}^{+42.9}$ & 14.6$_{-0.4}^{+0.7}$ & 121$_{-27}^{+31}$ & 0.79$_{-0.05}^{+0.06}$ & 7.8$_{-1.2}^{+1.6}$ \\
\hline
\multicolumn{11}{c}{\textbf{Silver sample}} \\
\hline
J1119+0219 & -- & -- & -- & -- & -- & -260.7$_{-36.4}^{+28.9}$ & 14.8$_{-1.2}^{+2.2}$ & 77$_{-33}^{+46}$ & 0.38$_{-0.08}^{+0.10}$ & 2.3$_{-0.7}^{+1.1}$ \\
J1137+5520 & -- & -- & -- & -- & -- & -85.9$_{-17.5}^{+25.9}$ & 13.2$_{-0.2}^{+0.3}$ & 76$_{-25}^{+19}$ & 1 & 2.5$_{-0.6}^{+1.1}$ \\
J1343+3934 & -- & -- & -- & -- & -- & -147.7$_{-3.4}^{+3.5}$ & 13.9$_{-0.6}^{+3.0}$ & 37$_{-19}^{+16}$ & 0.63$_{-0.04}^{+0.11}$ & 1.7$_{-0.2}^{+0.2}$ \\
J1909+5831 & -- & -- & -- & -- & -- & -24.0$_{-9.6}^{+8.1}$ & 13.7$_{-0.2}^{+0.2}$ & 158$_{-26}^{+28}$ & 1 & 7.1$_{-1.6}^{+2.2}$ \\
\hline
\end{tabular}

%% file: 05_Summary.tex
\section{Summary}

This paper is the second in our series on low-$z$ LRD surveys, extending the sample presented in \citetalias{Lin2025_localLRD}. We aim to build a statistical sample of low-$z$ LRDs to complement the high-$z$ JWST samples and to probe the cosmic evolution of this population. 

In this work, we present our selection of LRDs based on DESI DR1, currently the largest publicly available spectroscopic dataset. We compile {multiwavelength} photometry and conduct spectroscopic follow-up observations from Magellan/FIRE, LBT/LUCI and Keck/NIRES. Our findings are summarized as follows.

\medskip
\begin{itemize}
    \item We select 27 LRDs from DESI DR1 at $z = 0.2$–$0.9$. These sources are consistent with JWST-discovered high-$z$ LRDs in morphology, continuum properties, Balmer emission and absorption line properties, and narrow-line properties. Across all these diagnostics, they occupy distributions consistent with those of high-$z$ LRDs, suggesting that the same underlying physical processes are at work.  The discovery significantly expands the sample of known LRDs at $z < 1$.

    \item The selection in DESI DR1 yields an incomplete number density of $7.5 \times 10^{-10}$ cMpc$^{-3}$ at $z < 1$, which should be regarded as a conservative lower limit (Section~\ref{sec:number_density}).

    \item DESI LRDs exhibit a wide range of continuum shapes, with inflection points spanning from near the Balmer limit to just blueward of \ha\  ($\sim6200$\,\AA). Their luminosities ($M_{\rm UV}$, $L_{\rm 5100}$) occupy a distribution similar to that of high-$z$ LRDs discovered by JWST (Section~\ref{sec:continuum_luminosity}).

    \item The broad \ha\ and \hb\ luminosities are strongly correlated with $L_{\rm 5100}$. These relationships are well described by power laws but deviate from those observed in local type-1 AGNs. This discrepancy cautions against the direct application of local scaling relations when estimating the {BH} masses of LRDs. On the other  hand, it also suggests an alternative scaling relation may be {at play} (Section~\ref{sec:balmer_emission}).

    \item The Balmer decrement (\ha/\hb) of narrow lines has a median value of 3.4, while that of broad lines is 14.6. Such extreme broad-line Balmer decrements point to radiative transfer effects in dense gas rather than dust attenuation alone. No statistically significant correlation is found between the broad-line Balmer decrement and either $L_{\rm 5100}$ or the Balmer break strength (Section~\ref{sec:balmer_decrement}).

    \item Nineteen of the 27 LRDs (70\%) exhibit \ha\ and/or \hb\ absorption. The velocity shifts and EWs of \ha\ and \hb\ are highly correlated. However, the \hb\ EWs are systematically higher than those expected from the Voigt-profile parameters of the \ha\ absorber, under the assumption that a single cloud produces both transitions. This implies that the Balmer absorption in LRDs originates from multiple rather than single gas clouds.
   The absorbers span velocity shifts from $-368$ to $+165$\,km\,s$^{-1}$ (mostly blueshifted), column densities of $\log(N_{\rm H\beta}/\rm cm^{-2}) = 13.9$--15.5 and $\log(N_{\rm H\alpha}/\rm cm^{-2}) = 12.9$--18.5, and Doppler parameters $b \sim 21$--500\,km\,s$^{-1}$, revealing a wide range of absorber dynamical states. Most absorbers have covering factors near unity. We do not find a significant correlation between the Balmer absorber velocity or EW and either the Balmer break strength or {the Balmer decrement (Section}~\ref{sec:balmer_absorption}).

    \item All DESI LRDs exhibit low metallicity in their narrow-line regions or host galaxies, with a median value of $12 + \log(\mathrm{O/H}) = 7.8$. In the BPT diagrams, all DESI LRDs lie within the star-forming region or near the Seyfert--\ion{H}{2} boundary in the [\ion{N}{2}] and [\ion{S}{2}] diagrams, while most lie in the Seyfert region or near the Seyfert--\ion{H}{2} boundary in the [\ion{O}{1}] diagram. They occupy the same parameter space in the BPT diagram as emission-line galaxies at $z>1.4$ (Section~\ref{sec:emission_line_diagnostics}).

    \item In the \ion{He}{2} 4687 diagram, the \ion{He}{2}/\hb\ ratios of DESI LRDs lie between those of star-forming galaxies and AGNs. Their mean stacked \ion{He}{2}/\hb\ value ($\log$(\ion{He}{2}/\hb) $= -1.59\pm0.17$) exceeds that of most galaxies, including high-$z$ analogs and extremely metal-poor galaxies, but remains below typical AGNs. This indicates that DESI LRDs have an ionizing spectrum harder than most star-forming galaxies, yet softer than typical AGNs (Section~\ref{sec:emission_line_diagnostics}).

    \item Narrow \ion{Mg}{2} emission with FWHMs of 73--208\,km\,s$^{-1}$ is detected in six DESI LRDs. Their EWs range from {4–15}\,\AA, consistent with those of low-mass starburst galaxies (Section~\ref{sec:emission_line_diagnostics}).

    \item Broad [\ion{O}{3}] outflows are detected in 21 of 27 DESI LRDs (78\%), an incidence rate much higher than that of star-forming galaxies, implying an AGN origin. However, the outflow velocities are relatively modest ($v_{\rm out}\sim163$--644\,km\,s$^{-1}$, median 234\,km\,s$^{-1}$), lower than those of galactic outflows at $z>3$ and below most AGN-driven outflows (Section~\ref{sec:ionized_outflow}).

    \item Among sources with light curves of sufficient S/N for variability analysis, the majority show no significant variability or only tentative low-S/N variability requiring confirmation. The only robustly variable source is J1717+3807. Its $i$-band flux faded by $0.05 \pm 0.01$ mag over 3.3 yr in the rest frame, followed by a brightening of $0.02 \pm 0.01$ mag. Its WISE W1 and W2 light curves show monotonic fading of $0.123 \pm 0.006$ and $0.159 \pm 0.006$ mag, respectively, over a rest-frame baseline of 11.7 yr (Section \ref{sec:variability}).

    \item Twenty-one LRDs have sufficient wavelength coverage to characterize the optical-to-near-infrared continuum. Their continua are well described by modified blackbody-like spectra peaking at $0.6$--$1.4\,\micron$, corresponding to temperatures of $\sim$2000--4700\,K under Wien's law assuming a single blackbody. Five objects peak at $>1$\,\micron, corresponding to temperatures below 3000\,K.  The fraction of low-temperature LRDs among the 21 objects with measured peak wavelengths (24\%) is higher than that in $z=2.3$--$4.5$ LRDs (8\%), although the selection effect is not yet quantified (Section \ref{sec:hr_diagram}). 

    \item The {H-R} diagram of LRDs ($\lambda_{\rm peak}$ or $T_{\rm peak}$ versus $L_{\rm bb}$) does not reveal a clear sequence or correlation between $\lambda_{\rm peak}$ ($T_{\rm peak}$) and $L_{\rm bb}$. In the low-$T_{\rm peak}$ ($\lambda_{\rm peak}>1$\,\micron, $T_{\rm peak}<3000$ K) regime, the cooler envelopes imply larger envelope radii (Section \ref{sec:hr_diagram}).

    \item  A correlation between longer-wavelength SED inflection points and cooler envelopes is suggested for low-$z$ LRDs (Section~\ref{sec:long_wavelength_inflection}).

    \item The ionizing spectra of DESI LRDs, which are softer than those of typical AGNs but harder than those of star-forming galaxies, are most likely explained by a combination of AGN and host-galaxy emission. In the BH envelope framework, one possibility is that AGN photons are reprocessed by the envelope, or that they escape through low-opacity channels or a clumpy {envelope structure
 (Section}~\ref{sec:ionizing_source}).

    \item Finally, among all the DESI LRDs in this work, we highlight J1717+3807 as a particularly promising target for further follow-up studies. As the brightest and lowest-redshift object in our sample, it presents different properties from the other three bright $z<0.3$ LRDs: longer inflection points (near \hb+[\ion{O}{3}]), redder blackbody peaks ($\lambda_{\rm peak} \approx 1.14\,\micron$), indicative of a very cool BH envelope ($T_{\rm peak}\sim2500$\,K), and significant variability in $i$, WISE W1, and W2 bands over a baseline of $\sim 10$ yr (Section \ref{sec:highlight_J1717}). 
\end{itemize}

The most pressing question surrounding LRDs concerns the nature of their central engine, their BH masses, and their role in BH growth. BH mass estimates, as a key to deciphering LRDs' nature, remain uncertain.  The correlation between the Balmer luminosity and $L_{\rm 5100}$ suggests that empirical BH mass estimators calibrated for local type-1 AGNs may not be directly applicable, and instead points toward a potential new calibration relation for BH mass estimates in LRDs. Furthermore, establishing whether these systems trace a common pathway of BH assembly requires a comprehensive characterization of their structure and geometry, the physical conditions of the absorbing gas, their variability, and the interplay between the central engine and its circum-BH environment and host galaxy. Addressing these questions requires mapping the full diversity of the population and its evolution across cosmic time. Only by tracing these properties from $z > 6$ to $z \sim 0$ can we establish whether LRDs mark a transient yet universal phase of BH growth, or a phenomenon confined to specific physical conditions.

While JWST has delivered large statistical samples of LRDs at $z>2$, the low-redshift regime ($z<1$) remains a largely open frontier for the community.  Our current sample of 27 objects from DESI DR1, along with those from SDSS in \citetalias{Lin2025_localLRD}, does not yet span the full parameter space.  Future DESI data releases ($\sim$5$\times$ more spectra), ongoing and upcoming spectroscopic surveys (e.g., 4MOST, PFS, MUST), wide-field imaging surveys (e.g., CSST, Euclid, Roman, LSST, SPHEREx), and growing JWST/NIRSpec archival datasets at $z > 2$, will collectively enable the assembly of LRD samples across cosmic time. In addition to the LRD population, our understanding of the low-$z$ AGN population remains limited, particularly for low-metallicity and low-luminosity systems, as well as the broader population that shares only some of the properties of LRDs.  It is also critical to understand how these characteristics arise, how LRDs connect to other AGN populations, and how they fit into the broader picture of BH growth.

As the second paper in our low-redshift LRD survey series, (LRDs)$^2$, this work establishes the selection framework in DESI DR1 and presents an initial census. In subsequent papers, we will expand to larger and more complete samples, extend wavelength coverage through ongoing follow-up observations, and develop physically motivated models.

%% file: 99_Appendix.tex
\clearpage
\appendix
\counterwithin{figure}{section}
\counterwithin{table}{section}
\counterwithin{equation}{section}

\section{Sample Accounting {through} the Selection Pipeline}
\label{appendix:selection_accounting}

{
Table~\ref{tab:selection_accounting} summarizes the sample size after each major selection stage described in Section~\ref{sec:selection}. The red optical-continuum and emission-line criteria based on the DESI value-added catalogs are grouped into a single catalog-based stage because they are evaluated simultaneously.
 The subsequent selection is applied to the surviving sources using Legacy Survey DR10 catalog and direct measurements from the DESI spectra. The complete selection workflow leaves {5562} sources for visual inspection.

\begin{center}
\centering
\captionof{table}{Sample Accounting {through} the Selection Pipeline}
\label{tab:selection_accounting}

\noindent\makebox[\linewidth][c]{%
\begin{tabular}{lrr}
\hline
\hline
Selection stage & Sources remaining & Sources removed \\
\hline
DESI DR1 input
    & 27,547,223 & --- \\
$z>0.001$
    & 22,751,337 & 4,795,886 \\
Value-added-catalog-based red optical-continuum
(Section~\ref{sec:selection_red_continuum})
    & \multirow{2}{*}{2,989,065}
    & \multirow{2}{*}{19,762,272} \\
+ emission-line criteria
(Section~\ref{sec:emline_criteria})
    & & \\
PSF-like morphology (Section \ref{sec:morph})
    & 434,098 & 2,554,967 \\
Declining near-IR: W1 versus DESI red-edge criterion (Section \ref{sec:selection_nearIR})
    & 433,485 & 613 \\
Spectrum-integrated line S/N (Section \ref{sec:manual_criteria})
    & 90,940 & 342,545 \\
Blue UV-continuum measurement (Section \ref{sec:manual_criteria})
    & 83,414 & 7,526 \\
Non-parametric Balmer vs. [\ion{O}{3}] broad-line filtering (Section \ref{sec:manual_criteria})
    & 22,083 & 61,331 \\
Strong-[\ion{N}{2}] and/or [\ion{Ne}{5}] veto (Section \ref{sec:manual_criteria})
    & 21,028 & 1,055 \\
Spectrum-based blue optical-continuum rejection (Section \ref{sec:manual_criteria})
    & 5,562 & 15,466 \\
\hline
\end{tabular}%
}\par

\begin{flushleft}
\footnotesize
\textit{Note.} Each row corresponds to a major selection stage and indicates the relevant section of the manuscript. Theses stages are shown schematically by the corresponding boxes in Figure~\ref{fig:selection}. The table reports the number of sources retained after each stage and the number removed relative to the preceding stage. Because the value-added-catalog-based red optical-continuum and emission-line criteria are evaluated simultaneously, they are reported as a single selection stage.
\end{flushleft}

\vspace{0.5em}
\end{center}
}

\clearpage
\section{DESI DR1 LRD sample}\label{appendix:sample}

Figures~\ref{fig:sample_gold} and \ref{fig:sample_silver} present the SEDs and emission-line fits of the entire DESI DR1 LRD sample, excluding the two sources already shown in Figures~\ref{fig:example1}. The \textsc{Gold} sample (Figure~\ref{fig:sample_gold}) comprises sources with $k_{\rm red} > 0$ at $\geq 3\sigma$ significance, while the \textsc{Silver} sample (Figure~\ref{fig:sample_silver}) includes those with $k_{\rm red} > 0$ at $< 3\sigma$ significance (see Table~\ref{tab:continuum_property}).

\vspace*{12pt}
\sedpair
{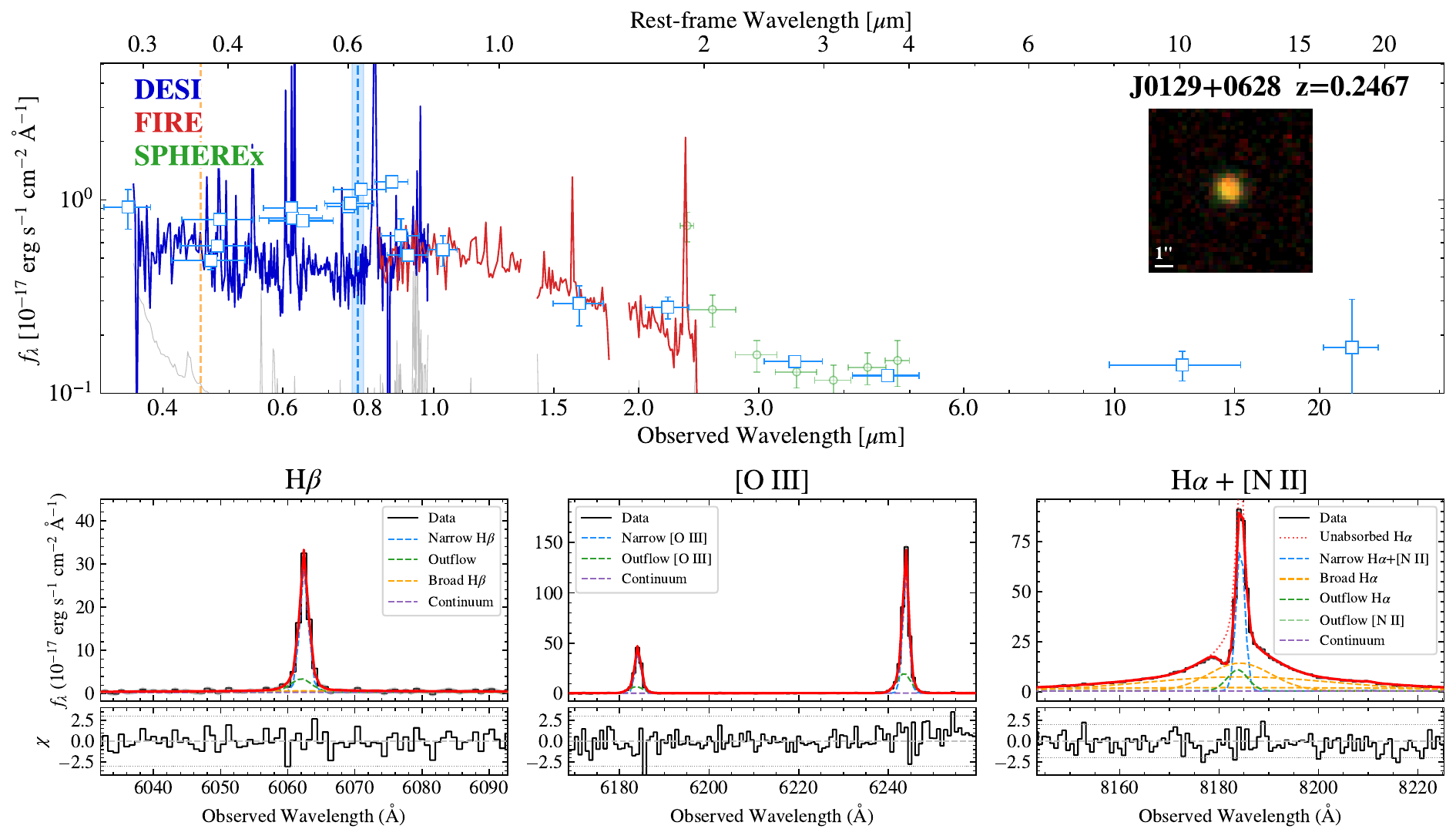}
{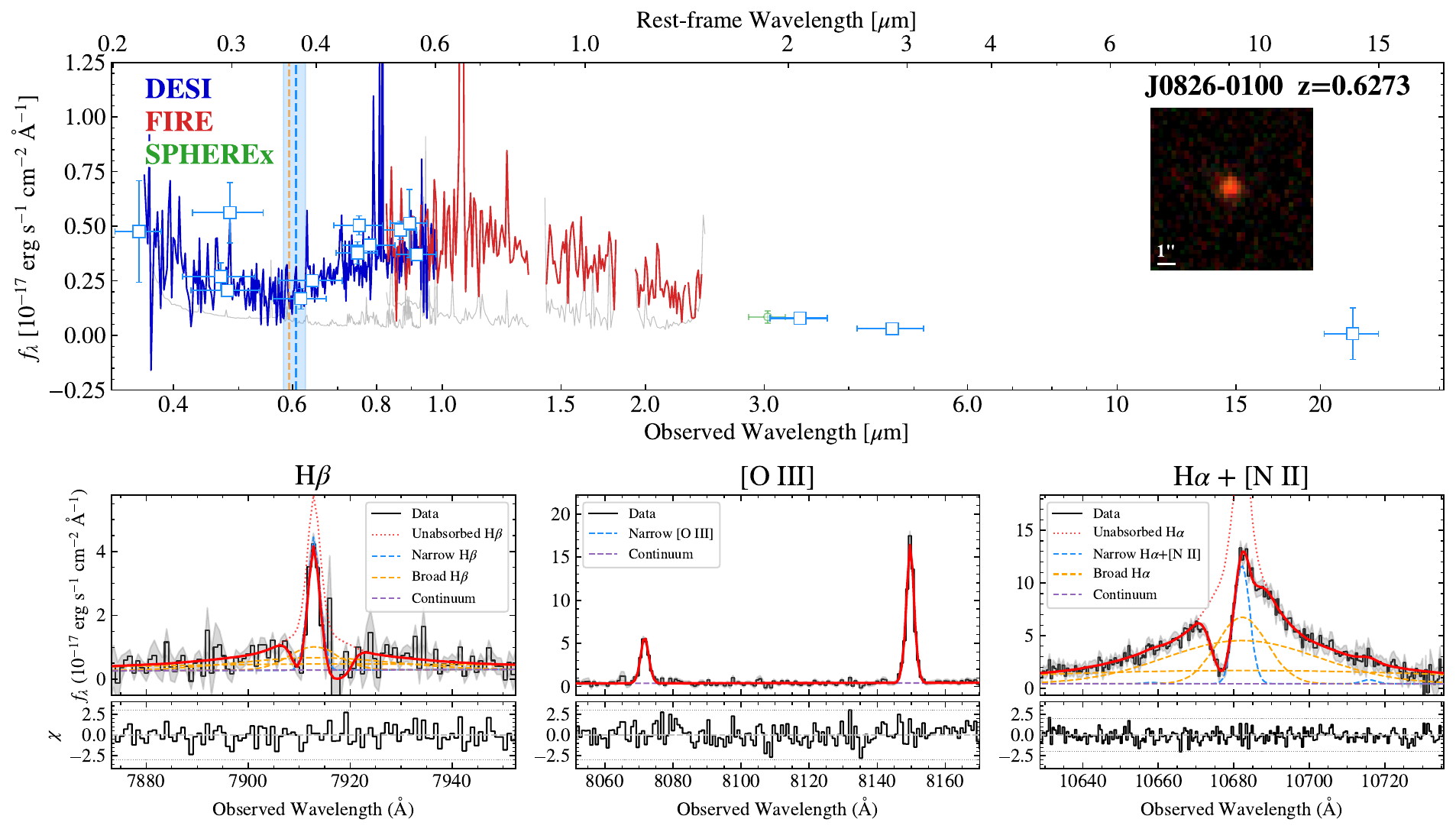}

\sedpair
{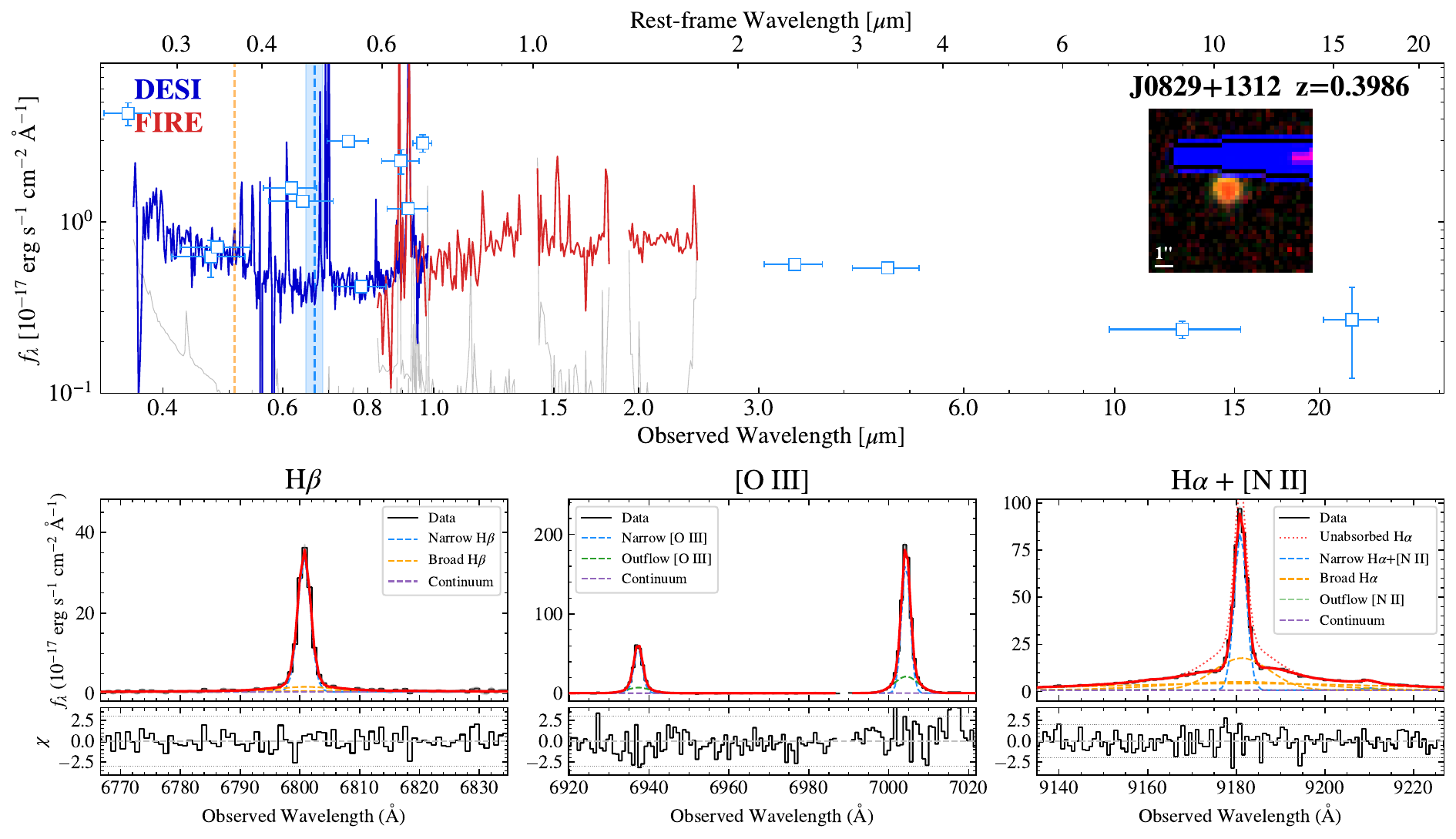}
{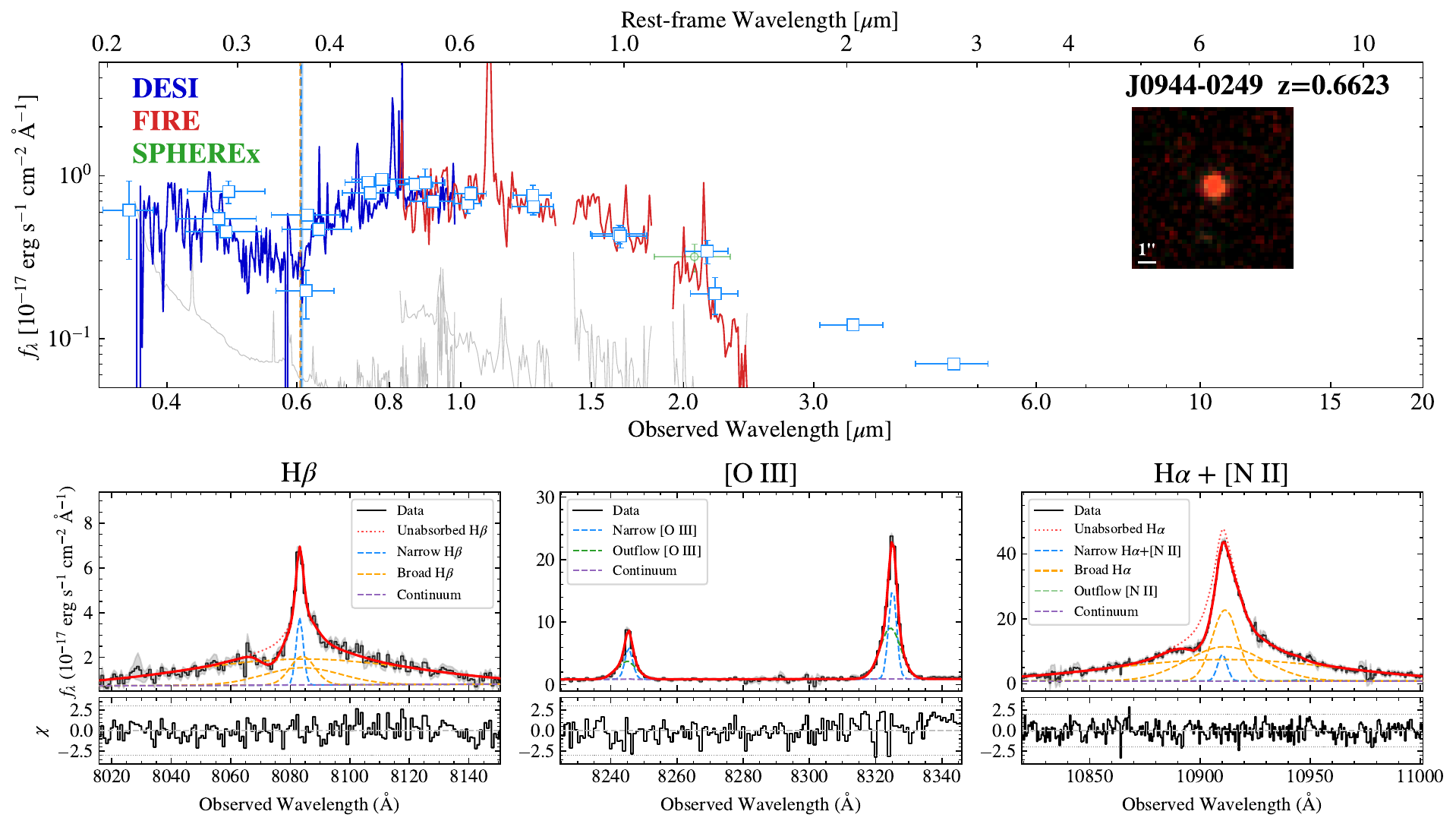}

\sedpair
{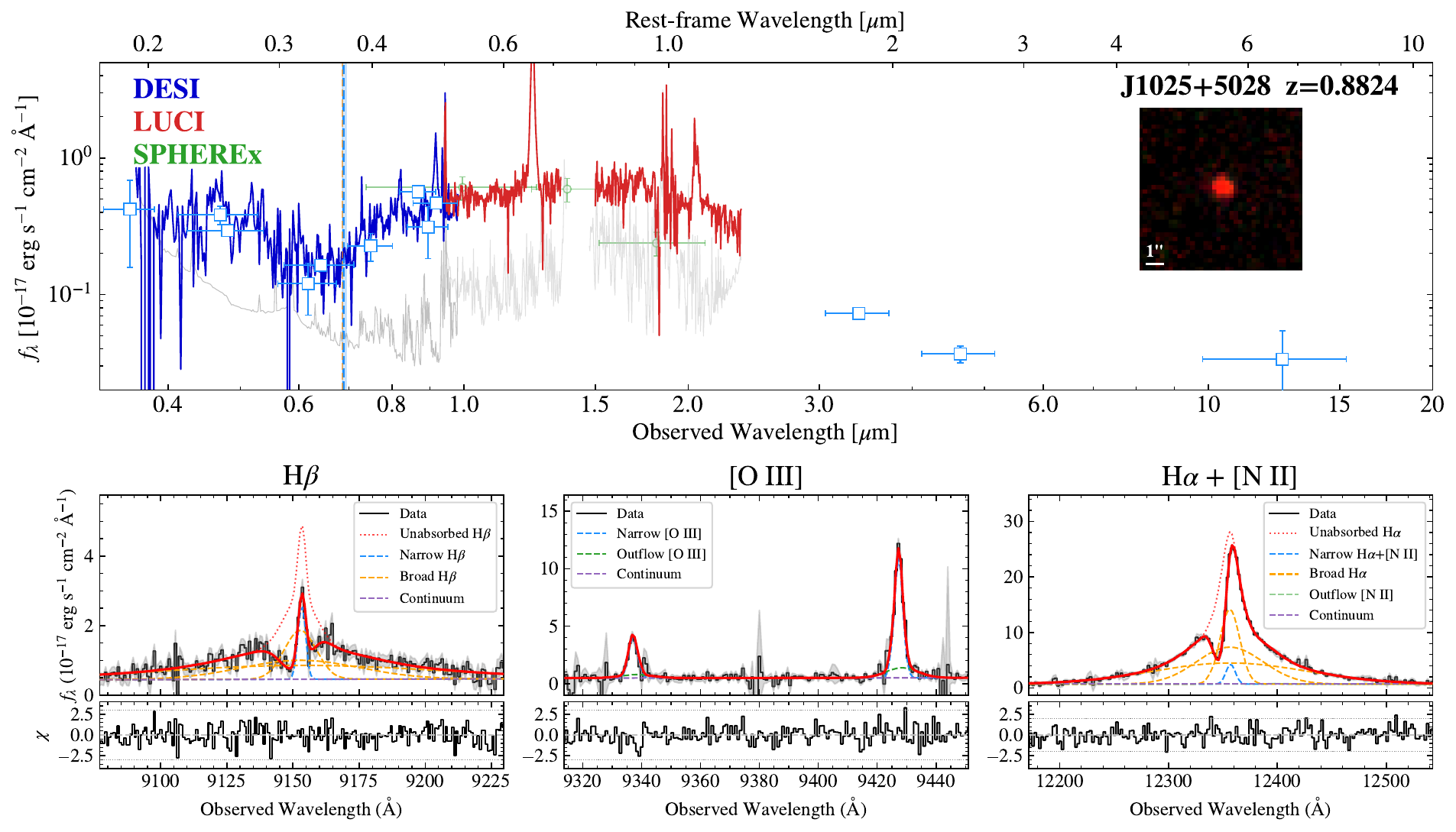}
{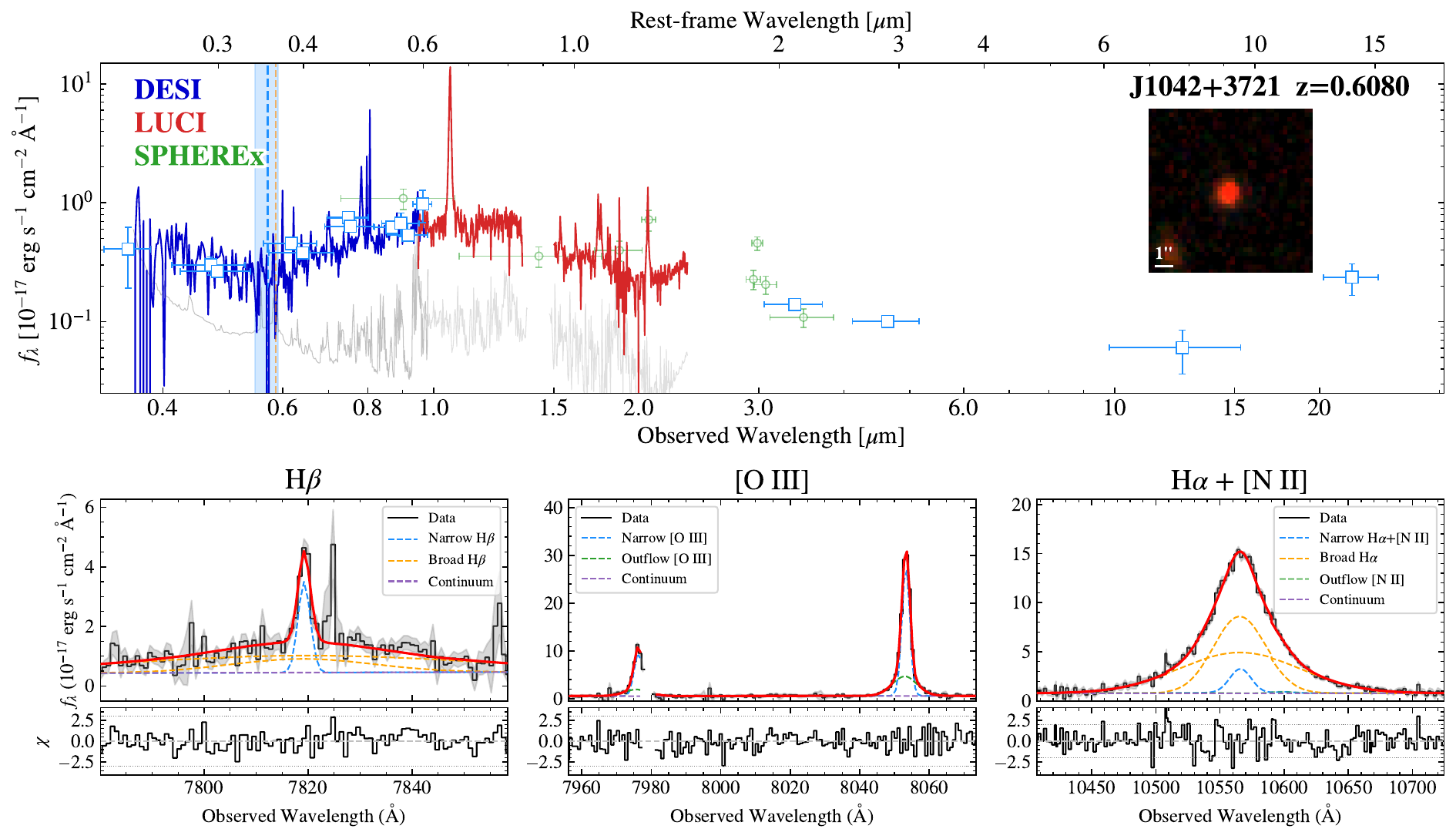}

\sedpair
{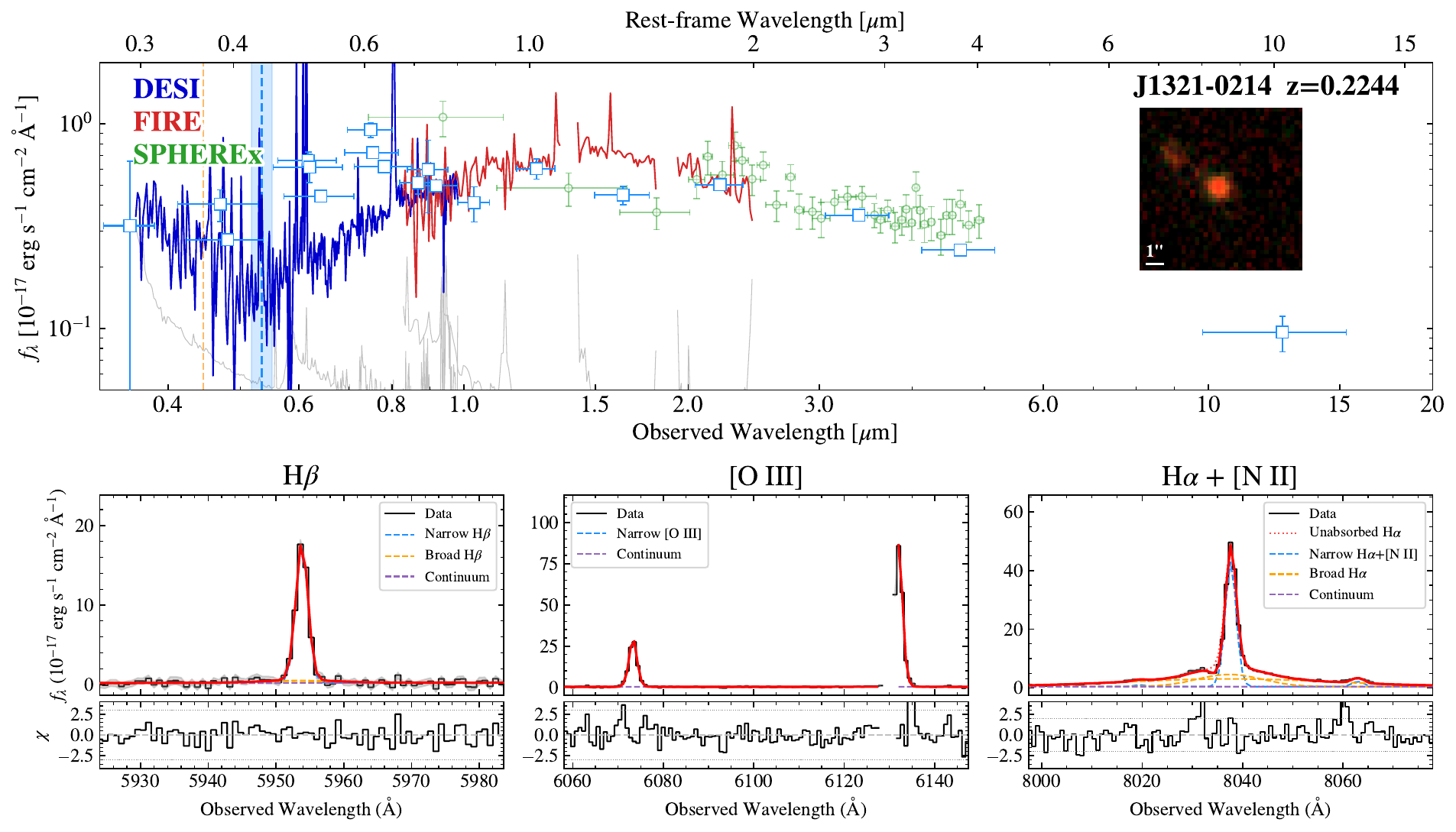}
{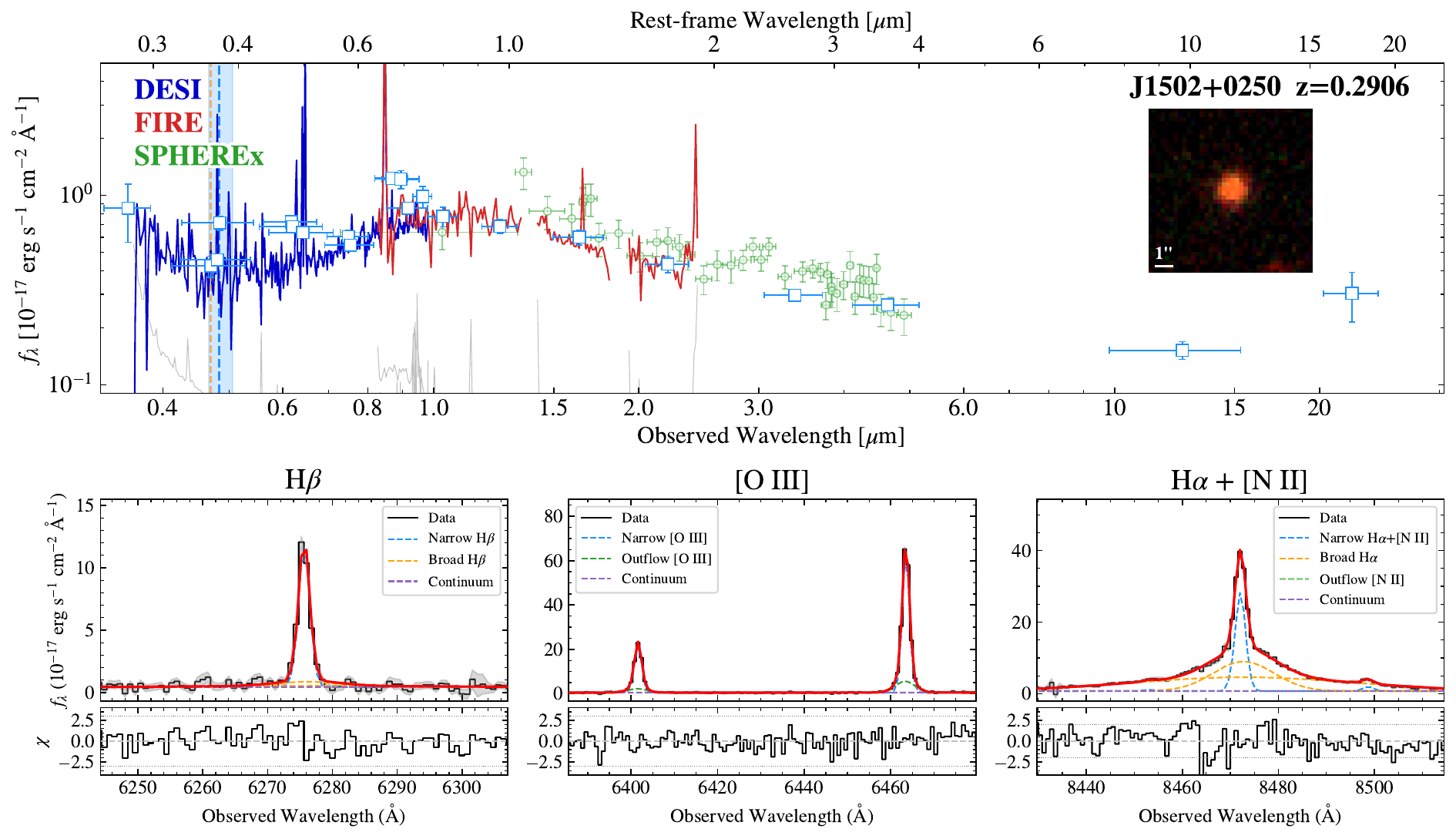}

\sedpair
{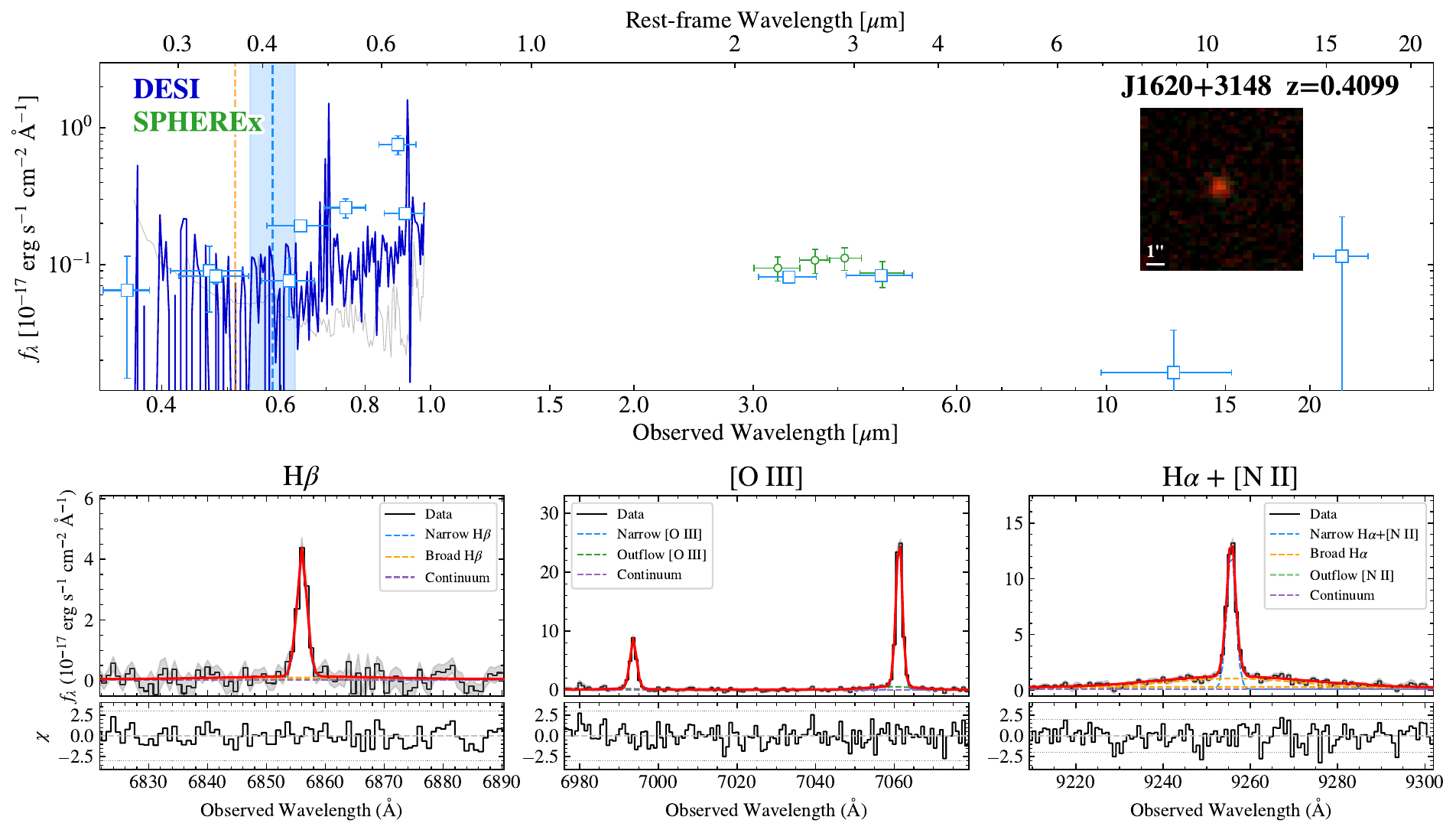}
{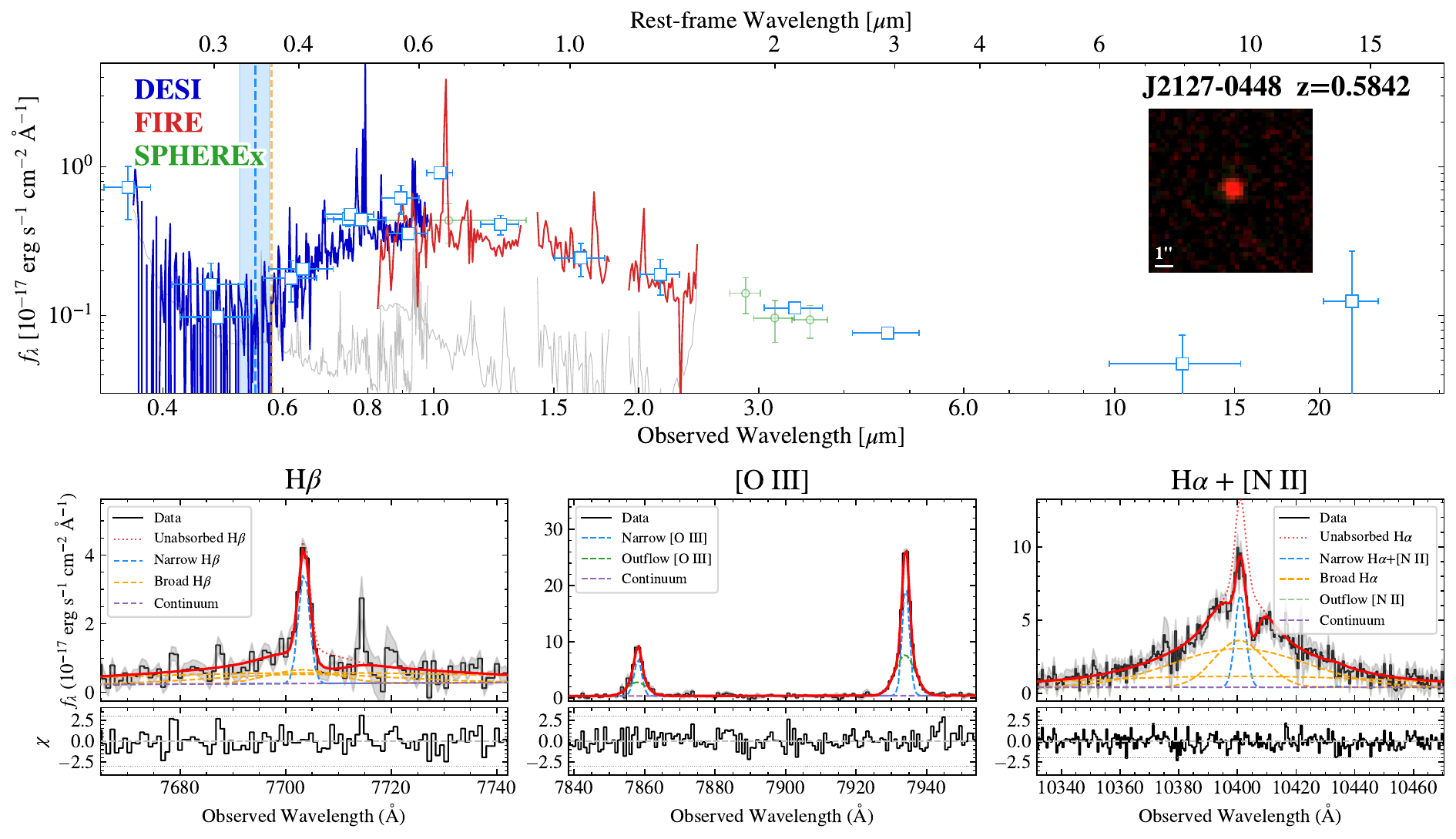}

\sedpair
{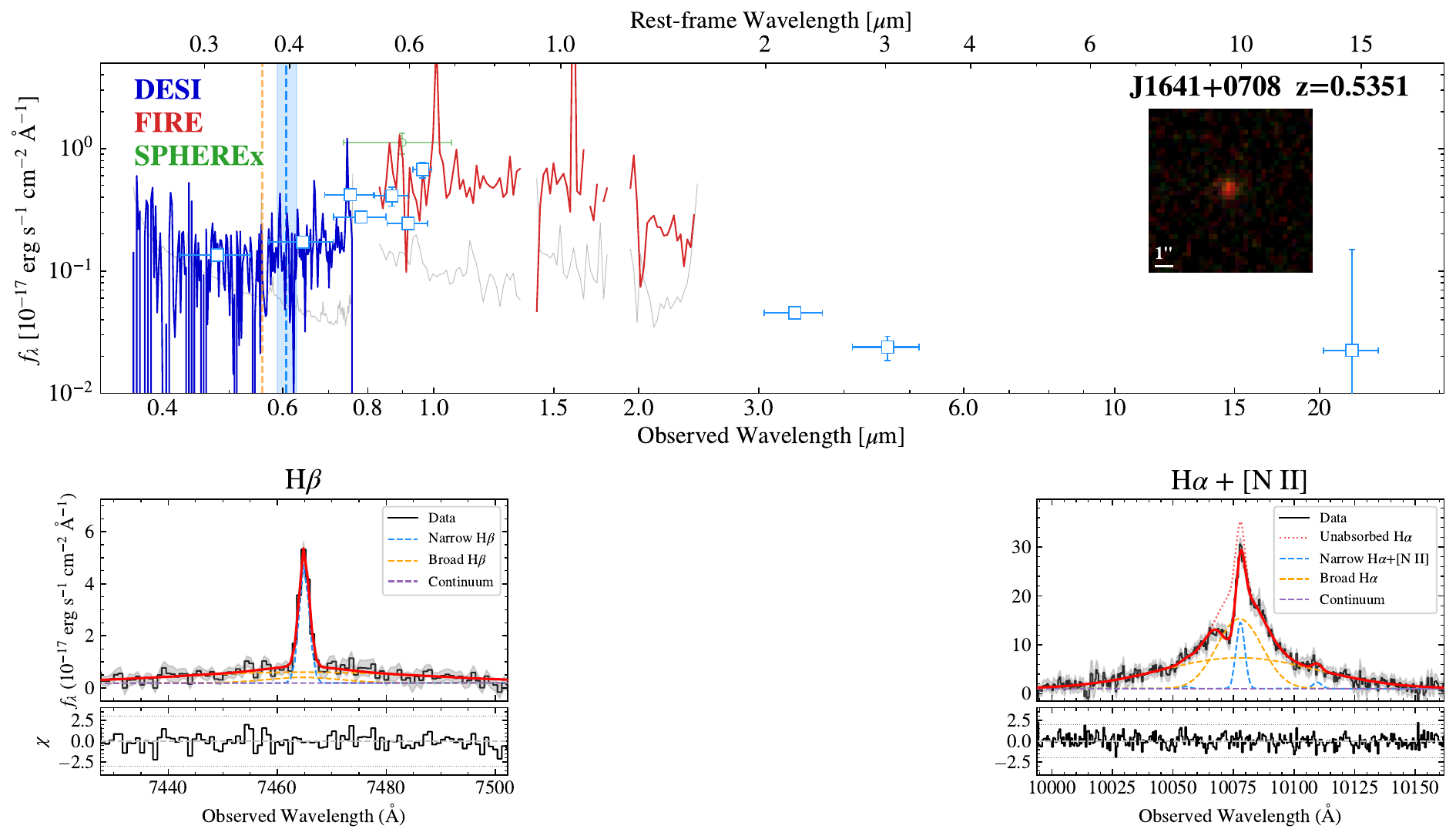}
{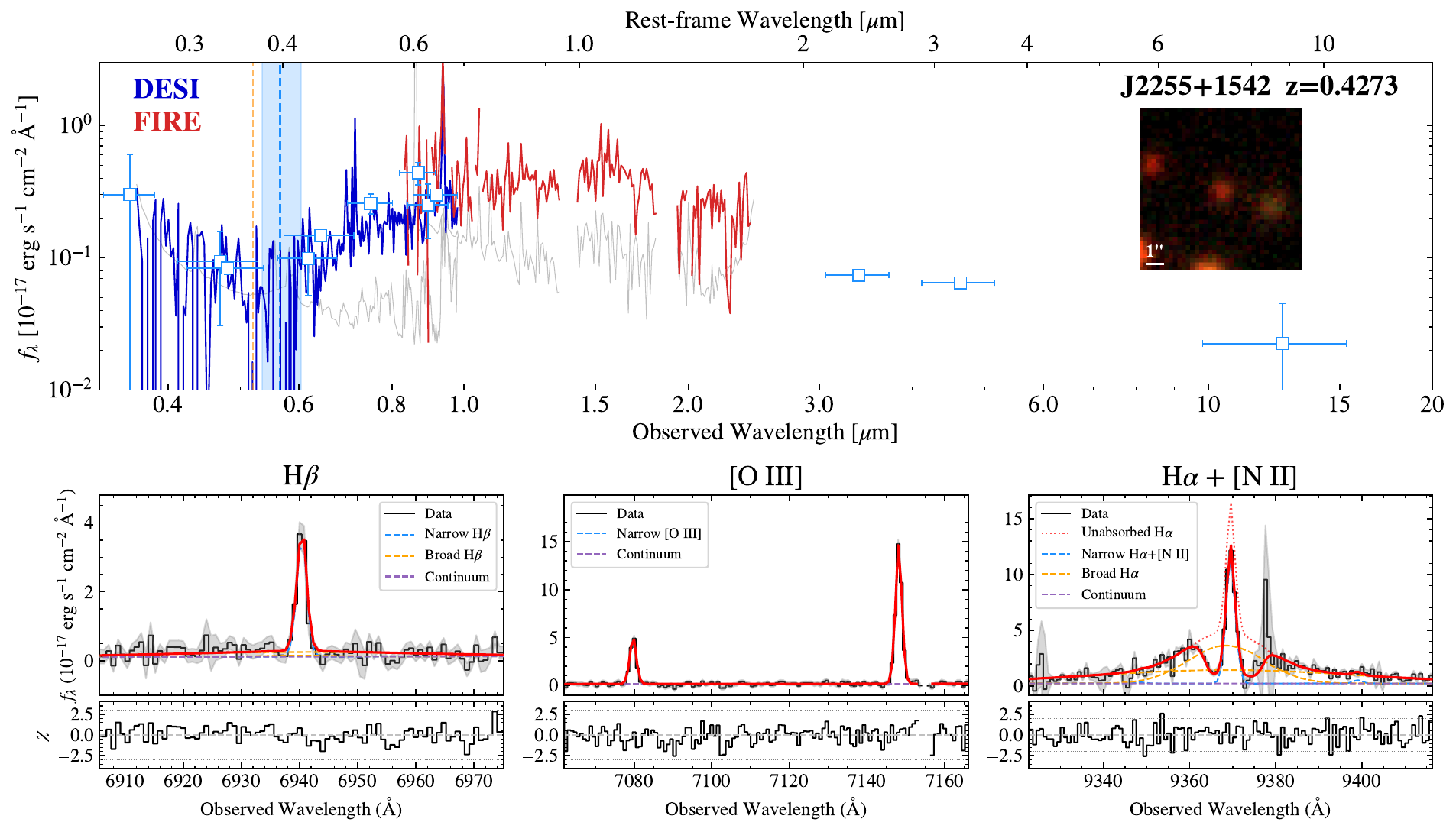}

\sedpair
{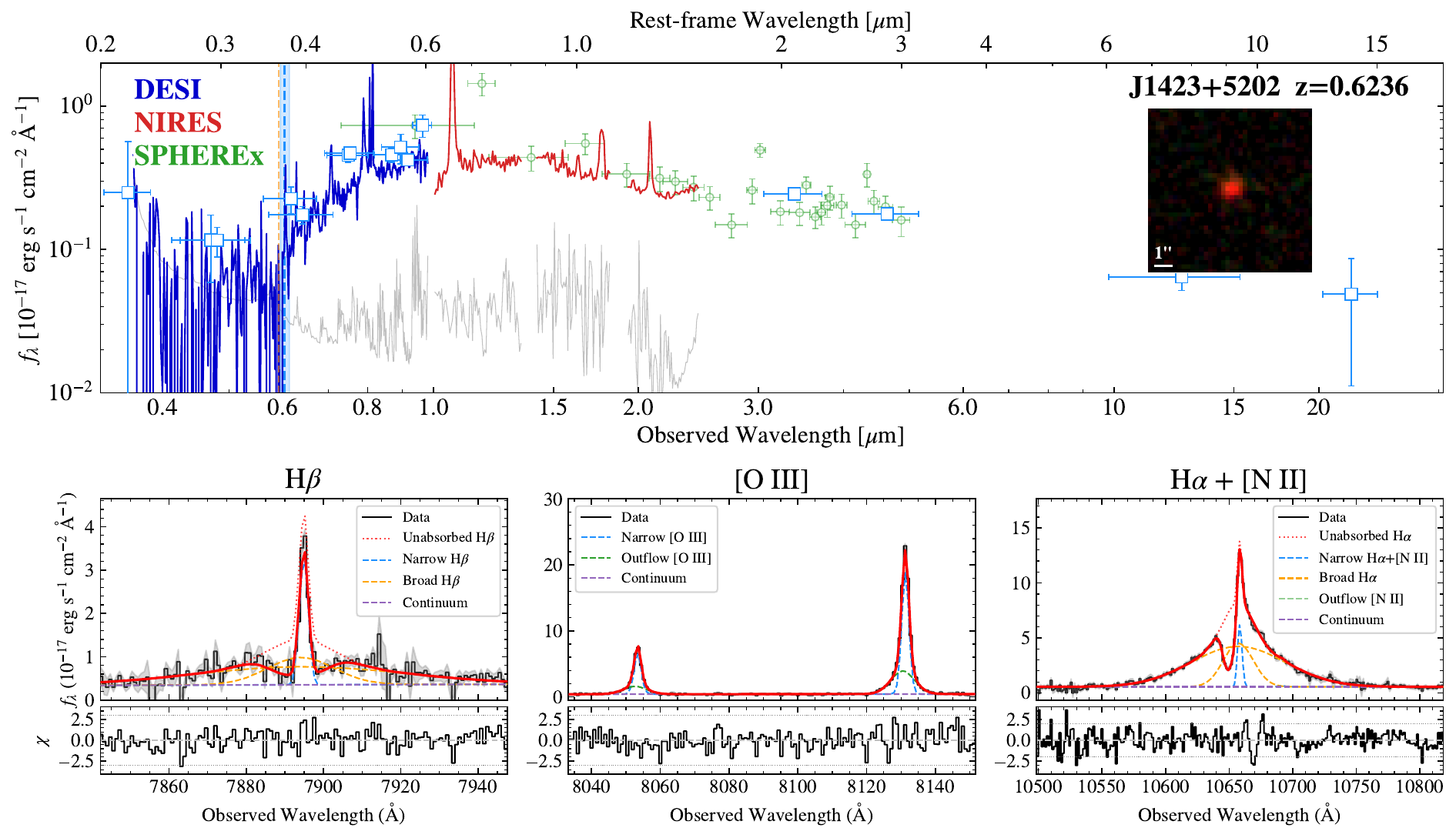}
{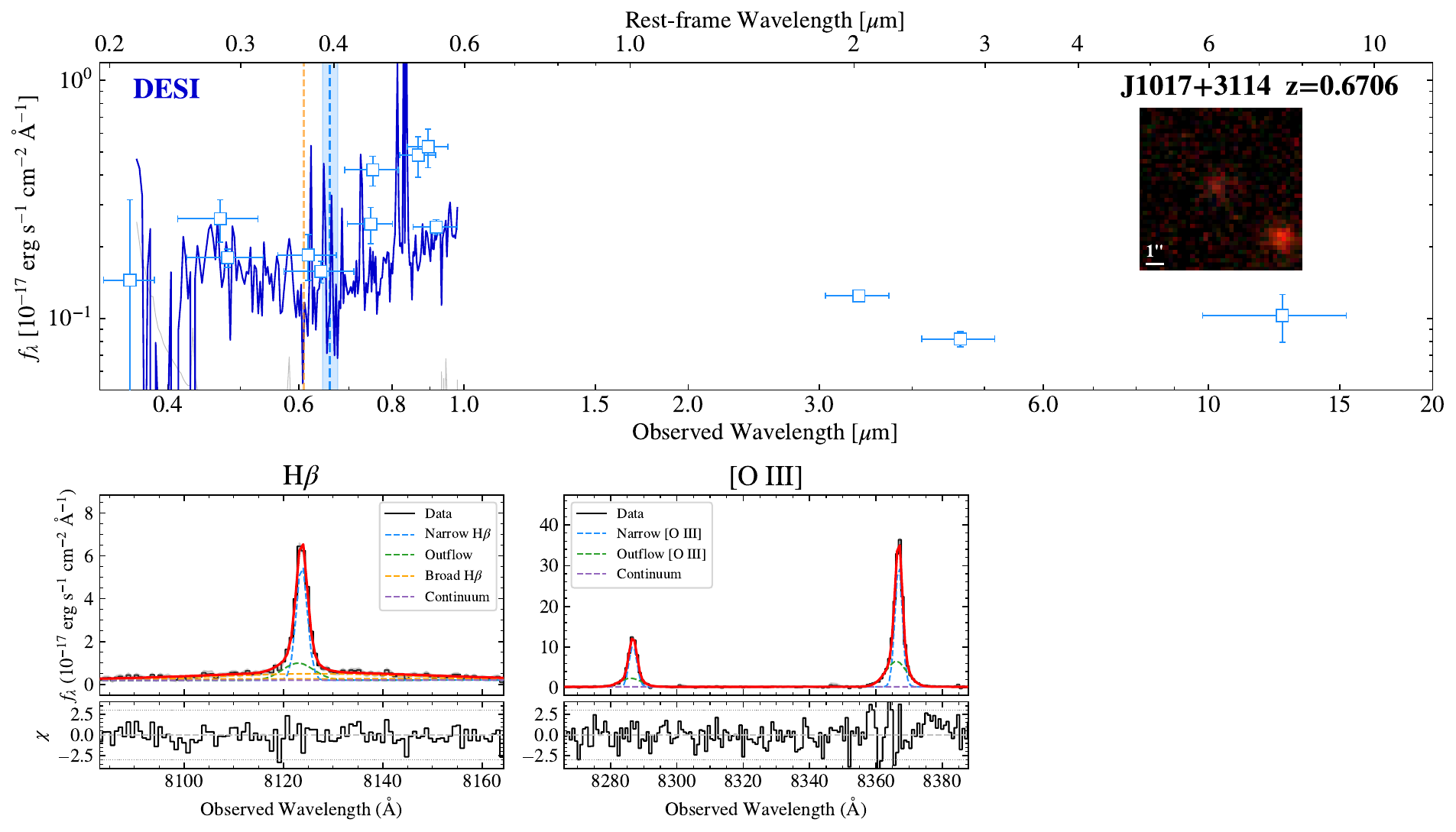}

\sedpair
{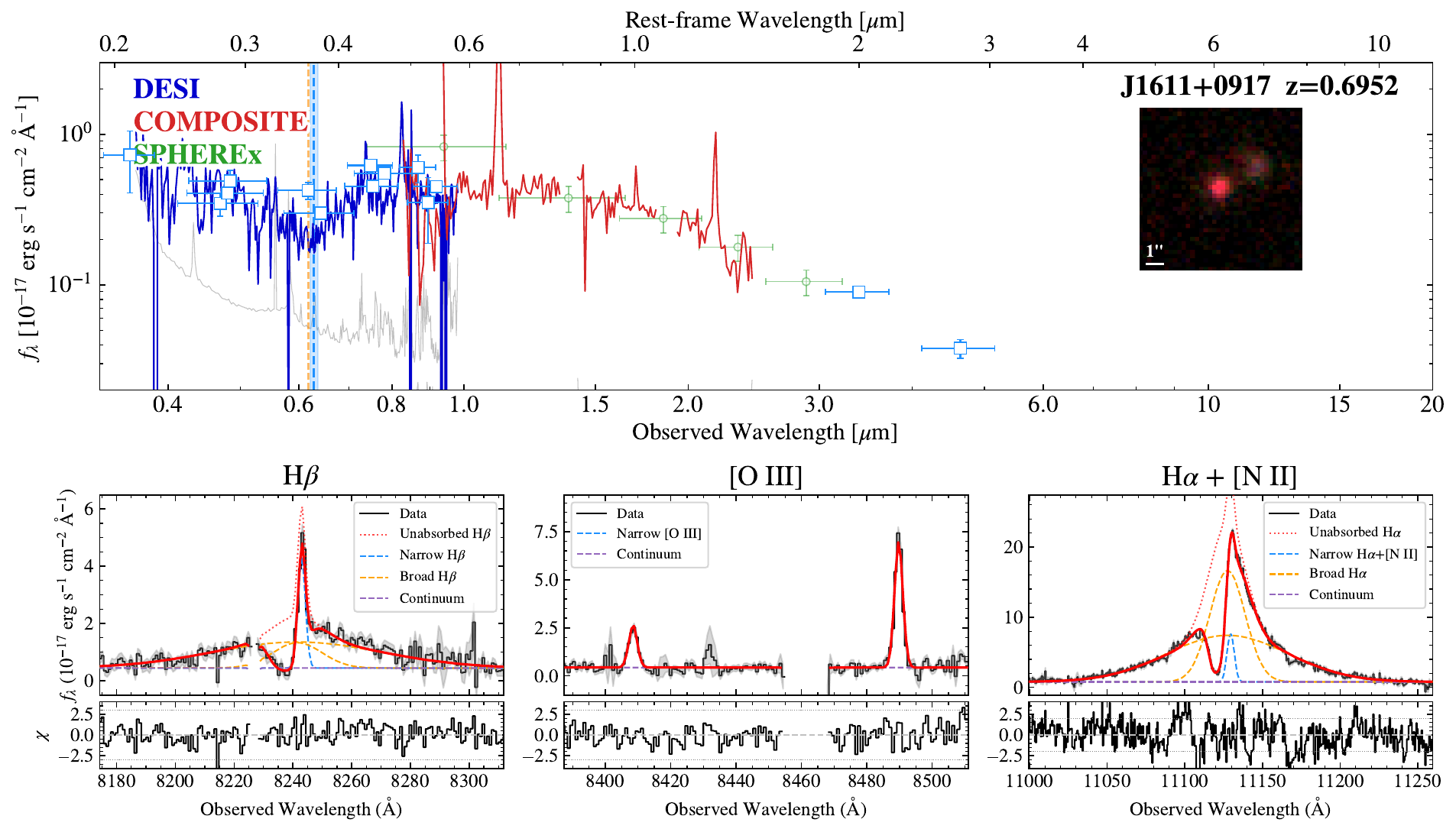}
{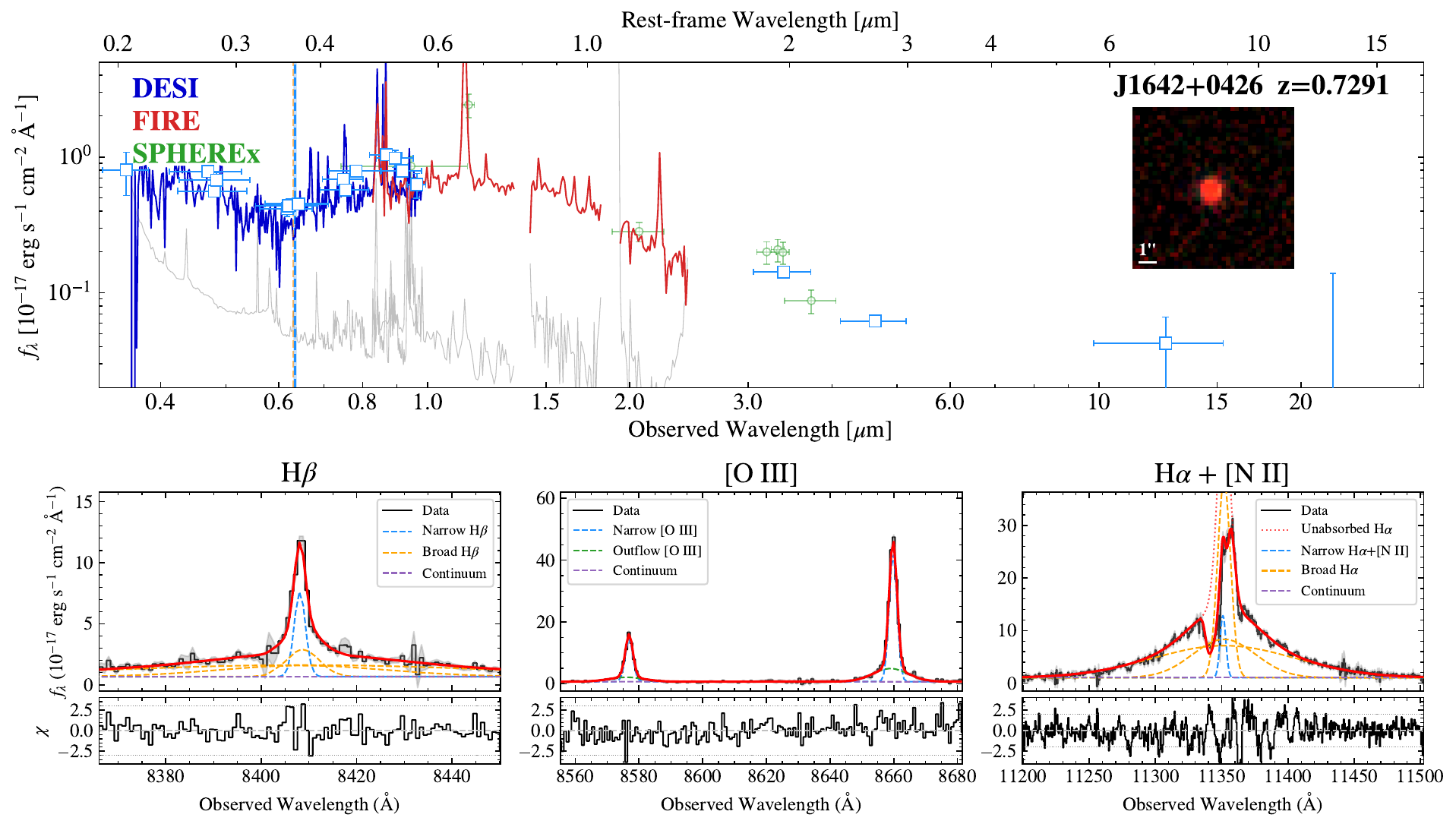}

\noindent
\includegraphics[width=0.49\textwidth]
{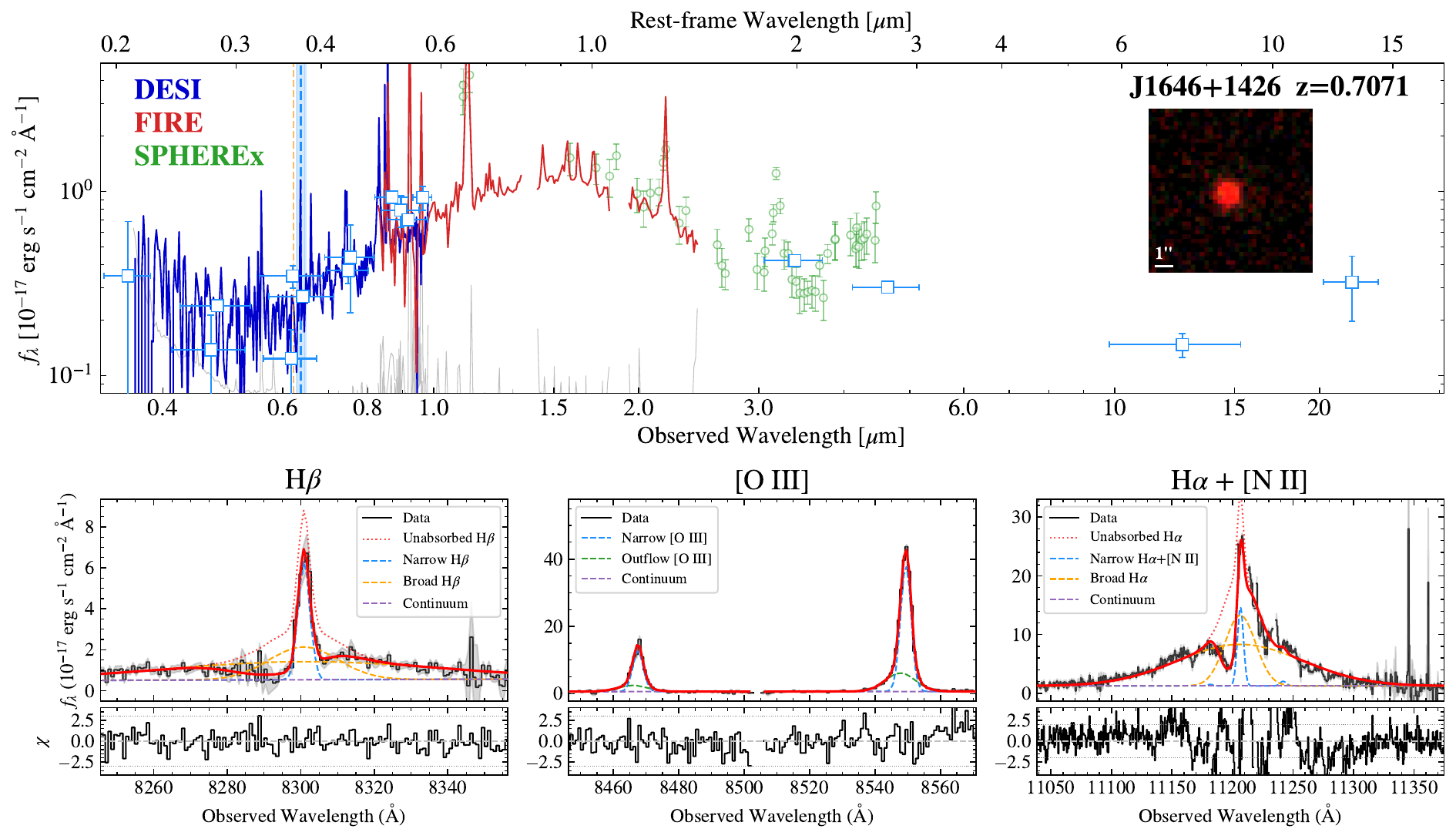}
\par
\vspace{0.5em}

\figcaption{
\textsc{Gold} sample: DESI DR1 LRDs with
$k_{\rm red}>0$ at $\geq3\sigma$ significance.
Similar to Figure~\ref{fig:example1}.
\label{fig:sample_gold}
}

\clearpage
\sedpair
{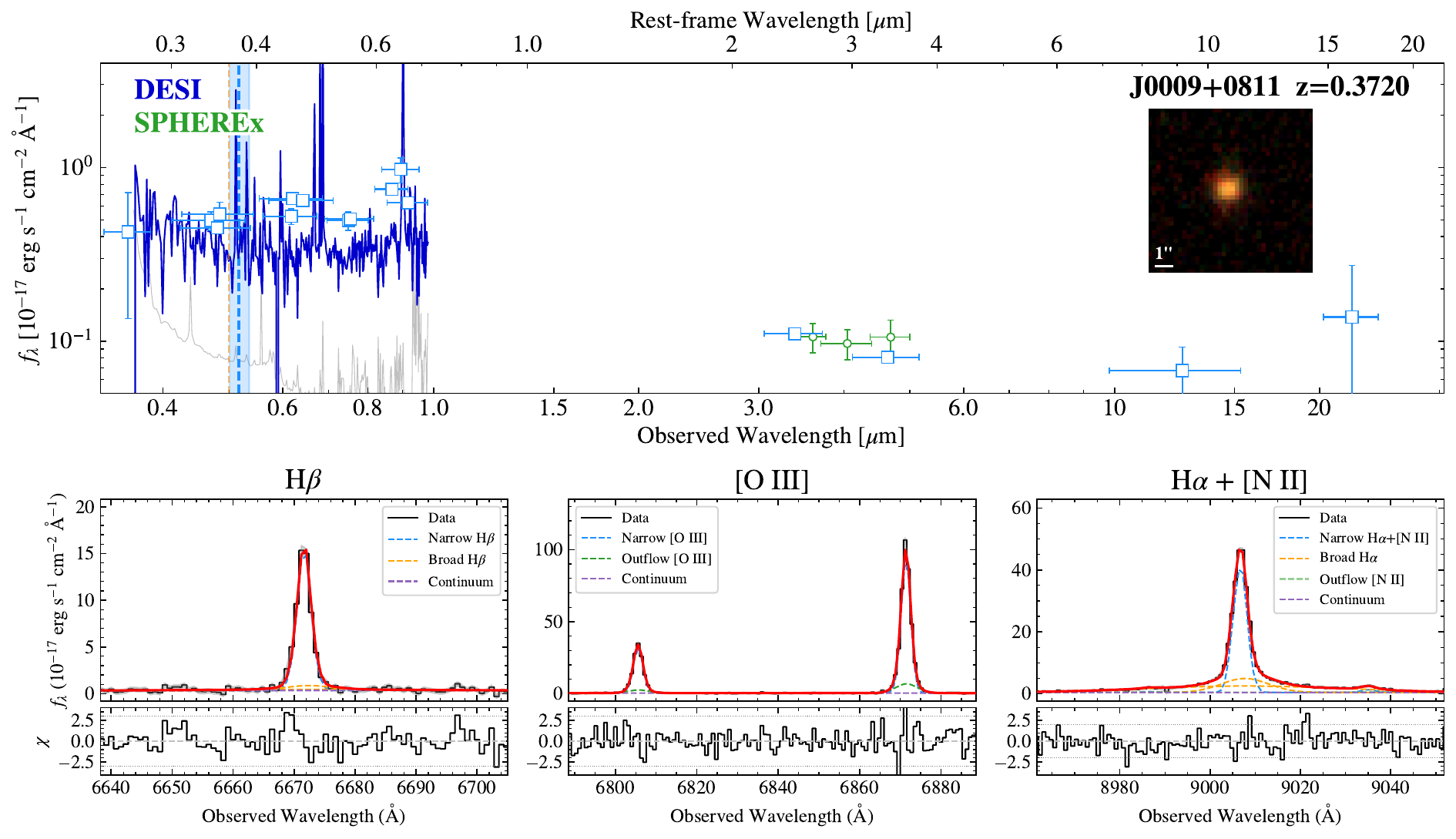}
{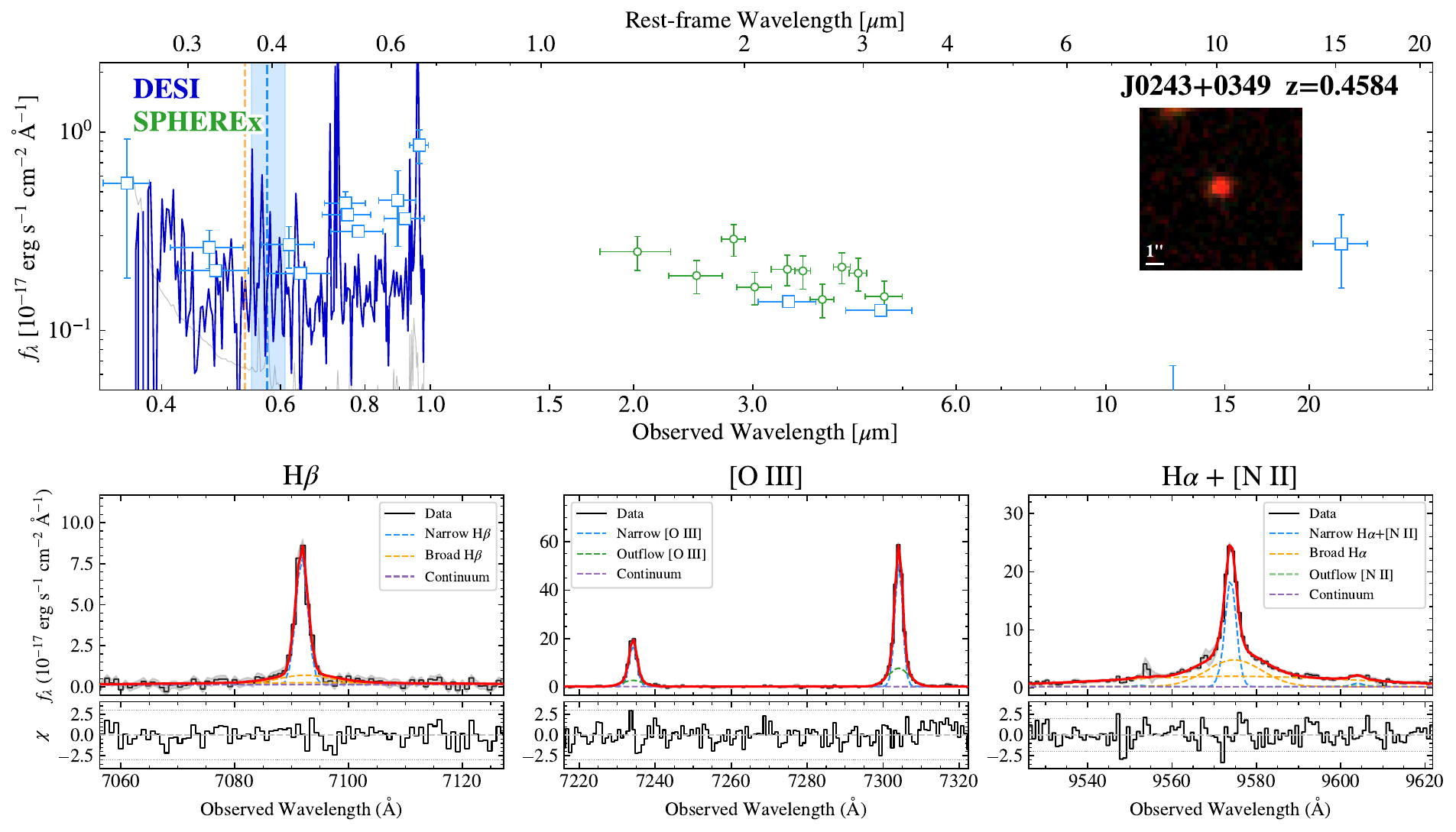}

\sedpair
{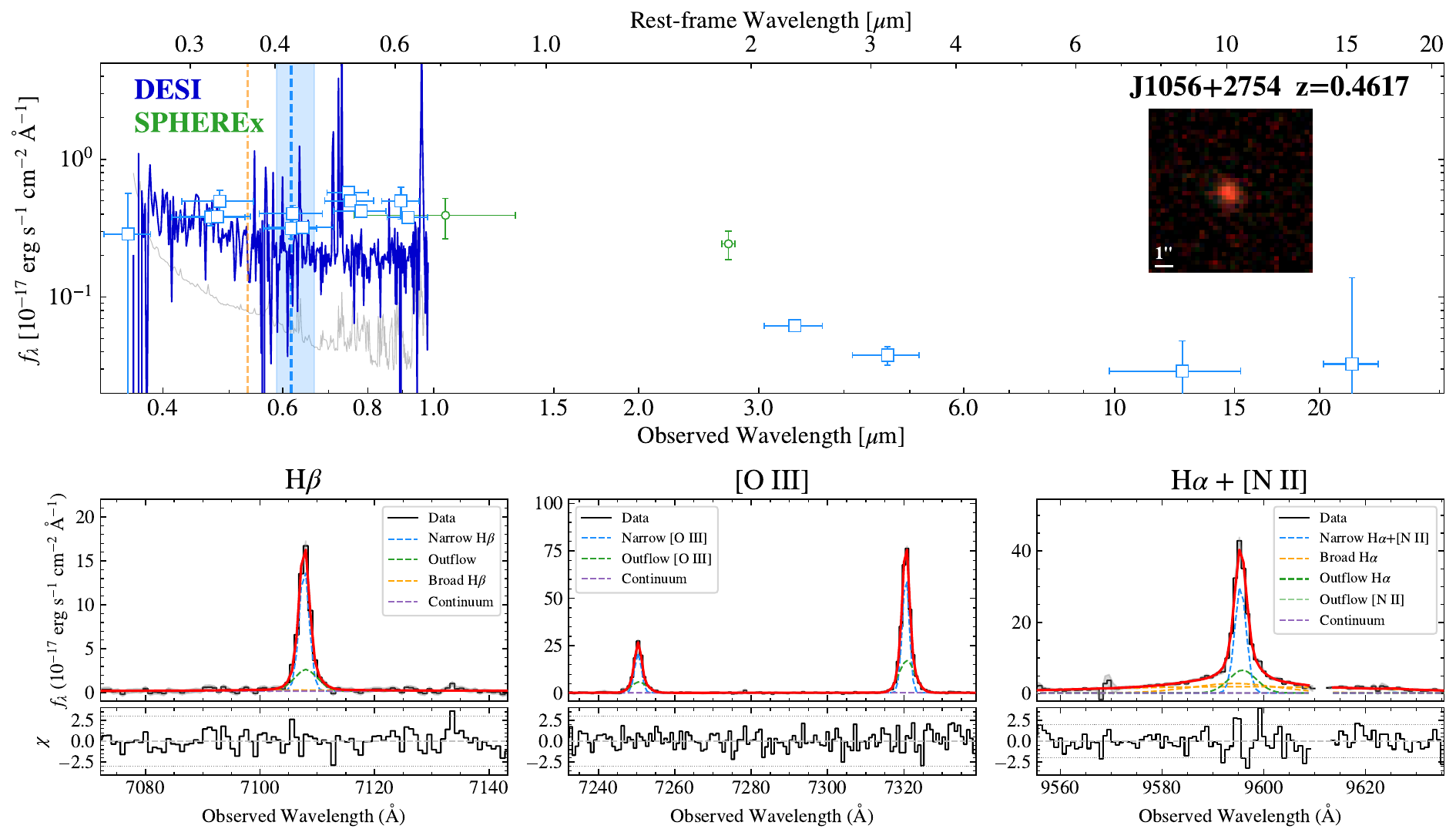}
{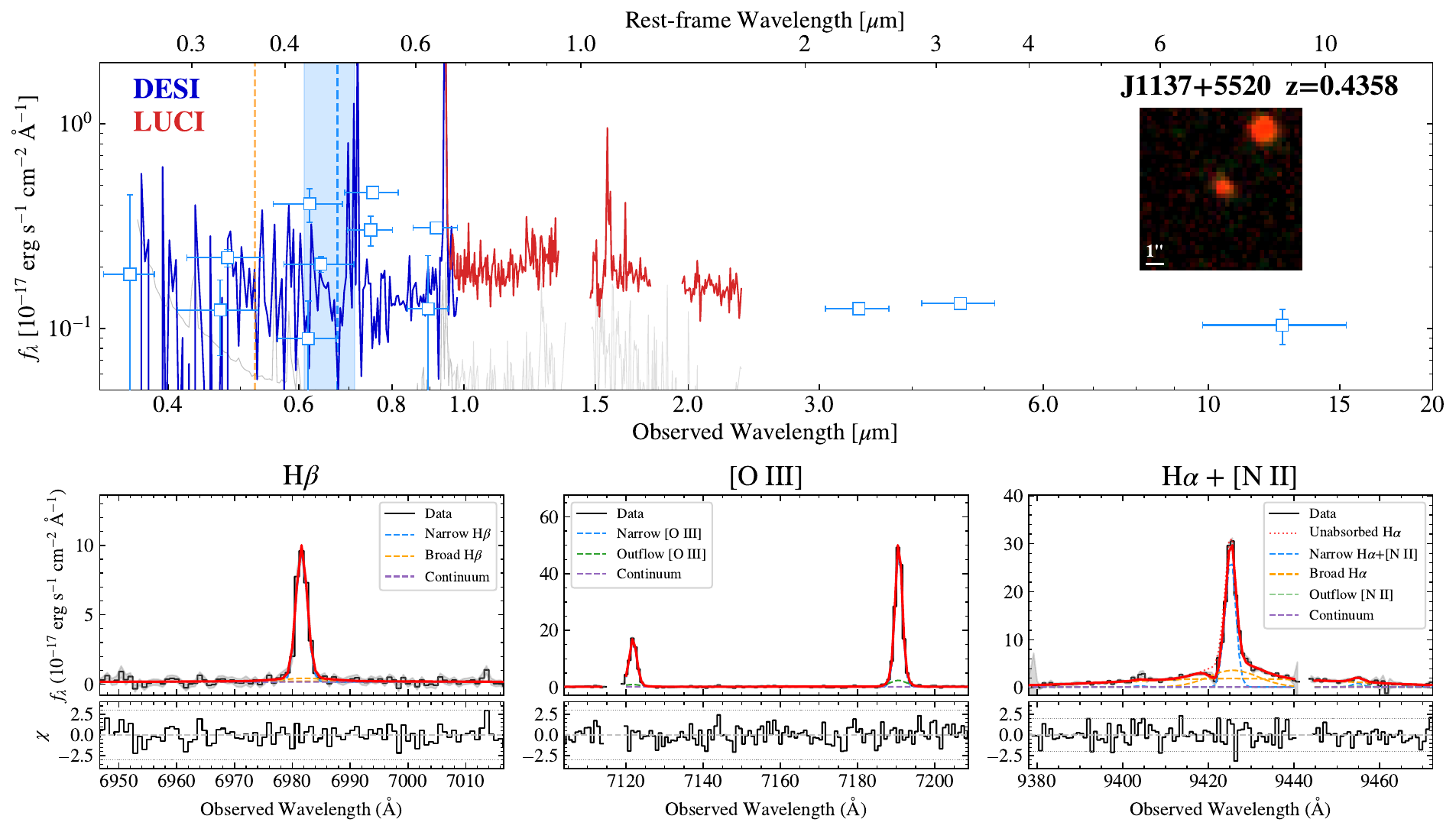}

\sedpair
{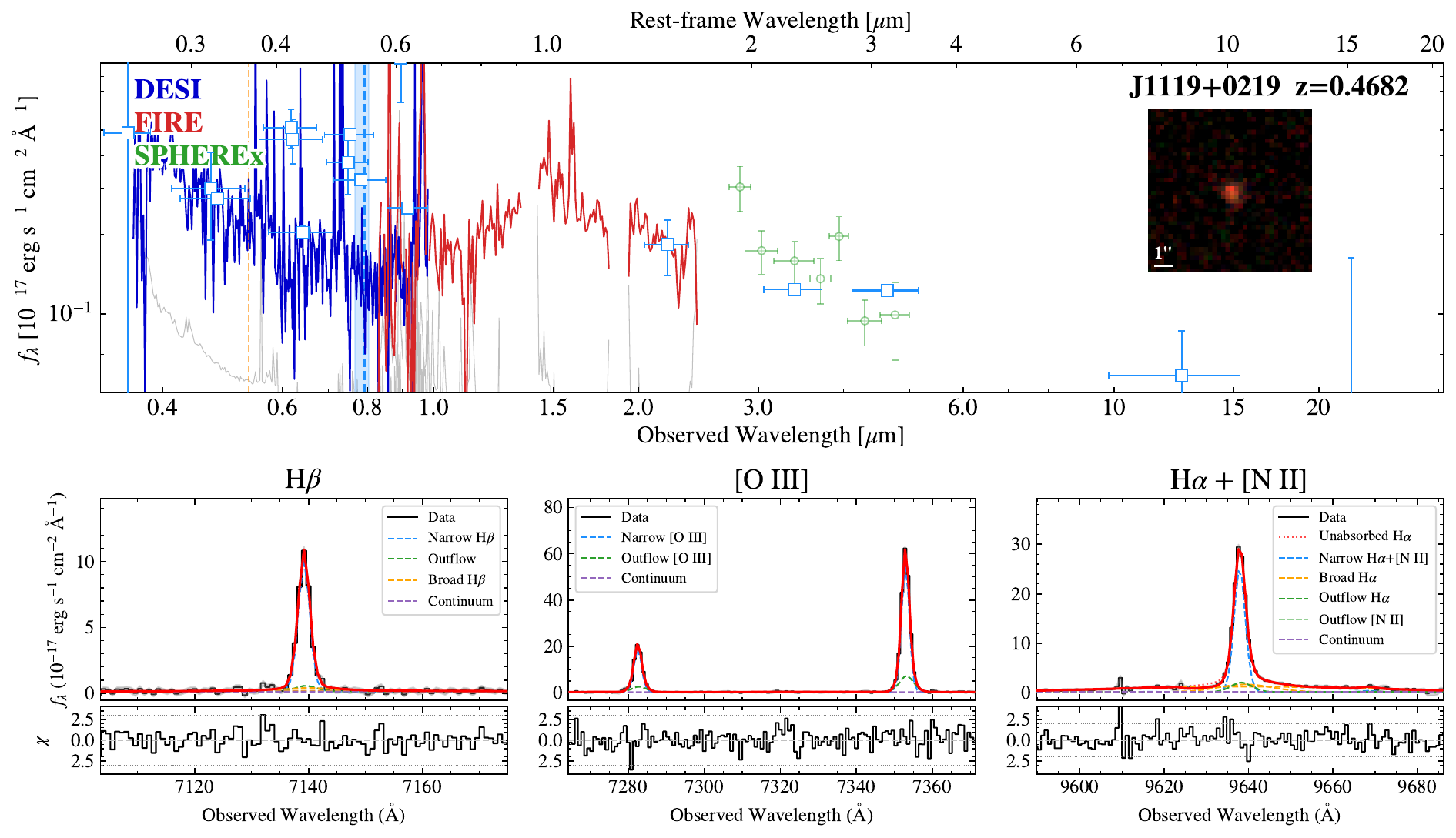}
{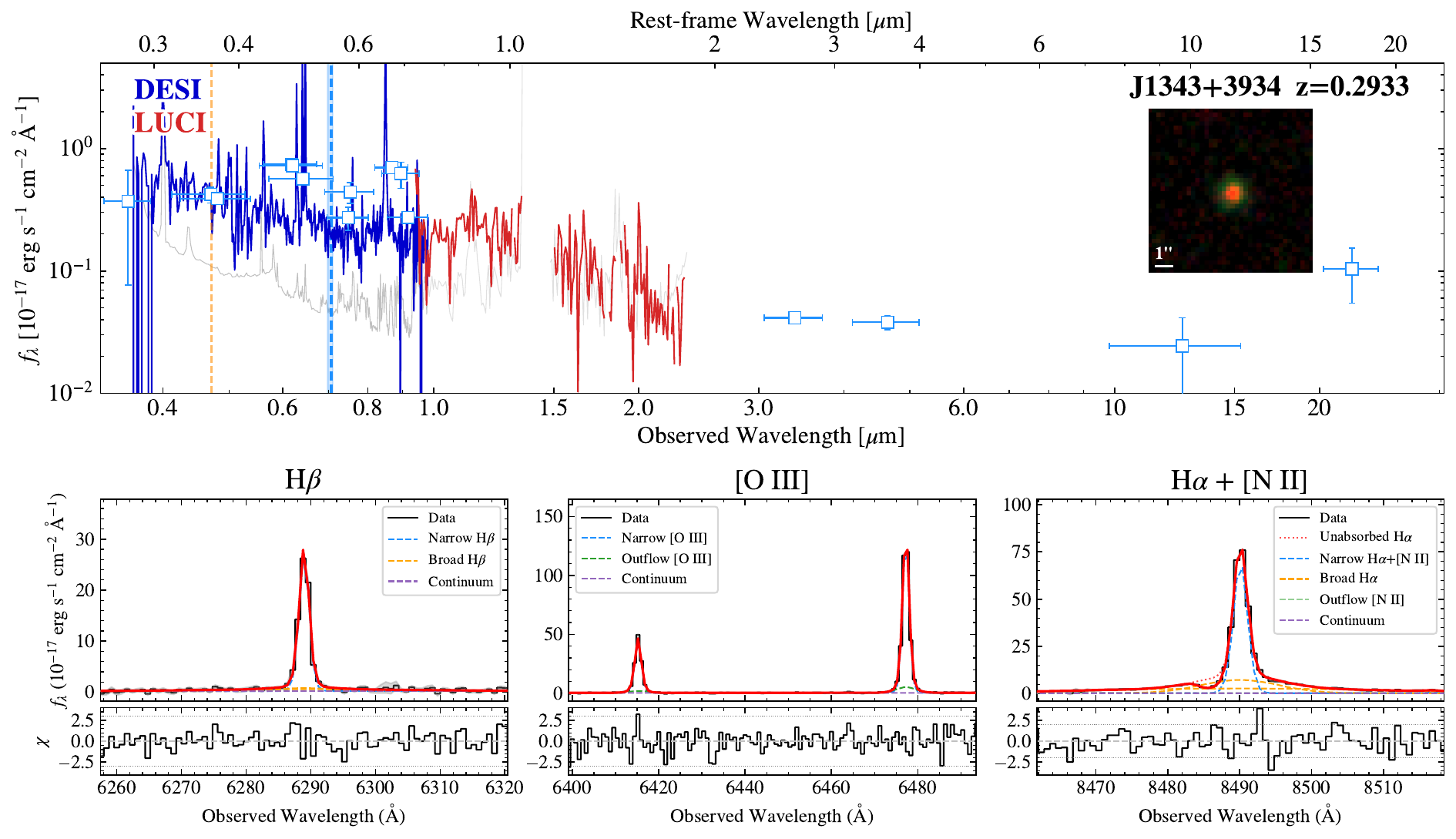}

\sedpair
{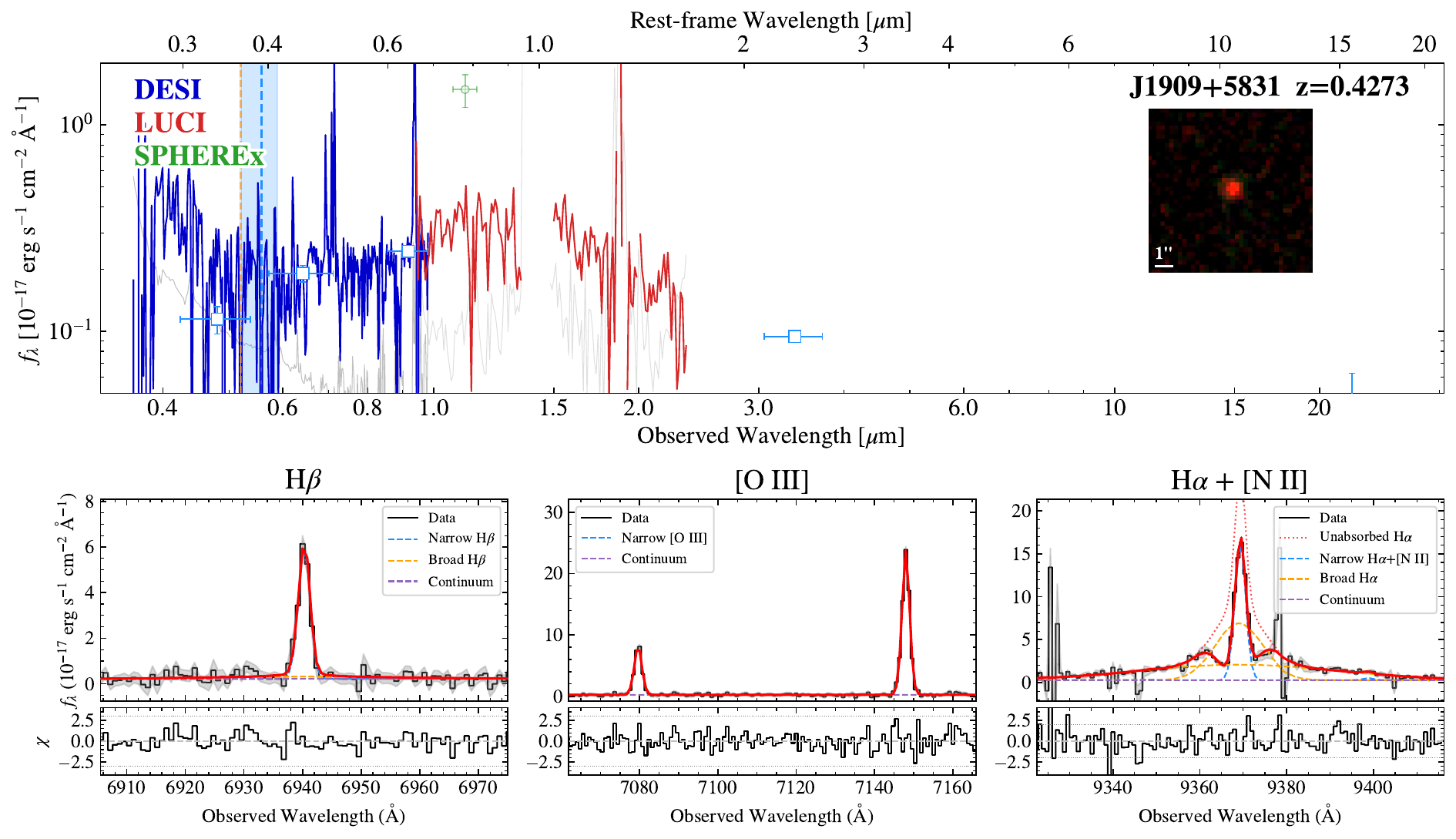}
{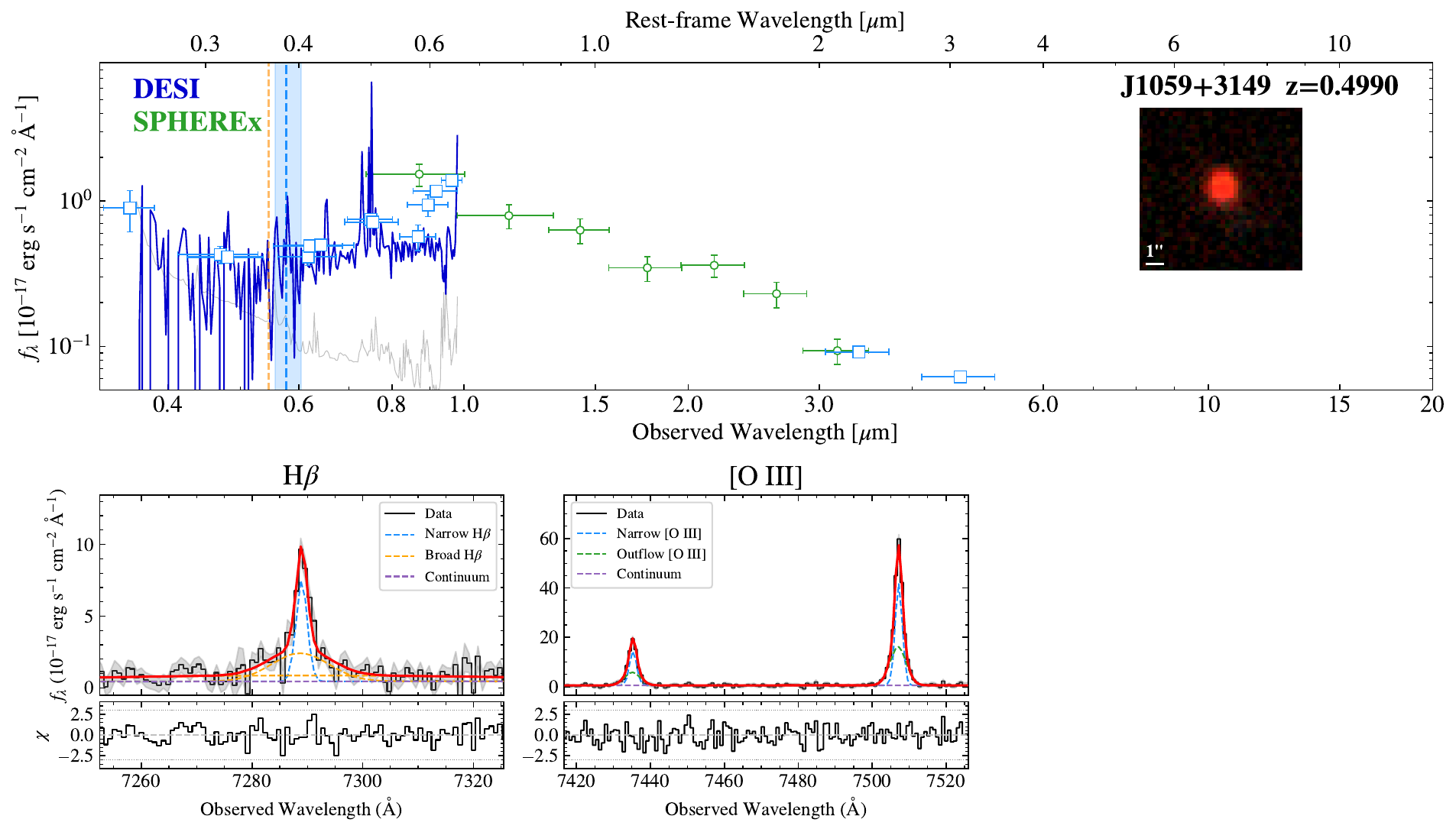}

\figcaption{
\textsc{Silver} sample: DESI DR1 LRDs with
$k_{\rm red}>0$ at $<3\sigma$ significance.
Similar to Figure~\ref{fig:example1}.
Classification into the \textsc{Silver} tier does not preclude LRD
nature; four of these eight sources exhibit Balmer absorption
features characteristic of LRDs.
\label{fig:sample_silver}
}

\clearpage
\section{The {Measured Properties}}\label{appendix:fitting_results}

\begin{figure*}
    \centering
    \includegraphics[width=0.93\linewidth]{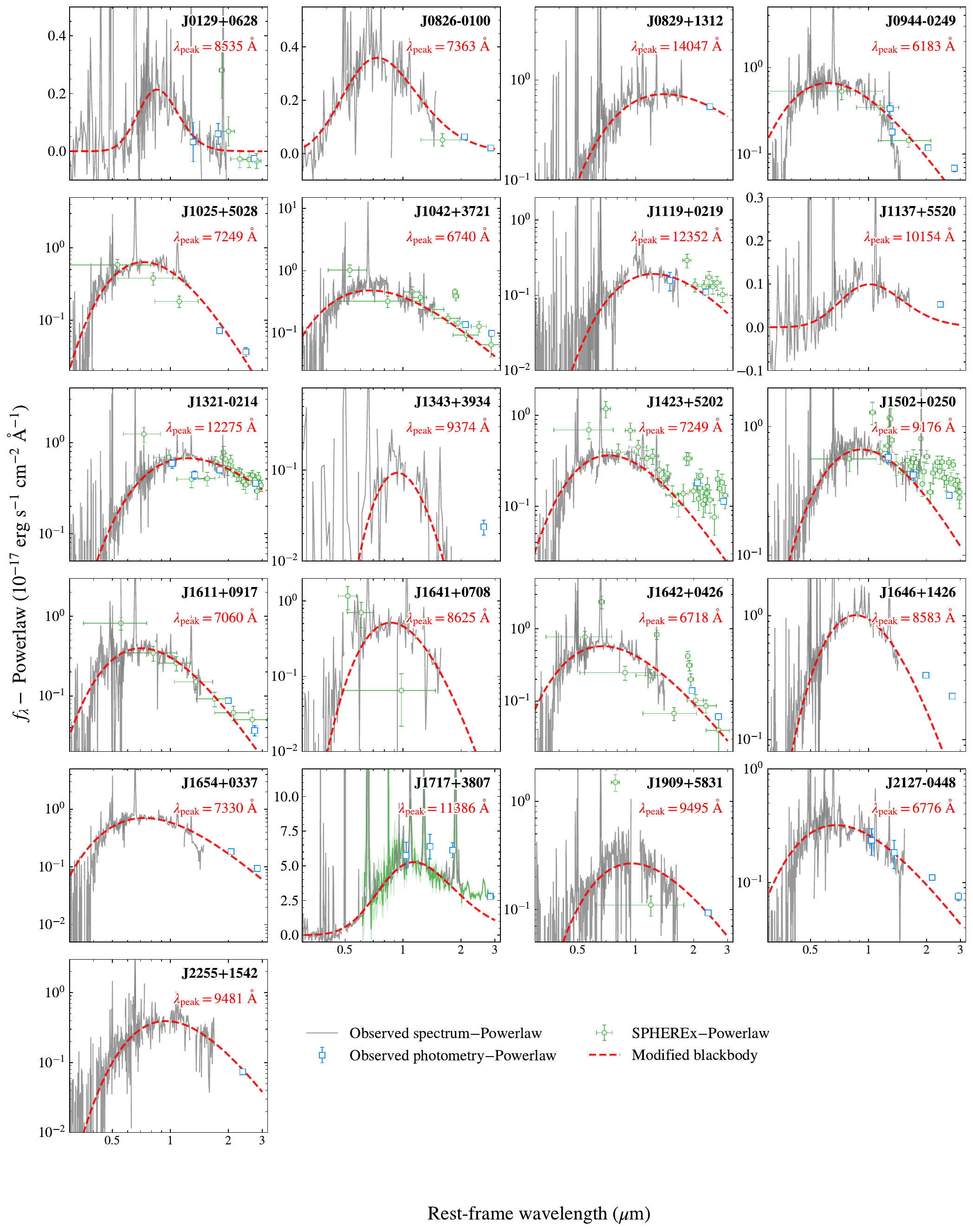}
    \caption{Targets in our sample with near-infrared spectra or photometry that sufficiently sample their blackbody-like continua and allow the peak wavelength to be determined. The spectrum and photometry shown for each object (gray lines and squares) have been subtracted by the power-law component fitted to the continuum blueward of the inflection points.}
    \label{fig:bb_fits}
\end{figure*}

{Table~\ref{tab:emission_line_properties} lists the measured emission-line properties of the DESI DR1 LRDs (see Section~\ref{sec:emission_line_measurement}), while Table~\ref{tab:outflow} gives the ionized gas outflow measurements (see Section~\ref{sec:ionized_outflow}). Table~\ref{tab:bb_fit} reports the modified blackbody fitting results, and Figure~\ref{fig:bb_fits} shows the corresponding fits for sources with sufficient near-IR coverage (see Section~\ref{sec:hr_diagram}).}

\begin{table*}[ht]
\centering
\setlength{\tabcolsep}{2.75pt}
\noindent\makebox[\linewidth][c]{\input{table/emission_line_properties.tex}\unskip}\par
\caption{Emission-line properties of DESI DR1 LRDs. $n$ and $b$ denote the narrow and broad components, respectively. For each target, the \hb\ and \ha\ emission lines are modeled with two to three broad components {with corresponding FWHMs tied between the two lines}. {The listed broad-line FWHMs and luminosities} denote the composite FWHM and luminosity of the broad components, defined as the summed luminosity and the corresponding FWHM of all broad components. {The machine-readable version is available on Zenodo as \texttt{main\_lines\_table.ecsv}.}}
\label{tab:emission_line_properties}
\end{table*}

\begin{table*}[ht]
\centering
\noindent\makebox[\linewidth][c]{\input{table/outflow_properties.tex}\unskip}\par
\caption{{Measured Properties of the Ionized Gas Outflows.} The columns show, for each source: outflow velocity {offset}, [O~III] outflow luminosity, FWHM of the outflow, H$\beta$ and H$\alpha$ outflow luminosities. {The machine-readable version is available on Zenodo as \texttt{main\_lines\_table.ecsv}.} \label{tab:outflow}}
\end{table*}

\begin{table*}[ht]
\centering
\noindent\makebox[\linewidth][c]{\input{table/blackbody_properties.tex}\unskip}\par
\caption{{Modified Blackbody Fitting Results.} $\lambda_{\rm peak}$ is the rest-frame peak wavelength of the modified blackbody (MBB) fit; $T_{\rm peak}$ is derived from $\lambda_{\rm peak}$ via Wien’s law. $\log L_{\rm bb}$ denotes the bolometric luminosity of the MBB. $T_{\rm MB}$ and $\beta_{\rm MB}$  are the fitted parameters of the modified blackbody (namely the characteristic temperature and emissivity index, respectively).  
Note: $T_{\rm MB}$ and $\beta_{\rm MB}$  are used here purely as a phenomenological description, {and they are not assigned} further physical interpretation. {The machine-readable version is available on Zenodo as \texttt{continuum\_table.ecsv}.} }
\label{tab:bb_fit}
\end{table*}

\clearpage

\clearpage
\section{Light curves of J1646+1426 and J1909+5831}\label{appendix:lc}

In Section \ref{sec:variability}, we present a variability analysis of our DESI LRD sample. In addition to the robustly variable source J1717+3807, J1646+1426 marginally satisfies the variability criteria in the ZTF $i$ band and WISE W2, but only at $\sim 3\sigma$ significance. Its variability requires further confirmation. Its light curves are shown in Figure \ref{fig:J1646_lc}.

J1909+5831’s WISE W1 light curve also passes the variability criteria. As shown in Figure \ref{fig:J1909_lc}, J1909+5831 exhibits a brightening by  $1.0\pm0.2$ mag from {2010–2014 to 2016–2019}. However, we note that the light curve is derived from forced photometry anchored to the source position in the Legacy Survey images and measured on the unWISE coadds. The source is too faint to be reliably detected in individual WISE single-exposure frames. The light curve should be {re-measured} using more robust photometric methods to ensure a reliable variability assessment.

\begin{center}
    \includegraphics[width=0.75\linewidth]{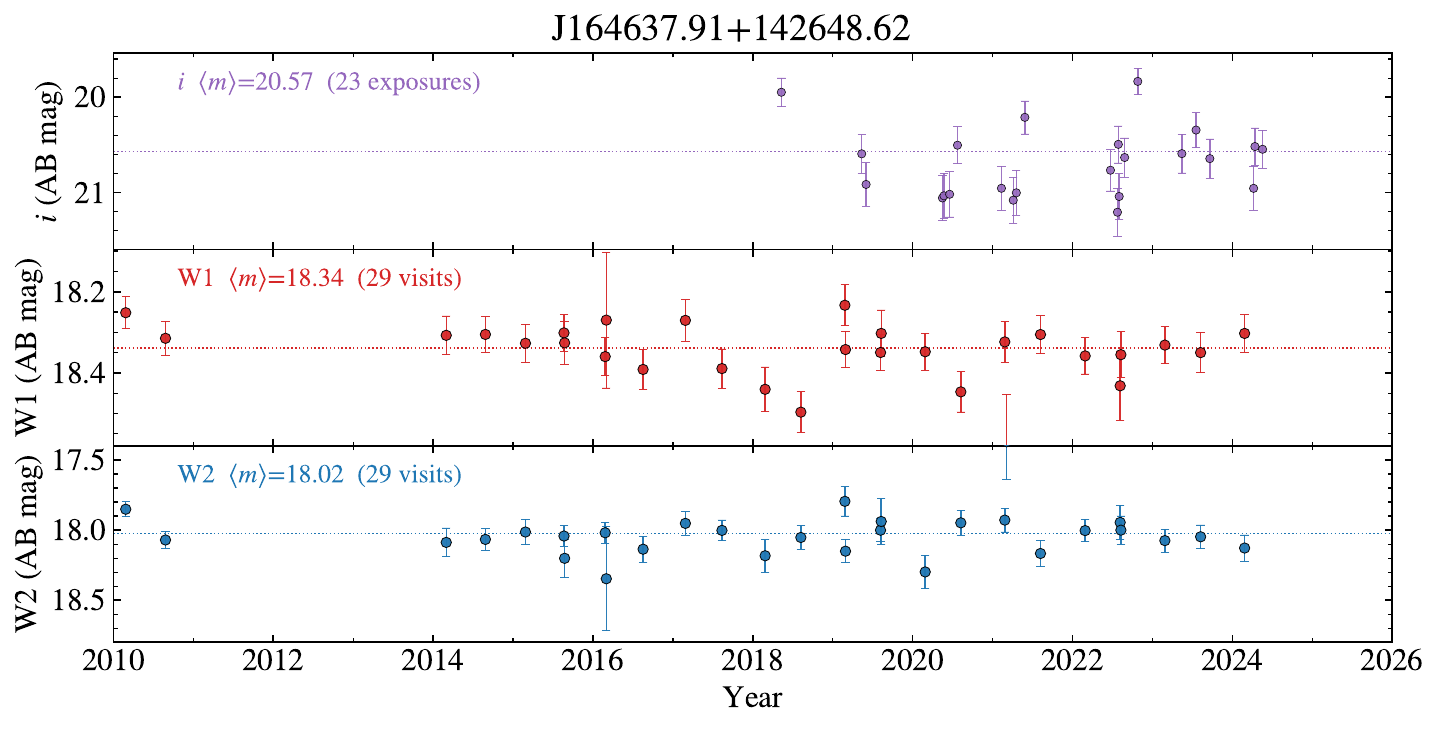}
    \captionof{figure}{The ZTF $i$-band and WISE W1 and W2 light curves of J1646+1426. The $i$-band variability is based on only 23 exposures {and, thus, lacks} statistical significance. The W2 light curve, although marginally passing our selection criteria, shows low significance (only $3\sigma$).}
    \label{fig:J1646_lc}
\end{center}

\begin{center}
    \includegraphics[width=0.75\linewidth]{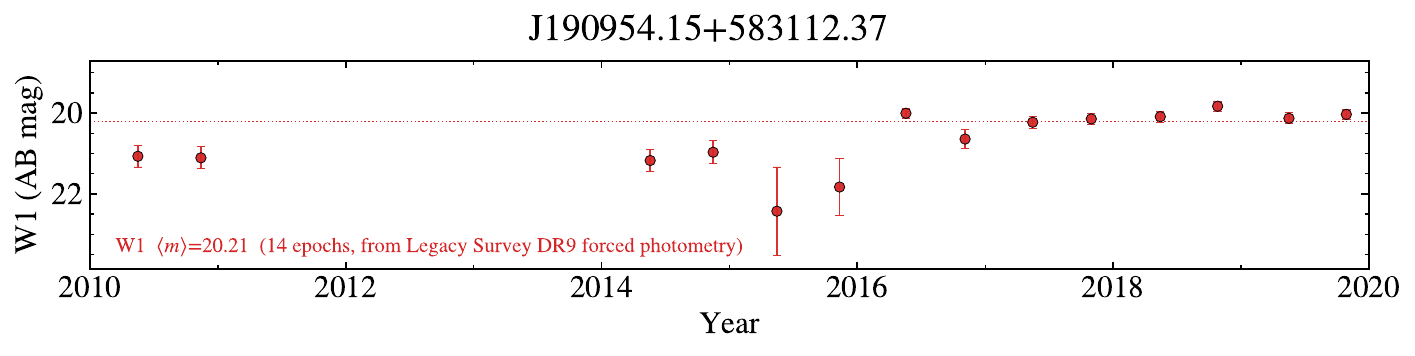}
\captionof{figure}{The W1 light curve of J1909+5831.}
\label{fig:J1909_lc}
\end{center}

\clearpage

%% file: table/emission_line_properties.tex
\begin{tabular}{lcccccccc}
\hline
Name & FWHM$_{\rm n}$ & $L_{\mathrm{[O\,III], n}}$ & $L_{\mathrm{H}\beta, {\rm n}}$ & FWHM$_{\mathrm{H}\beta, {\rm b}}$ & $L_{\mathrm{H}\beta, {\rm b}}$ & $L_{\mathrm{H}\alpha, {\rm n}}$ & FWHM$_{\mathrm{H}\alpha, {\rm b}}$ & $L_{\mathrm{H}\alpha, {\rm b}}$ \\
 & (km s$^{-1}$) & ($10^{40}$ erg s$^{-1}$) & ($10^{40}$ erg s$^{-1}$) & (km s$^{-1}$) & ($10^{40}$ erg s$^{-1}$) & ($10^{40}$ erg s$^{-1}$) & (km s$^{-1}$) & ($10^{40}$ erg s$^{-1}$) \\
\hline
\multicolumn{9}{c}{\textbf{Gold sample}} \\
\hline
J0129+0628 & 50$_{-1}^{+1}$ & 45.1$_{-1.3}^{+1.1}$ & 10.4$_{-0.4}^{+0.4}$ & 978$_{-252}^{+905}$ & 5.2$_{-1.3}^{+1.6}$ & 34.2$_{-1.0}^{+1.0}$ & 750$_{-30}^{+29}$ & 142.8$_{-4.1}^{+6.5}$ \\
J0826--0100 & 85$_{-2}^{+2}$ & 87.6$_{-1.4}^{+1.4}$ & 22.4$_{-0.9}^{+1.2}$ & 681$_{-148}^{+454}$ & 76.8$_{-13.9}^{+11.6}$ & 104.2$_{-8.4}^{+6.2}$ & 665$_{-91}^{+106}$ & 887.1$_{-88.8}^{+90.4}$ \\
J0829+1312 & 96$_{-1}^{+1}$ & 282.1$_{-3.7}^{+3.0}$ & 56.7$_{-0.9}^{+0.9}$ & 577$_{-47}^{+66}$ & 19.0$_{-1.8}^{+2.1}$ & 180.9$_{-2.0}^{+2.2}$ & 629$_{-79}^{+103}$ & 451.8$_{-48.9}^{+61.9}$ \\
J0944--0249 & 85$_{-5}^{+6}$ & 90.4$_{-6.1}^{+6.7}$ & 18.6$_{-1.6}^{+1.6}$ & 903$_{-116}^{+124}$ & 299.6$_{-9.3}^{+10.0}$ & 87.9$_{-11.8}^{+11.5}$ & 551$_{-20}^{+19}$ & 3275.8$_{-73.1}^{+78.5}$ \\
J1017+3114 & 78$_{-1}^{+1}$ & 184.6$_{-2.9}^{+2.9}$ & 32.8$_{-0.9}^{+0.9}$ & 1961$_{-174}^{+159}$ & 45.5$_{-5.0}^{+6.3}$ & -- & -- & -- \\
J1025+5028 & 97$_{-5}^{+5}$ & 159.5$_{-9.6}^{+8.5}$ & 31.3$_{-2.8}^{+2.7}$ & 698$_{-104}^{+402}$ & 415.7$_{-50.9}^{+59.7}$ & 157.3$_{-36.2}^{+26.7}$ & 706$_{-46}^{+43}$ & 5922.3$_{-110.8}^{+130.0}$ \\
J1042+3721 & 84$_{-3}^{+3}$ & 132.5$_{-4.1}^{+3.9}$ & 15.0$_{-1.0}^{+1.1}$ & 1948$_{-184}^{+236}$ & 104.4$_{-7.6}^{+7.5}$ & 85.3$_{-24.5}^{+20.5}$ & 1556$_{-89}^{+85}$ & 1432.4$_{-22.5}^{+25.1}$ \\
J1321--0214 & 79$_{-1}^{+1}$ & 32.4$_{-0.2}^{+0.3}$ & 6.4$_{-0.1}^{+0.2}$ & 630$_{-48}^{+175}$ & 1.1$_{-0.3}^{+0.4}$ & 19.9$_{-0.2}^{+0.2}$ & 785$_{-29}^{+25}$ & 32.7$_{-0.5}^{+0.4}$ \\
J1423+5202 & 72$_{-2}^{+2}$ & 90.7$_{-2.2}^{+2.1}$ & 14.6$_{-0.6}^{+0.6}$ & 1099$_{-158}^{+347}$ & 73.8$_{-9.0}^{+13.6}$ & 43.7$_{-1.8}^{+2.0}$ & 1218$_{-39}^{+39}$ & 796.4$_{-10.6}^{+12.1}$ \\
J1502+0250 & 73$_{-2}^{+2}$ & 41.5$_{-1.0}^{+0.9}$ & 7.5$_{-0.2}^{+0.2}$ & 599$_{-50}^{+110}$ & 2.6$_{-0.6}^{+0.7}$ & 22.5$_{-0.4}^{+0.4}$ & 690$_{-15}^{+15}$ & 101.6$_{-1.0}^{+1.0}$ \\
J1611+0917 & 91$_{-4}^{+4}$ & 49.2$_{-1.3}^{+1.3}$ & 29.4$_{-2.4}^{+2.0}$ & 1233$_{-120}^{+165}$ & 193.0$_{-7.6}^{+8.8}$ & 115.3$_{-6.6}^{+5.9}$ & 939$_{-12}^{+12}$ & 2268.3$_{-18.0}^{+16.9}$ \\
J1620+3148 & 63$_{-2}^{+2}$ & 36.8$_{-0.7}^{+0.8}$ & 6.0$_{-0.4}^{+0.4}$ & 1588$_{-419}^{+1026}$ & 3.9$_{-1.6}^{+3.2}$ & 21.9$_{-0.5}^{+0.5}$ & 1397$_{-149}^{+145}$ & 37.9$_{-4.6}^{+10.4}$ \\
J1641+0708 & 66$_{-6}^{+6}$ & -- & 13.1$_{-0.8}^{+0.8}$ & 1467$_{-342}^{+381}$ & 34.6$_{-4.6}^{+4.9}$ & 73.8$_{-9.5}^{+9.4}$ & 830$_{-50}^{+47}$ & 1012.8$_{-36.8}^{+41.0}$ \\
J1642+0426 & 82$_{-2}^{+3}$ & 331.7$_{-6.2}^{+7.4}$ & 54.7$_{-3.2}^{+7.9}$ & 486$_{-42}^{+483}$ & 388.4$_{-22.2}^{+16.5}$ & 163.8$_{-10.7}^{+24.1}$ & 357$_{-18}^{+34}$ & 4340.8$_{-109.2}^{+125.1}$ \\
J1646+1426 & 124$_{-3}^{+3}$ & 369.1$_{-8.6}^{+10.9}$ & 56.9$_{-3.3}^{+2.4}$ & 1110$_{-43}^{+56}$ & 288.3$_{-13.3}^{+10.3}$ & 199.7$_{-10.0}^{+10.1}$ & 1125$_{-15}^{+16}$ & 3046.1$_{-51.0}^{+38.8}$ \\
J1654+0337 & 60$_{-2}^{+3}$ & 99.2$_{-3.4}^{+3.9}$ & 19.2$_{-1.0}^{+1.1}$ & 1297$_{-304}^{+480}$ & 140.1$_{-14.4}^{+11.0}$ & 86.9$_{-7.4}^{+7.2}$ & 980$_{-46}^{+47}$ & 1805.7$_{-40.9}^{+38.6}$ \\
J1717+3807 & 127$_{-1}^{+1}$ & 224.9$_{-2.1}^{+2.0}$ & 42.4$_{-0.9}^{+0.9}$ & 892$_{-87}^{+121}$ & 20.1$_{-1.6}^{+1.7}$ & 136.3$_{-1.5}^{+1.5}$ & 1326$_{-58}^{+60}$ & 427.7$_{-5.3}^{+5.5}$ \\
J2127--0448 & 70$_{-5}^{+5}$ & 78.2$_{-4.8}^{+4.3}$ & 13.2$_{-1.1}^{+1.1}$ & 1144$_{-306}^{+362}$ & 72.9$_{-9.1}^{+12.9}$ & 43.5$_{-4.5}^{+6.9}$ & 937$_{-163}^{+137}$ & 474.6$_{-33.2}^{+40.1}$ \\
J2255+1542 & 62$_{-2}^{+2}$ & 23.2$_{-0.4}^{+0.5}$ & 5.8$_{-0.4}^{+0.5}$ & 1923$_{-555}^{+345}$ & 5.9$_{-1.5}^{+1.6}$ & 22.9$_{-1.0}^{+1.0}$ & 812$_{-94}^{+96}$ & 117.0$_{-6.3}^{+10.3}$ \\
\hline
\multicolumn{9}{c}{\textbf{Silver sample}} \\
\hline
J0009+0811 & 104$_{-2}^{+2}$ & 149.5$_{-3.6}^{+3.3}$ & 23.4$_{-0.6}^{+0.6}$ & 466$_{-52}^{+83}$ & 4.7$_{-0.9}^{+1.0}$ & 81.4$_{-1.7}^{+1.5}$ & 538$_{-39}^{+41}$ & 87.5$_{-2.2}^{+2.2}$ \\
J0243+0349 & 85$_{-2}^{+2}$ & 114.7$_{-2.9}^{+2.8}$ & 17.6$_{-0.6}^{+0.5}$ & 552$_{-41}^{+54}$ & 11.5$_{-1.9}^{+1.9}$ & 53.4$_{-1.4}^{+1.3}$ & 612$_{-27}^{+27}$ & 175.4$_{-4.1}^{+4.2}$ \\
J1056+2754 & 69$_{-3}^{+2}$ & 125.0$_{-5.6}^{+5.3}$ & 28.1$_{-1.3}^{+1.3}$ & 969$_{-261}^{+726}$ & 5.6$_{-1.8}^{+1.9}$ & 75.9$_{-3.4}^{+3.1}$ & 900$_{-76}^{+84}$ & 160.2$_{-5.3}^{+5.3}$ \\
J1059+3149 & 77$_{-8}^{+7}$ & 107.8$_{-14.6}^{+14.1}$ & 18.5$_{-2.7}^{+2.4}$ & 585$_{-144}^{+172}$ & 67.9$_{-14.4}^{+19.3}$ & -- & -- & -- \\
J1119+0219 & 85$_{-2}^{+2}$ & 130.7$_{-7.0}^{+6.7}$ & 23.2$_{-1.2}^{+1.0}$ & 573$_{-123}^{+233}$ & 6.3$_{-1.6}^{+1.4}$ & 76.4$_{-4.2}^{+3.0}$ & 730$_{-132}^{+176}$ & 93.8$_{-5.1}^{+4.7}$ \\
J1137+5520 & 75$_{-2}^{+2}$ & 91.0$_{-4.6}^{+3.0}$ & 17.3$_{-0.7}^{+0.6}$ & 497$_{-64}^{+126}$ & 4.1$_{-1.2}^{+1.5}$ & 63.0$_{-3.2}^{+5.2}$ & 589$_{-52}^{+57}$ & 125.4$_{-4.6}^{+4.9}$ \\
J1343+3934 & 64$_{-1}^{+1}$ & 83.9$_{-1.0}^{+1.0}$ & 17.2$_{-0.4}^{+0.4}$ & 627$_{-91}^{+199}$ & 5.6$_{-0.9}^{+0.8}$ & 54.6$_{-0.7}^{+0.6}$ & 550$_{-21}^{+20}$ & 65.8$_{-1.3}^{+1.4}$ \\
J1909+5831 & 80$_{-1}^{+1}$ & 42.0$_{-0.5}^{+0.5}$ & 10.3$_{-0.5}^{+0.5}$ & 740$_{-252}^{+663}$ & 3.9$_{-1.5}^{+1.6}$ & 36.1$_{-1.0}^{+1.0}$ & 488$_{-95}^{+139}$ & 136.5$_{-23.1}^{+59.9}$ \\
\hline
\end{tabular}

%% file: table/outflow_properties.tex
\begin{tabular}{lccccc}
\hline
Name & $\Delta v_{\rm out}$ & $L_{\rm [OIII], out}$ & FWHM$_{\rm out}$ & $L_{\rm H\beta, out}$ & $L_{\rm H\alpha, out}$ \\
 & (km s$^{-1}$) & ($10^{40}$ erg s$^{-1}$) & (km s$^{-1}$) & ($10^{40}$ erg s$^{-1}$) & ($10^{40}$ erg s$^{-1}$) \\
\hline
\multicolumn{6}{c}{\textbf{Gold sample}} \\
\hline
J0129+0628 & -21$_{-3}^{+2}$ & 16.0$_{-1.0}^{+1.1}$ & 173$_{-7}^{+7}$ & 2.5$_{-0.5}^{+0.4}$ & 11.2$_{-2.0}^{+2.0}$ \\
J0826--0100 & -- & -- & -- & -- & -- \\
J0829+1312 & 5$_{-2}^{+2}$ & 95.6$_{-2.7}^{+3.2}$ & 302$_{-8}^{+6}$ & -- & -- \\
J0944--0249 & -24$_{-3}^{+3}$ & 130.5$_{-6.3}^{+6.0}$ & 262$_{-8}^{+8}$ & -- & -- \\
J1017+3114 & -27$_{-2}^{+2}$ & 105.8$_{-2.5}^{+2.7}$ & 268$_{-5}^{+5}$ & 13.2$_{-1.3}^{+1.2}$ & -- \\
J1025+5028 & 37$_{-27}^{+35}$ & 47.8$_{-10.4}^{+10.9}$ & 397$_{-67}^{+89}$ & -- & -- \\
J1042+3721 & -13$_{-5}^{+5}$ & 60.7$_{-3.7}^{+3.6}$ & 311$_{-16}^{+18}$ & -- & -- \\
J1321--0214 & -- & -- & -- & -- & -- \\
J1423+5202 & -28$_{-3}^{+3}$ & 53.4$_{-1.8}^{+2.0}$ & 296$_{-10}^{+10}$ & -- & -- \\
J1502+0250 & -9$_{-6}^{+5}$ & 8.5$_{-0.9}^{+0.9}$ & 244$_{-19}^{+21}$ & -- & -- \\
J1611+0917 & -- & -- & -- & -- & -- \\
J1620+3148 & -85$_{-61}^{+48}$ & 5.6$_{-1.2}^{+1.5}$ & 658$_{-179}^{+325}$ & -- & -- \\
J1641+0708 & -- & -- & -- & -- & -- \\
J1642+0426 & -28$_{-6}^{+6}$ & 148.6$_{-7.3}^{+6.9}$ & 445$_{-20}^{+21}$ & -- & -- \\
J1646+1426 & -59$_{-5}^{+5}$ & 148.4$_{-10.3}^{+8.1}$ & 376$_{-18}^{+23}$ & -- & -- \\
J1654+0337 & -37$_{-10}^{+6}$ & 31.4$_{-3.8}^{+3.0}$ & 216$_{-13}^{+20}$ & -- & -- \\
J1717+3807 & 31$_{-1}^{+1}$ & 110.4$_{-1.7}^{+1.9}$ & 348$_{-4}^{+3}$ & 11.4$_{-1.3}^{+1.3}$ & 37.5$_{-3.6}^{+4.0}$ \\
J2127--0448 & -8$_{-3}^{+3}$ & 77.2$_{-4.7}^{+4.7}$ & 245$_{-10}^{+8}$ & -- & -- \\
J2255+1542 & -- & -- & -- & -- & -- \\
\hline
\multicolumn{6}{c}{\textbf{Silver sample}} \\
\hline
J0009+0811 & 7$_{-5}^{+5}$ & 22.4$_{-3.0}^{+3.4}$ & 268$_{-21}^{+25}$ & -- & -- \\
J0243+0349 & -0.5$_{-3}^{+3}$ & 44.0$_{-2.5}^{+2.7}$ & 263$_{-12}^{+13}$ & -- & -- \\
J1056+2754 & 13$_{-2}^{+3}$ & 75.8$_{-5.3}^{+5.7}$ & 195$_{-7}^{+8}$ & 10.7$_{-1.4}^{+1.4}$ & 37.2$_{-4.1}^{+3.9}$ \\
J1059+3149 & -7$_{-5}^{+4}$ & 87.6$_{-13.6}^{+13.9}$ & 197$_{-15}^{+18}$ & -- & -- \\
J1119+0219 & 13$_{-3}^{+3}$ & 29.5$_{-5.7}^{+7.1}$ & 177$_{-19}^{+19}$ & 1.8$_{-1.4}^{+1.3}$ & 10.6$_{-3.3}^{+4.2}$ \\
J1137+5520 & -3$_{-8}^{+10}$ & 9.5$_{-2.7}^{+4.1}$ & 215$_{-44}^{+58}$ & -- & -- \\
J1343+3934 & 0.0$_{-4}^{+3}$ & 8.3$_{-0.9}^{+0.9}$ & 245$_{-20}^{+23}$ & -- & -- \\
J1909+5831 & -- & -- & -- & -- & -- \\
\hline
\end{tabular}

%% file: table/blackbody_properties.tex
\begin{tabular}{lccccc}
\hline
Name & $\lambda_{\rm peak}$ & $T_{\rm peak}$ & $\log L_{\rm bb}$ & $T_{\rm MB}$ & $\beta_{\rm MB}$ \\
 & (\AA) & (K) & ($\mathrm{erg\,s^{-1}}$) & (K) &  \\
\hline
\multicolumn{6}{c}{\textbf{Gold sample}} \\
\hline
J0129+0628 & 8535$_{-81}^{+83}$ & 3396$_{-33}^{+33}$ & 42.46$_{-0.02}^{+0.02}$ & 979$_{-70}^{+82}$ & 12.22$_{-1.30}^{+1.25}$ \\
J0826--0100 & 7363$_{-62}^{+62}$ & 3936$_{-33}^{+33}$ & 44.05$_{-0.01}^{+0.01}$ & 3791$_{-132}^{+133}$ & 0.19$_{-0.16}^{+0.16}$ \\
J0829+1312 & 14047$_{-109}^{+114}$ & 2063$_{-17}^{+16}$ & 44.45$_{-0.02}^{+0.02}$ & 3734$_{-83}^{+83}$ & -2.07$_{-0.06}^{+0.07}$ \\
J0944--0249 & 6183$_{-39}^{+37}$ & 4687$_{-28}^{+30}$ & 44.36$_{-0.01}^{+0.01}$ & 5135$_{-125}^{+122}$ & -0.42$_{-0.09}^{+0.09}$ \\
J1025+5028 & 7249$_{-33}^{+35}$ & 3998$_{-19}^{+18}$ & 44.63$_{-0.01}^{+0.01}$ & 3096$_{-64}^{+66}$ & 1.42$_{-0.13}^{+0.12}$ \\
J1042+3721 & 6740$_{-44}^{+41}$ & 4300$_{-26}^{+28}$ & 44.22$_{-0.01}^{+0.01}$ & 5403$_{-146}^{+148}$ & -0.97$_{-0.09}^{+0.09}$ \\
J1321--0214 & 12275$_{-48}^{+48}$ & 2361$_{-9}^{+9}$ & 43.64$_{-0.01}^{+0.01}$ & 3760$_{-51}^{+52}$ & -1.74$_{-0.04}^{+0.04}$ \\
J1423+5202 & 7249$_{-30}^{+28}$ & 3998$_{-16}^{+17}$ & 44.08$_{-0.01}^{+0.01}$ & 4298$_{-94}^{+92}$ & -0.34$_{-0.09}^{+0.09}$ \\
J1502+0250 & 9176$_{-33}^{+33}$ & 3158$_{-11}^{+11}$ & 43.63$_{-0.01}^{+0.01}$ & 3894$_{-75}^{+74}$ & -0.90$_{-0.07}^{+0.07}$ \\
J1611+0917 & 7060$_{-32}^{+28}$ & 4105$_{-16}^{+18}$ & {44.203$_{-0.004}^{+0.004}$} & 3997$_{-65}^{+71}$ & 0.13$_{-0.07}^{+0.07}$ \\
J1641+0708 & 8625$_{-72}^{+73}$ & 3360$_{-28}^{+28}$ & 43.85$_{-0.02}^{+0.02}$ & 1583$_{-70}^{+71}$ & 5.53$_{-0.43}^{+0.50}$ \\
J1642+0426 & 6718$_{-28}^{+28}$ & 4314$_{-18}^{+18}$ & 44.44$_{-0.01}^{+0.01}$ & 4563$_{-88}^{+92}$ & -0.26$_{-0.08}^{+0.09}$ \\
J1646+1426 & 8583$_{-26}^{+23}$ & 3376$_{-9}^{+10}$ & {44.614$_{-0.005}^{+0.005}$} & 2518$_{-39}^{+42}$ & 1.66$_{-0.11}^{+0.11}$ \\
J1654+0337 & 7330$_{-26}^{+26}$ & 3954$_{-14}^{+14}$ & 44.45$_{-0.01}^{+0.01}$ & 4594$_{-67}^{+67}$ & -0.67$_{-0.06}^{+0.06}$ \\
J1717+3807 & 11386$_{-20}^{+18}$ & 2545$_{-4}^{+5}$ & {44.110$_{-0.004}^{+0.004}$} & 2562$_{-20}^{+21}$ & -0.03$_{-0.04}^{+0.04}$ \\
J2127--0448 & 6776$_{-61}^{+65}$ & 4277$_{-41}^{+39}$ & 44.09$_{-0.01}^{+0.02}$ & 6261$_{-211}^{+231}$ & -1.49$_{-0.10}^{+0.10}$ \\
J2255+1542 & 9481$_{-68}^{+68}$ & 3057$_{-22}^{+22}$ & 43.72$_{-0.01}^{+0.01}$ & 2834$_{-93}^{+100}$ & 0.38$_{-0.17}^{+0.18}$ \\
\hline
\multicolumn{6}{c}{\textbf{Silver sample}} \\
\hline
J1119+0219 & 12352$_{-143}^{+156}$ & 2346$_{-29}^{+28}$ & 43.72$_{-0.02}^{+0.02}$ & 2635$_{-98}^{+107}$ & -0.52$_{-0.18}^{+0.19}$ \\
J1137+5520 & 10154$_{-169}^{+187}$ & 2854$_{-52}^{+48}$ & 43.08$_{-0.03}^{+0.03}$ & 1984$_{-193}^{+237}$ & 2.15$_{-0.80}^{+0.81}$ \\
J1343+3934 & 9374$_{-211}^{+204}$ & 3092$_{-66}^{+71}$ & 42.32$_{-0.05}^{+0.04}$ & 881$_{-83}^{+172}$ & 12.48$_{-2.93}^{+1.81}$ \\
J1909+5831 & 9495$_{-93}^{+91}$ & 3052$_{-29}^{+30}$ & 43.71$_{-0.01}^{+0.02}$ & 3938$_{-162}^{+186}$ & -1.07$_{-0.15}^{+0.15}$ \\
\hline
\end{tabular}